\documentclass[useAMS,usenatbib,usegraphicx]{mn2e}
\usepackage{times}
\newcommand\kms{$\rm km\,s^{-1}$}
\newcommand{\sauron}{{\texttt {SAURON}}}

\newcommand{\Oiii}{[{\sc O$\,$iii}]}
\newcommand{\Ha}{H$\alpha$}
\newcommand{\Hb}{H$\beta$}
\newcommand{\Hi}{{\sc H$\,$i}}
\newcommand{\Hii}{{\sc H$\,$ii}}
\newcommand{\Ni}{[{\sc N$\,$i}]}
\newcommand{\Nii}{[{\sc N$\,$ii}]}

\newcommand{\Vg}{$V_{\rm gas}$}
\newcommand{\Sg}{$\sigma_{\rm gas}$}
\newcommand{\HST}{{\it HST\/}}
\newcommand{\eg}{e.g.,}

\newcommand{\apj}{ApJ}
\newcommand{\apjl}{ApJ}
\newcommand{\apjs}{ApJS}
\newcommand{\aj}{AJ}
\newcommand{\mnras}{MNRAS}
\newcommand{\aaps}{A\&AS}
\newcommand{\aap}{A\&A}
\newcommand{\pasp}{PASP}

\newcounter{subfigure}
\newcommand{\placefigMasknoMask}{
\begin{figure*}
\begin{center}
  \includegraphics[width=\textwidth,angle=0]{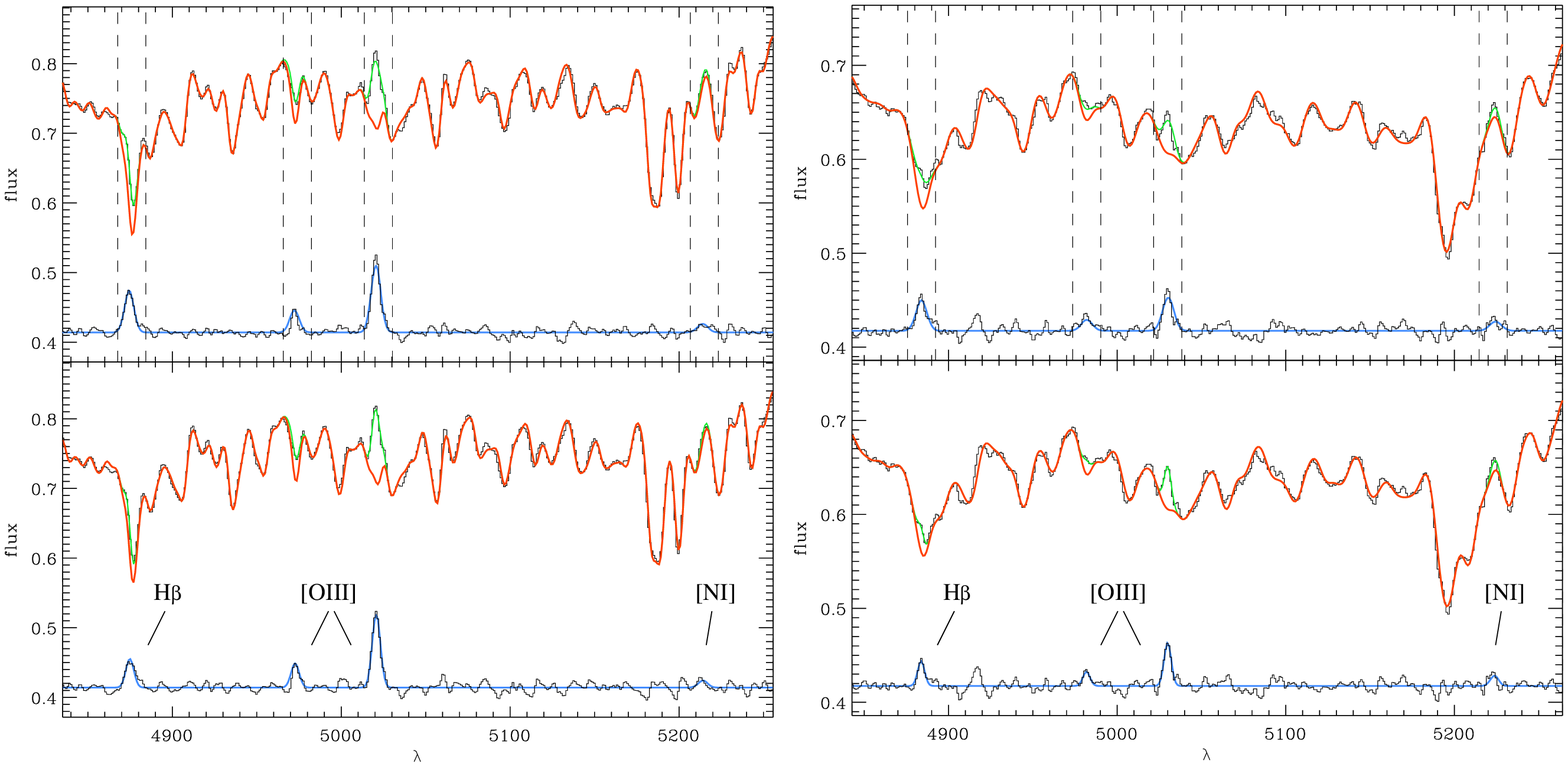}
\end{center}

\caption[]{Examples of the difference between the emission-line
measurements obtained when the lines are fitted to the residuals of
the stellar-continuum fit, itself derived while masking regions
contaminated by emission (top panels), or when both the stellar
continuum and the emission lines are fitted simultaneously (lower
panels).
At the top of each panel the black histogram shows the galaxy
spectrum, the red line the best fitting stellar template, and the
green line the best fit when the emission lines are added. These are
shown at the bottom of each panel by the blue line, along with a
second histogram showing the difference between the galaxy spectrum
and the best stellar template. A constant has been added to both.
The dashed vertical lines show the spectral regions that have been
masked.
Limiting the spectral range available to the template-fitting by
masking introduces spurious features in the \Hb\ and \Ni\
regions that leads to overestimated amplitudes for these lines (top
panels).
The line widths are also overestimated, particularly when the \Oiii\ lines
are weak and no longer dominate the resulting gas kinematics (right).
Both biases, in the line amplitudes and kinematics, are overcome when
the stellar continuum and emission lines are fitted together, and all
the information in the spectra is used (lower panels).
}
\label{fig:MasknoMask}
\end{figure*}
}

\newcommand{\placefigHbProblem}{
\begin{figure*}
\begin{center}
  \includegraphics[width=\textwidth]{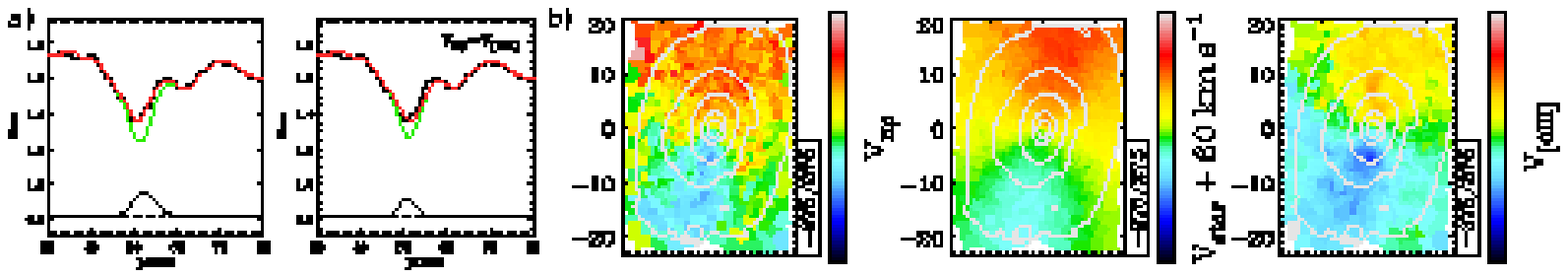}
\end{center}
\caption[]{Example of template-mismatch contamination in the \Hb\
measurement, in the S0 galaxy NGC~3489.
(a) Spectral region around \Hb.
The solid black histograms show the galaxy spectrum, the green lines
the best fitting stellar template, and the red lines the best fit when
the \Hb\ emission, indicated below by the solid lines, is added.  When
the velocity of \Hb\ emission is free to vary (left) the two main
absorption features are best fit by a combination of templates and a
spurious \Hb\ emission feature that contributes to the `bump' in the
spectrum between the two absorptions. When the velocity of \Hb\
emission is constrained to that of \Oiii\ (right), the best fit
includes a weaker, narrower emission feature closer to the \Hb\
absorption.
(b) Velocity maps. 
As the unconstrained \Hb\ emission is a spurious feature roughly
always at the same relative position with respect to the \Hb\
absorption feature, approximately 2/3 of a pixel (60km/s) to the red,
the corresponding velocity field (left) resembles the stellar velocity
field when a constant is added to it (center, note the curved
zero-velocity region in green).
The right panel shows the \Oiii\ emission velocities, which trace the
real gas velocities.}
\label{fig:HbProblem}
\end{figure*}
}

\newcommand{\placefigSauronVsPalomar}{
\begin{figure}
\begin{center}
  \includegraphics[width=\columnwidth,bb= 86 215 514 574]{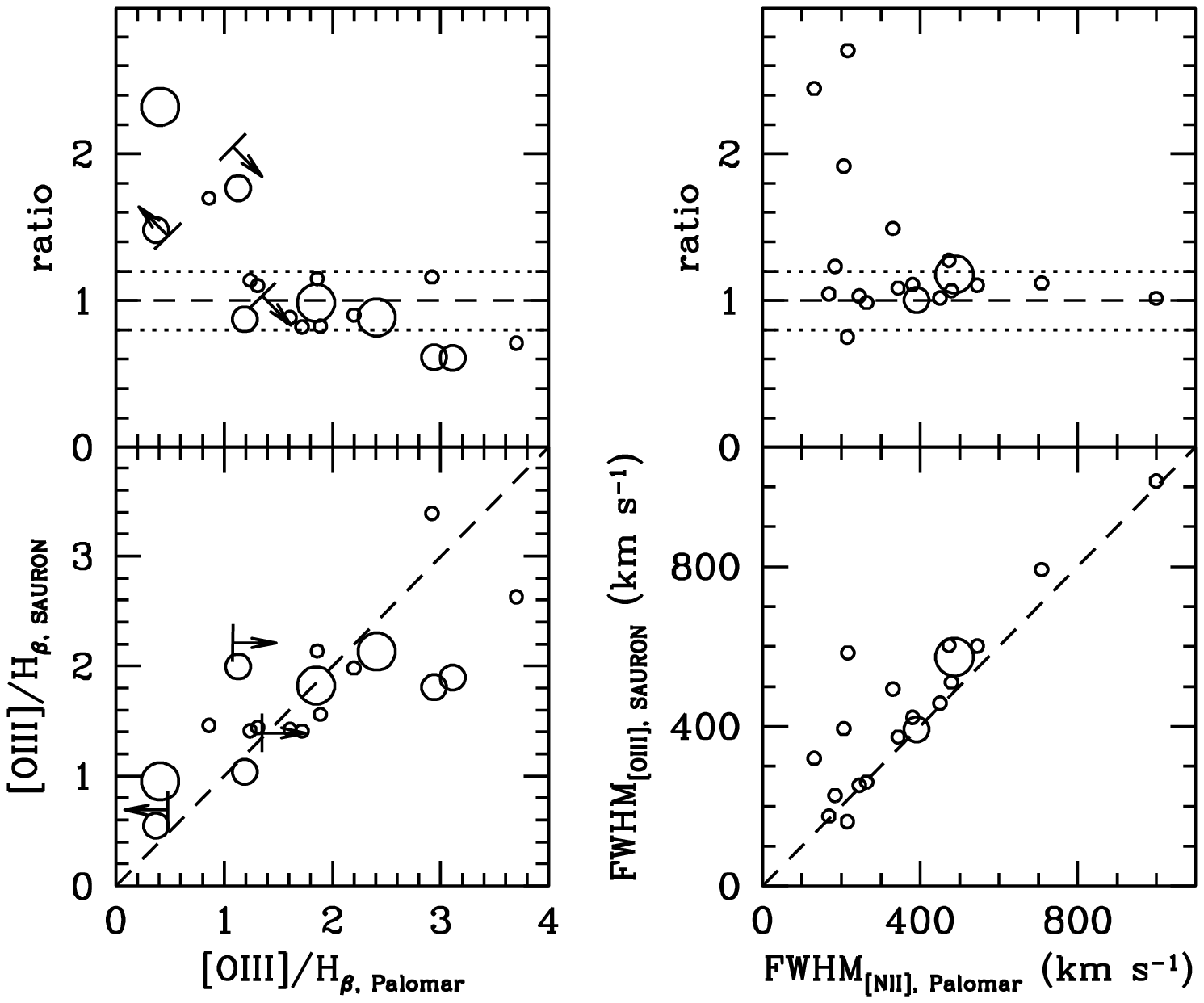}
\end{center}
\caption[]{\sauron\ vs. Palomar. 
\Oiii/\Hb\ line ratio (left) and intrinsic width (right) of the [{\sc
N$\,$ii}] lines from the nuclear spectra of \citet{Ho97b} compared
to \Oiii/\Hb\ ratio and intrinsic width of the \Oiii\ lines measured
with \sauron\ in the same apertures.
The lower panels show the two sets of data directly against each
other, while the upper panels show their ratio.
Larger symbols correspond to less accurate measurements from Ho et
al., while the arrows correspond to upper and lower limits from the
same work. The dashed lines show the identity relation, while the
dotted lines in the top panels define the region where the \sauron\
and Palomar measurements agree to within 20\%.
}
\label{fig:SauronVsPalomar}
\end{figure}
}

\newcommand{\placefigHbfaster}{
\begin{figure}
\begin{center}
  \includegraphics[width=\columnwidth]{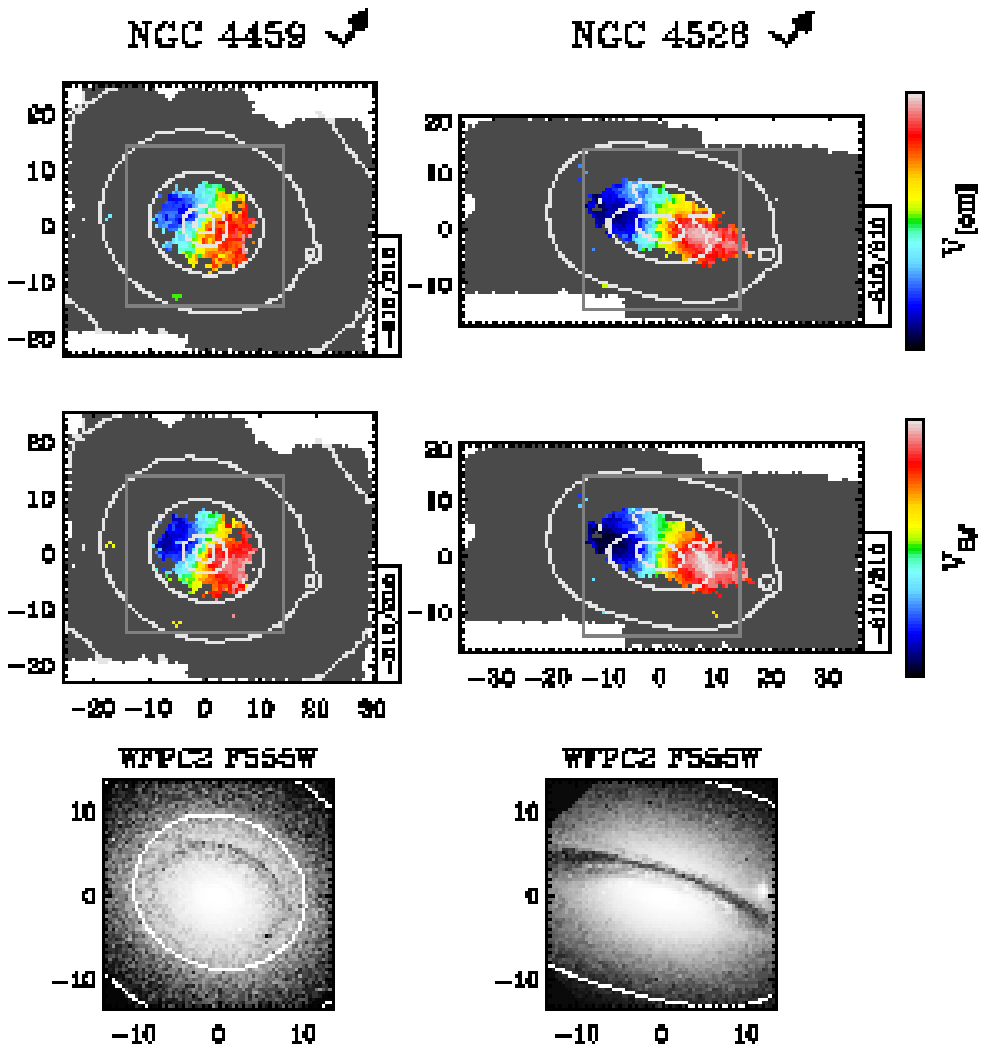}
\end{center}
\caption[]{Faster rotating \Hb\ kinematics in galaxies with perfect dusty disks.
Two examples of galaxies with very regular gas kinematics and
circularly symmetric dust lanes where the \Oiii\ and \Hb\ kinematics
could be derived independently. The velocity maps for the \Oiii\ (top)
and \Hb\ lines (middle) are as in Fig\ref{fig:Mapsa}, and show
regions where both \Oiii\ and \Hb\ lines were detected following our
standard approach. Lighter and darker colors on the receding and
approaching side, respectively, of the \Hb\ velocity maps, illustrate
how the \Hb\ kinematics show faster rotation velocities than the
\Oiii\ kinematics. The grey boxes in the velocity maps indicate 
the field of the \HST\ images (bottom), showing the dust-lane
morphology (see also Fig\ref{fig:Mapsb}).
}
\label{fig:Hbfaster}
\end{figure}
}

\newcommand{\placefigHistMisalESO}{
\begin{figure}
\begin{center}
  \includegraphics[width=\columnwidth,bb= 77 215 515 640]{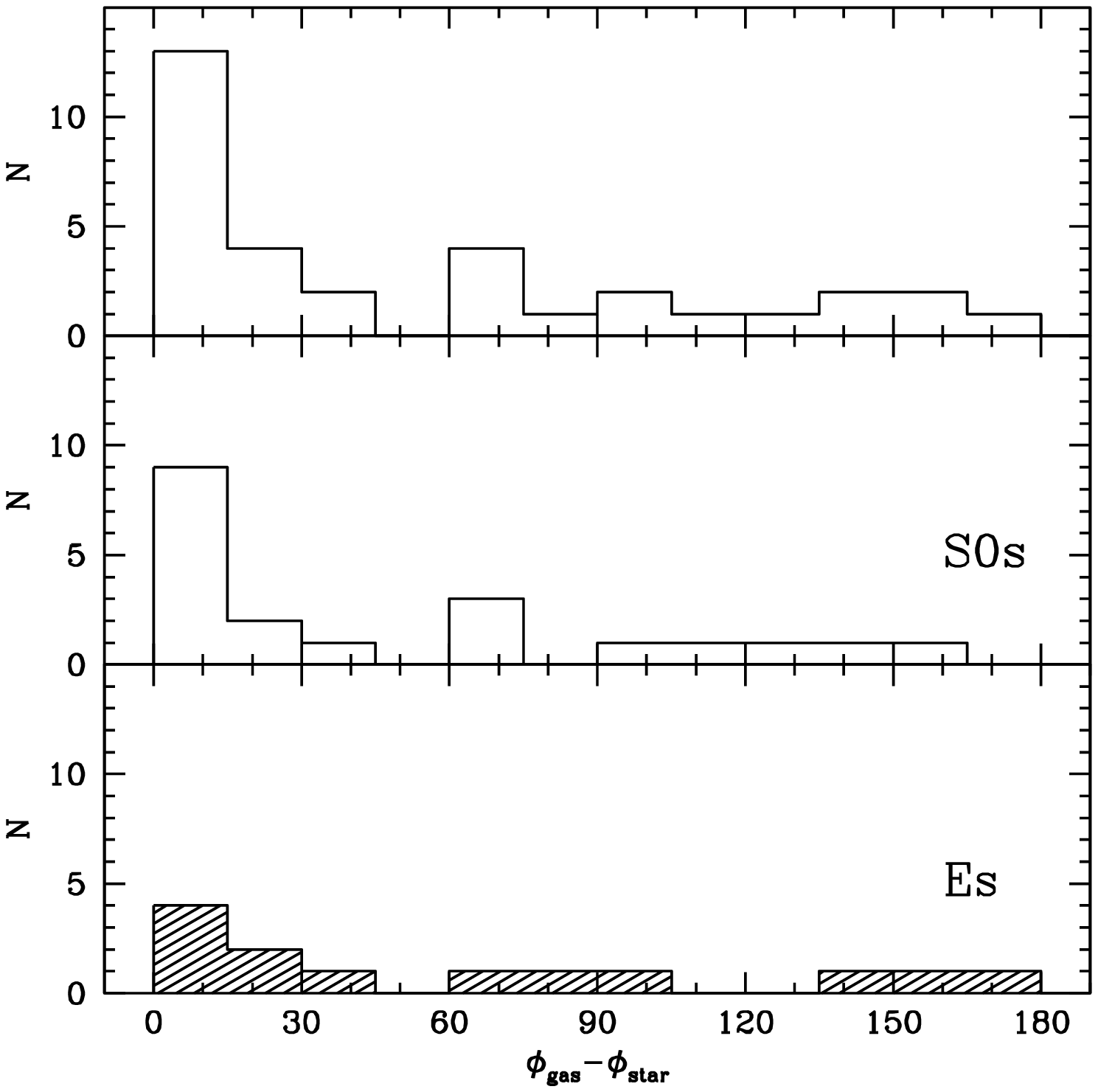}
\end{center}
\caption[]{Distribution of the median values for the kinematic
misalignments between star and gas for all galaxies listed in
Table~\ref{tab:misal} (top) or only for lenticular (middle) and
elliptical (bottom) galaxies. The values of Table~\ref{tab:misal} are
now shown between 0\degr\ and 180\degr.
}
\label{fig:HistMisalES0}
\end{figure}
}

\newcommand{\placefigHistMisalFlRo}{
\begin{figure}
\begin{center}
  \includegraphics[width=\columnwidth,bb= 77 215 515 640]{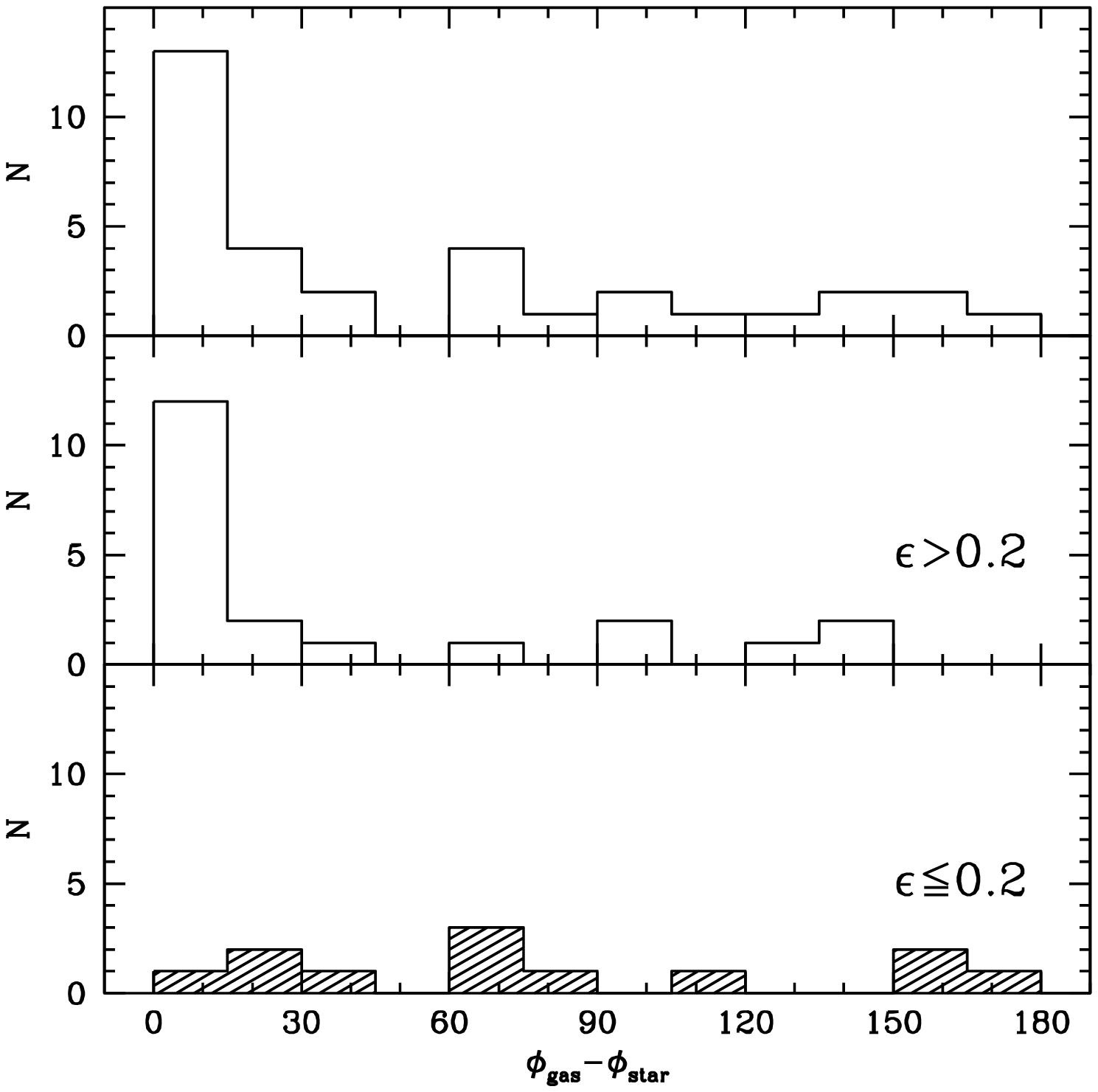}
\end{center}
\caption[]{Same as Figure~\ref{fig:HistMisalES0} but now showing 
flatter and rounder galaxies in the middle and bottom panel,
respectively.
}
\label{fig:HistMisalFlRo}
\end{figure}
}

\newcommand{\placefigHistMisalRnR}{
\begin{figure}
\begin{center}
  \includegraphics[width=\columnwidth,bb= 77 215 515 640]{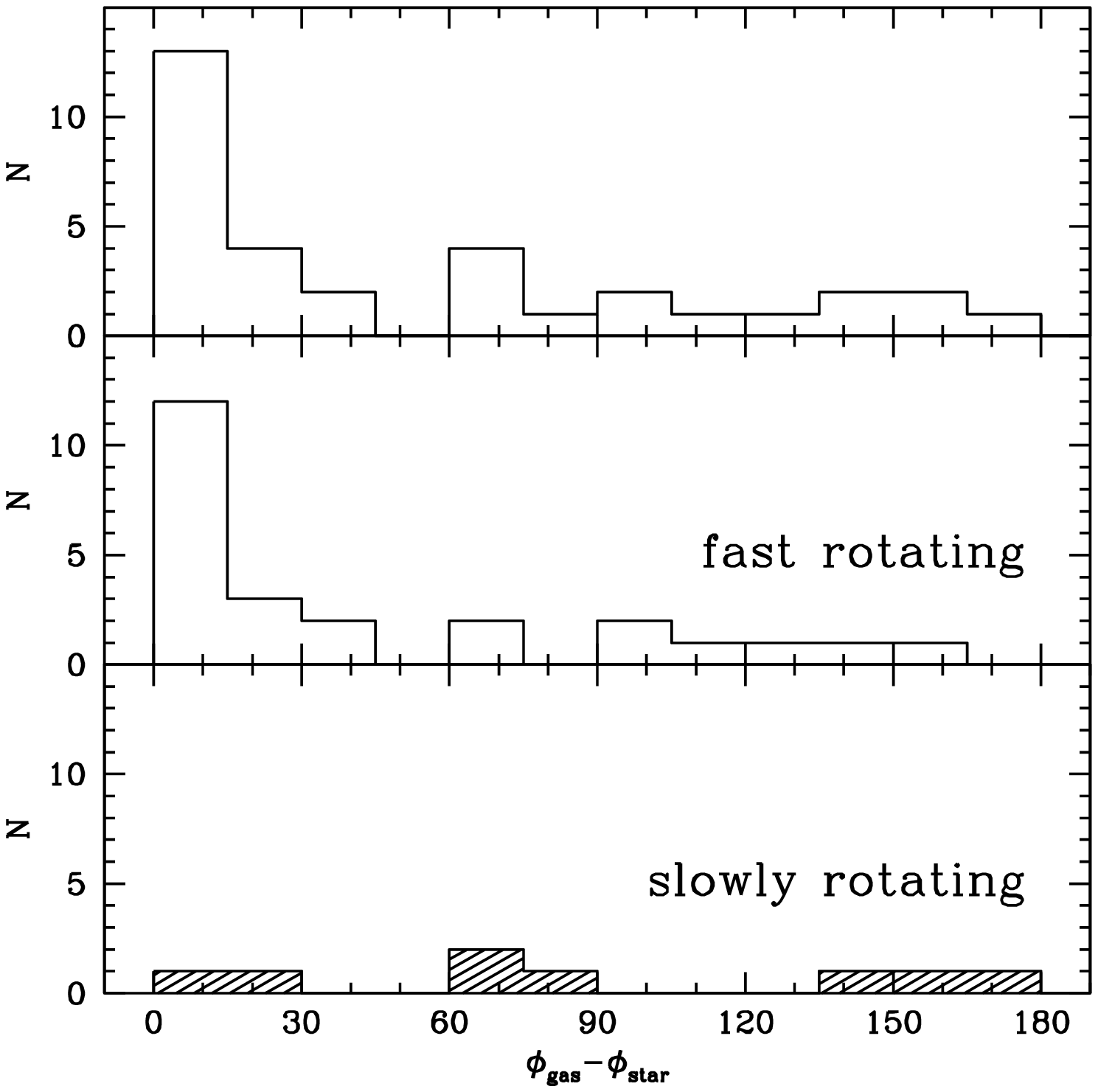}
\end{center}
\caption[]{Same as Figure~\ref{fig:HistMisalES0} but now showing 
fast and slowly rotating galaxies in the middle and bottom panel,
respectively.
}
\label{fig:HistMisalRnR}
\end{figure}
}

\newcommand{\placefigMaps}{
\renewcommand{\thefigure}{\arabic{figure}\alph{subfigure}}
\setcounter{subfigure}{1}

\begin{figure*}
\begin{center}
\includegraphics[width=0.98\textwidth, trim=0cm 0cm 0cm  0cm]{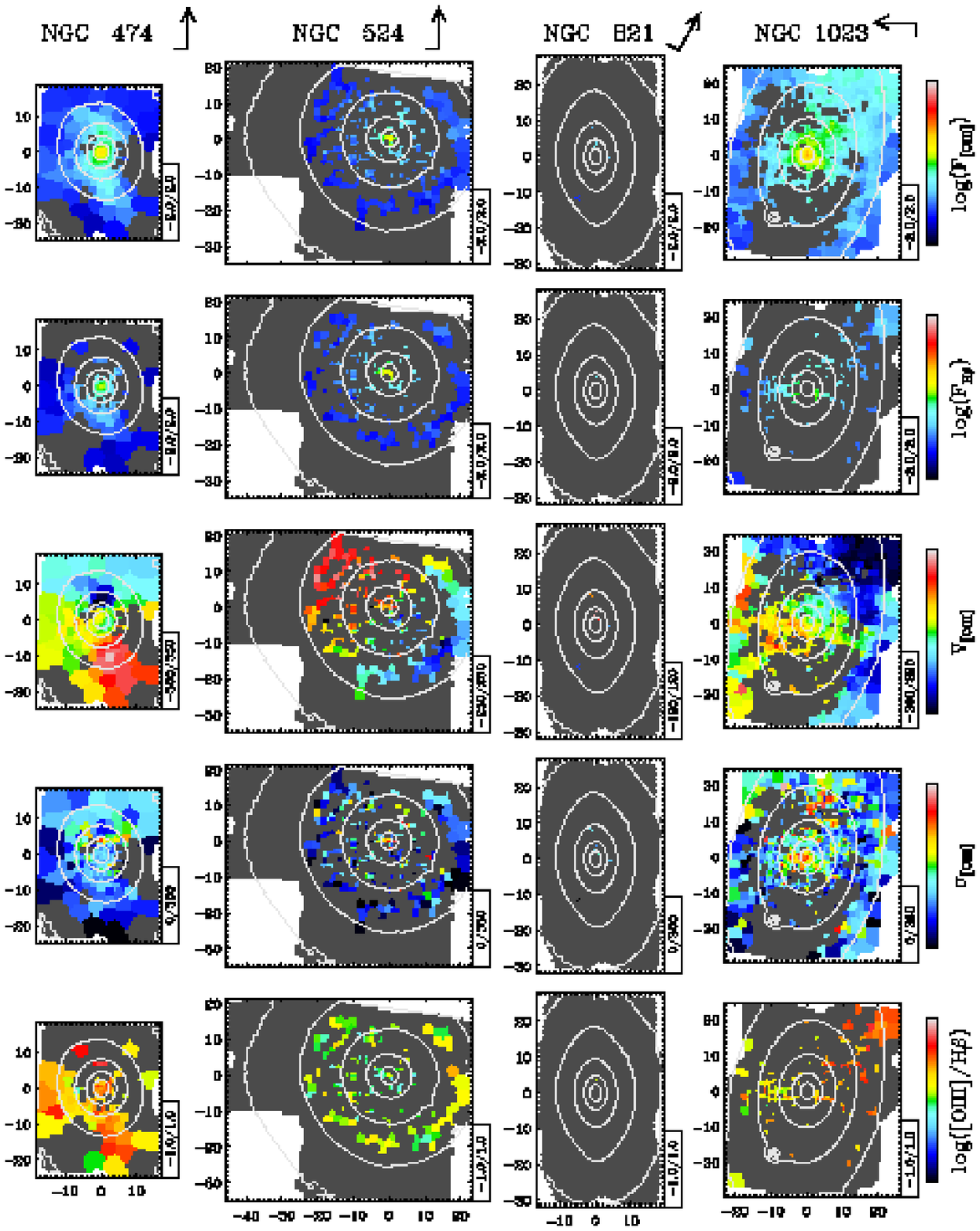}
\end{center}
\caption[]{Maps of the ionised-gas distribution and kinematics of the
48 E and S0 galaxies in the \sauron\ representative sample. 
The \sauron\ spectra have been spatially binned to a minimum
$S/N$ of 60 in the stellar continuum, consistent with Paper~III.
All maps are plotted on the same angular scale, in arcseconds. The
arrow and its associated dash at the top of each column mark the North
and East directions, respectively. From top to bottom: i) and ii) flux
of the \Oiii$\lambda5007$ and \Hb\ emission line, in $10^{-16}\rm
erg\,s^{-1}cm^{-2} arcsec^{-2}$ and in a logarithmic scale, iii) and
iv) gas mean velocity and intrinsic velocity dispersion in \kms, as
traced by the \Oiii$\lambda\lambda4959,5007$ lines, v) values of the
\Oiii$\lambda$5007/\Hb\ ratio. The cuts levels are indicated in a
box on the right hand side of each map. Gas velocities are shown with
respect to the stellar systemic velocity. Regions without detected
emission are shown in dark grey.}
\label{fig:Mapsa}
\end{figure*}
\addtocounter{figure}{-1}
\addtocounter{subfigure}{1}
\begin{figure*}
\begin{center}
\includegraphics[width=0.98\textwidth, trim=0cm 0cm 0cm  0cm]{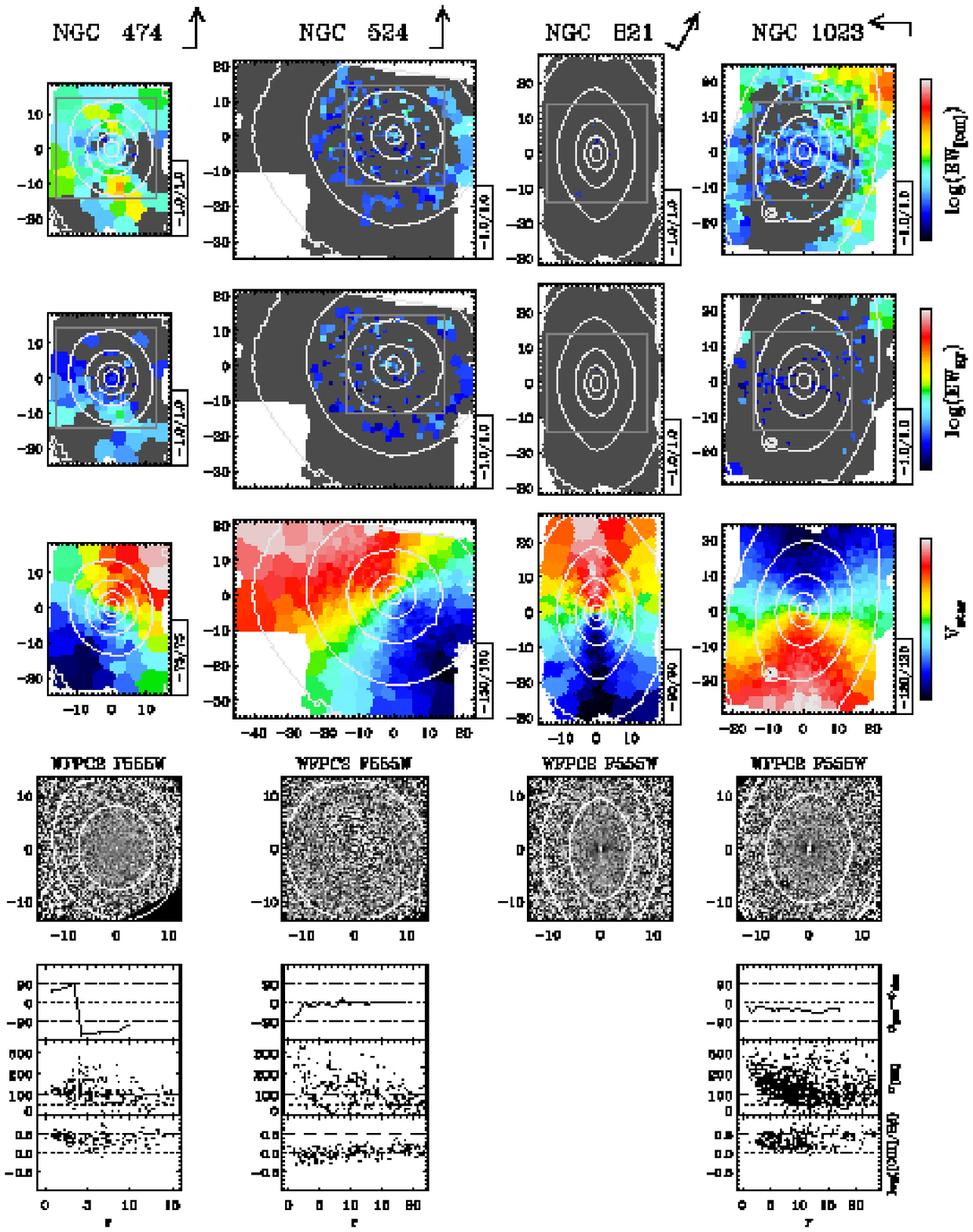}
\end{center}
\caption[]{From top to bottom: i) and ii) equivalent width of the
\Oiii$\lambda5007$ and \Hb\ emission lines, in \AA\ and in a
logarithmic scale, iii) stellar mean velocity in \kms\ from Paper~III,
iv) unsharp-masked images obtained from \HST\ observations or
\sauron\ reconstructed intensity maps, and v) radial profiles for the
misalignment between the kinematics of gas and stars (up), for the
velocity dispersion of the \Oiii\ lines (middle) and for the
\Oiii$\lambda$5007/\Hb\ ratio (down). The \sauron\ maps are as in
Figure~\ref{fig:Mapsa}. The grey boxes in the top two maps indicate the
field of the \HST\ images. 
}
\label{fig:Mapsb}
\end{figure*}

\clearpage
\addtocounter{figure}{-1}
\addtocounter{subfigure}{-1}
\begin{figure*}
\begin{center}
  \includegraphics[width=0.98\textwidth, trim=0cm 0cm 0cm 0cm]{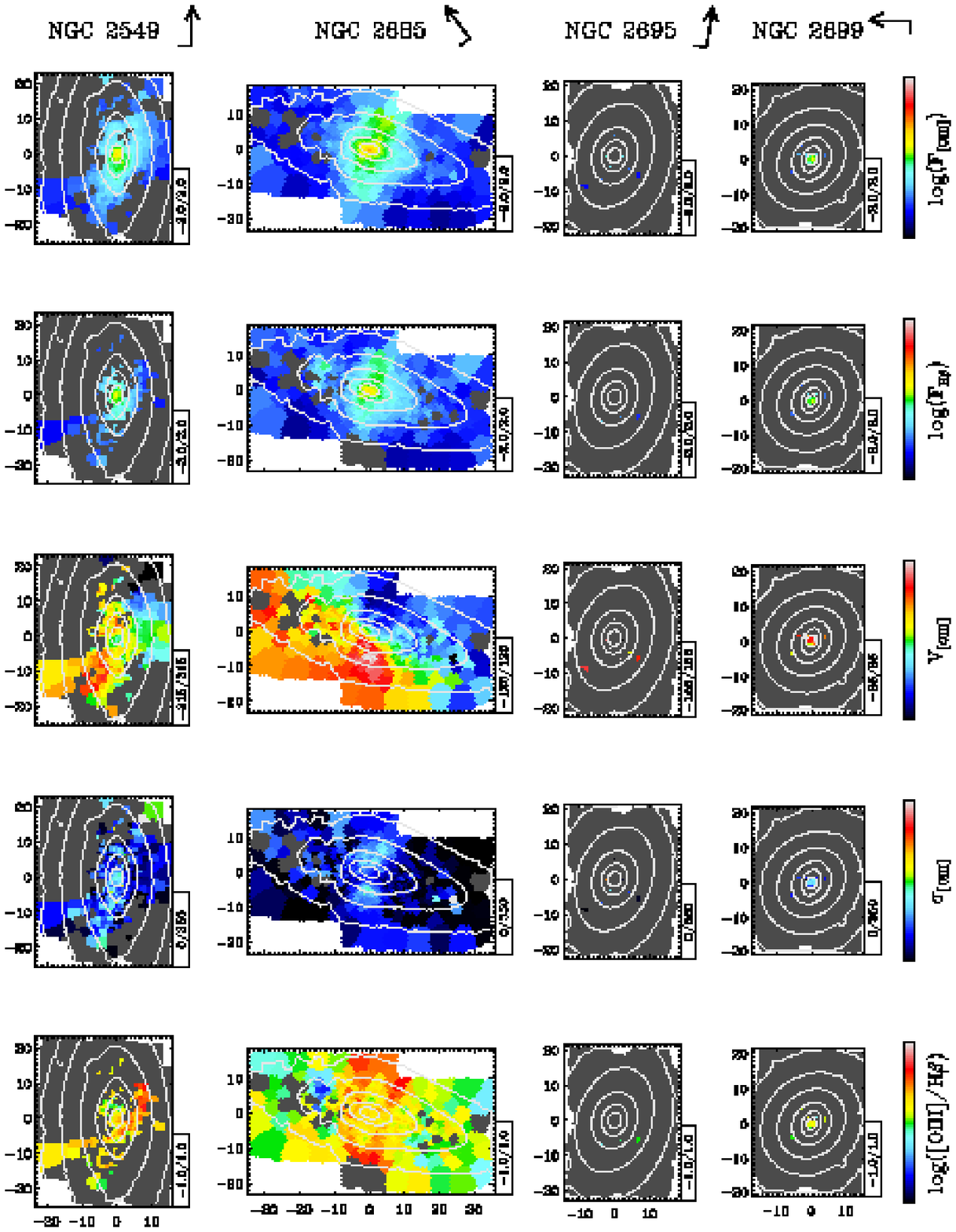}
\end{center}
\caption[]{Continue}
\end{figure*}
\addtocounter{figure}{-1}
\addtocounter{subfigure}{1}
\begin{figure*}
\begin{center}
  \includegraphics[width=0.98\textwidth, trim=0cm 0cm 0cm 0cm]{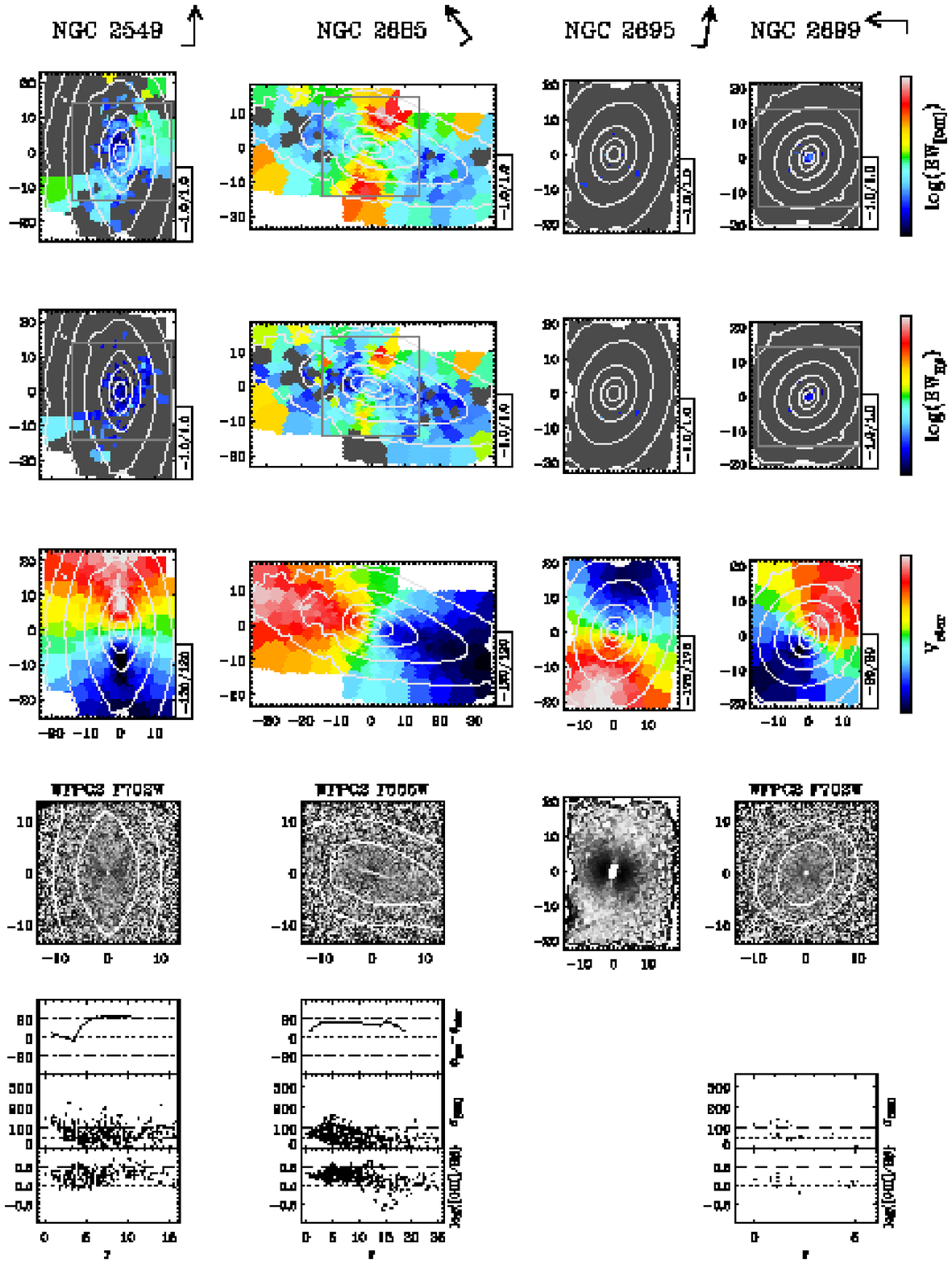}
\end{center}
\caption[]{Continue}
\end{figure*}

\clearpage
\addtocounter{figure}{-1}
\addtocounter{subfigure}{-1}
\begin{figure*}
\begin{center}
  \includegraphics[width=0.98\textwidth, trim=0cm 0cm 0cm 0cm]{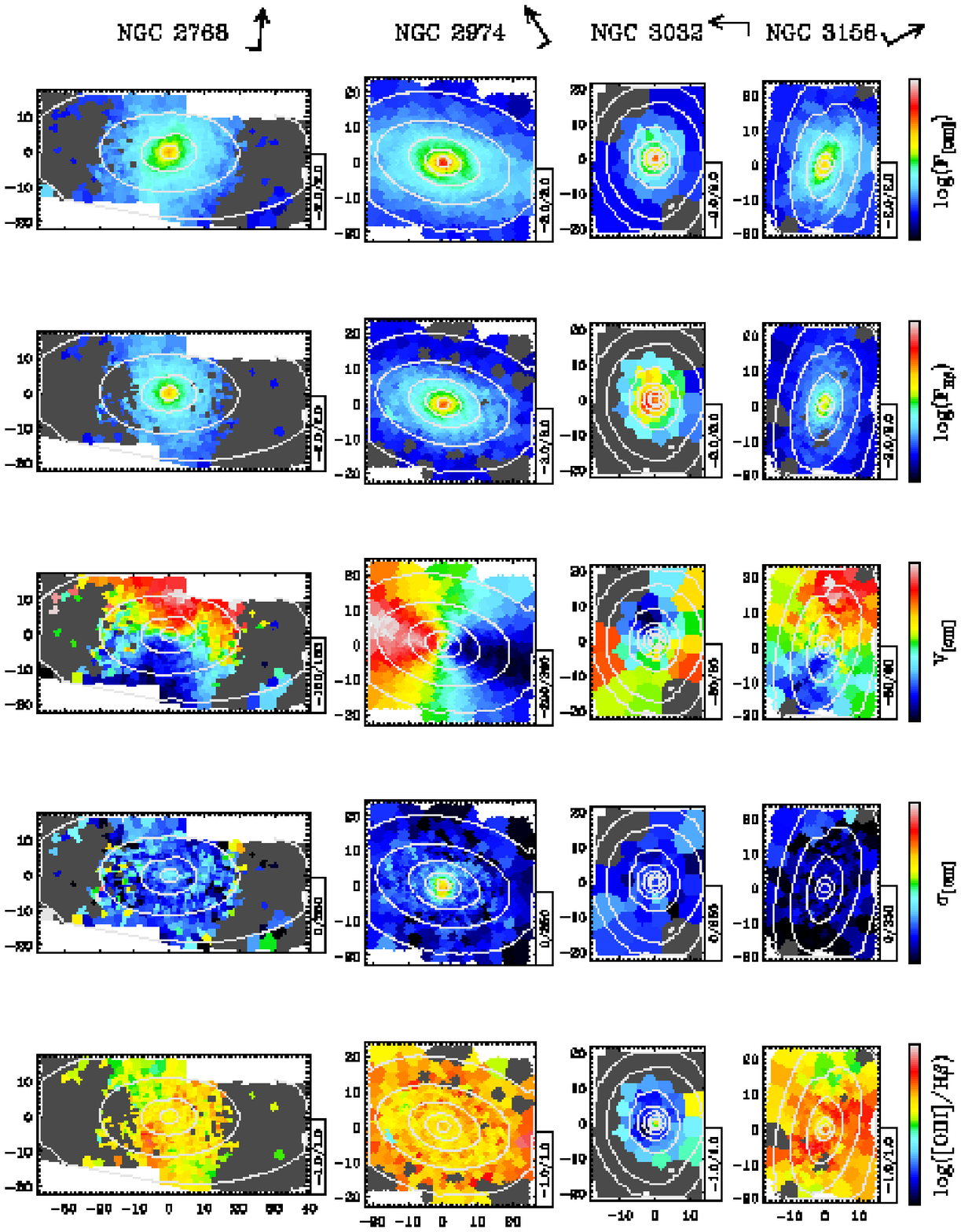}
\end{center}
\caption[]{Continue}
\end{figure*}
\addtocounter{figure}{-1}
\addtocounter{subfigure}{1}
\begin{figure*}
\begin{center}
  \includegraphics[width=0.98\textwidth, trim=0cm 0cm 0cm 0cm]{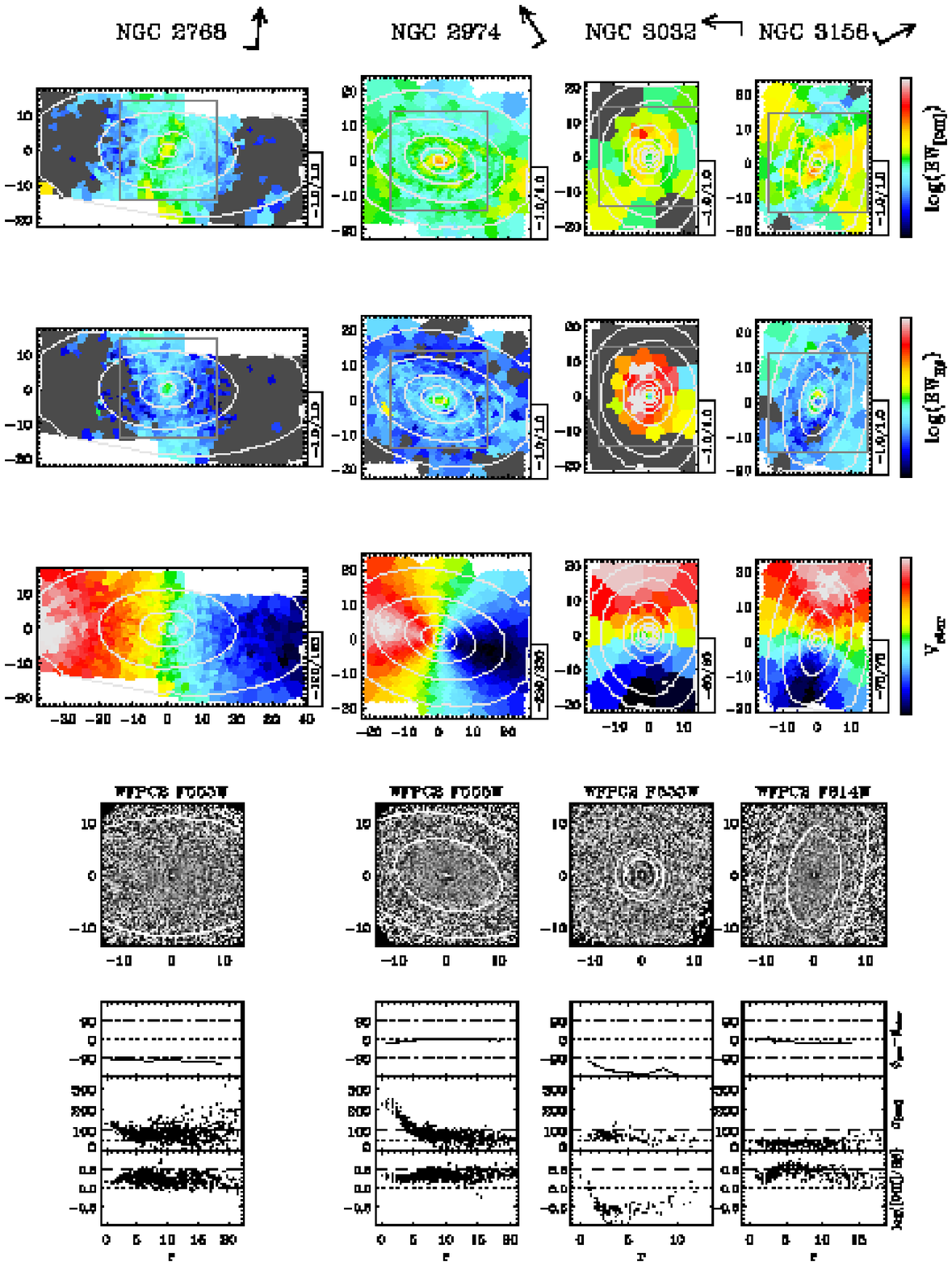}
\end{center}
\caption[]{Continue}
\end{figure*}

\clearpage
\addtocounter{figure}{-1}
\addtocounter{subfigure}{-1}
\begin{figure*}
\begin{center}
  \includegraphics[width=0.98\textwidth, trim=0cm 0cm 0cm 0cm]{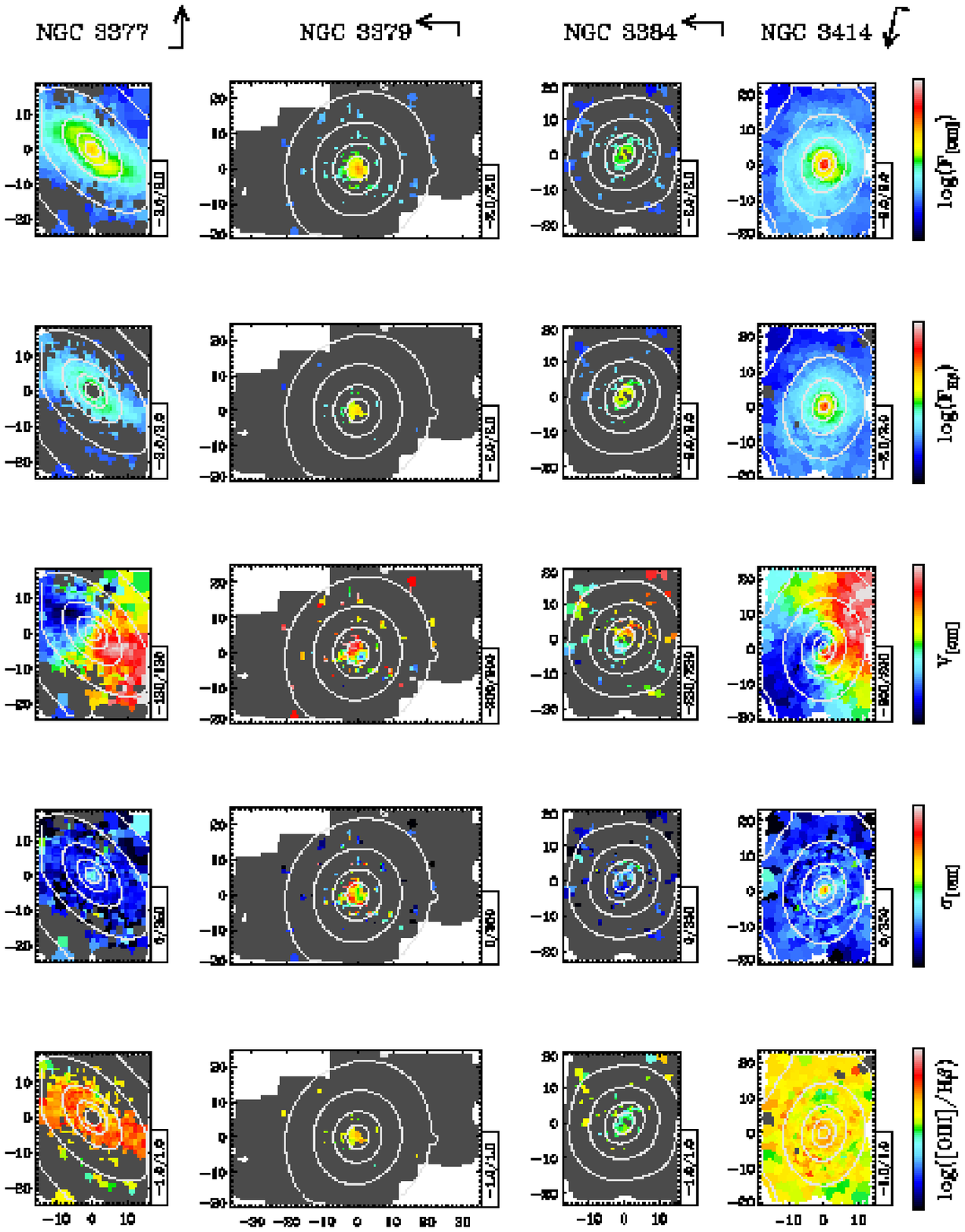}
\end{center}
\caption[]{Continue}
\end{figure*}
\addtocounter{figure}{-1}
\addtocounter{subfigure}{1}
\begin{figure*}
\begin{center}
  \includegraphics[width=0.98\textwidth, trim=0cm 0cm 0cm 0cm]{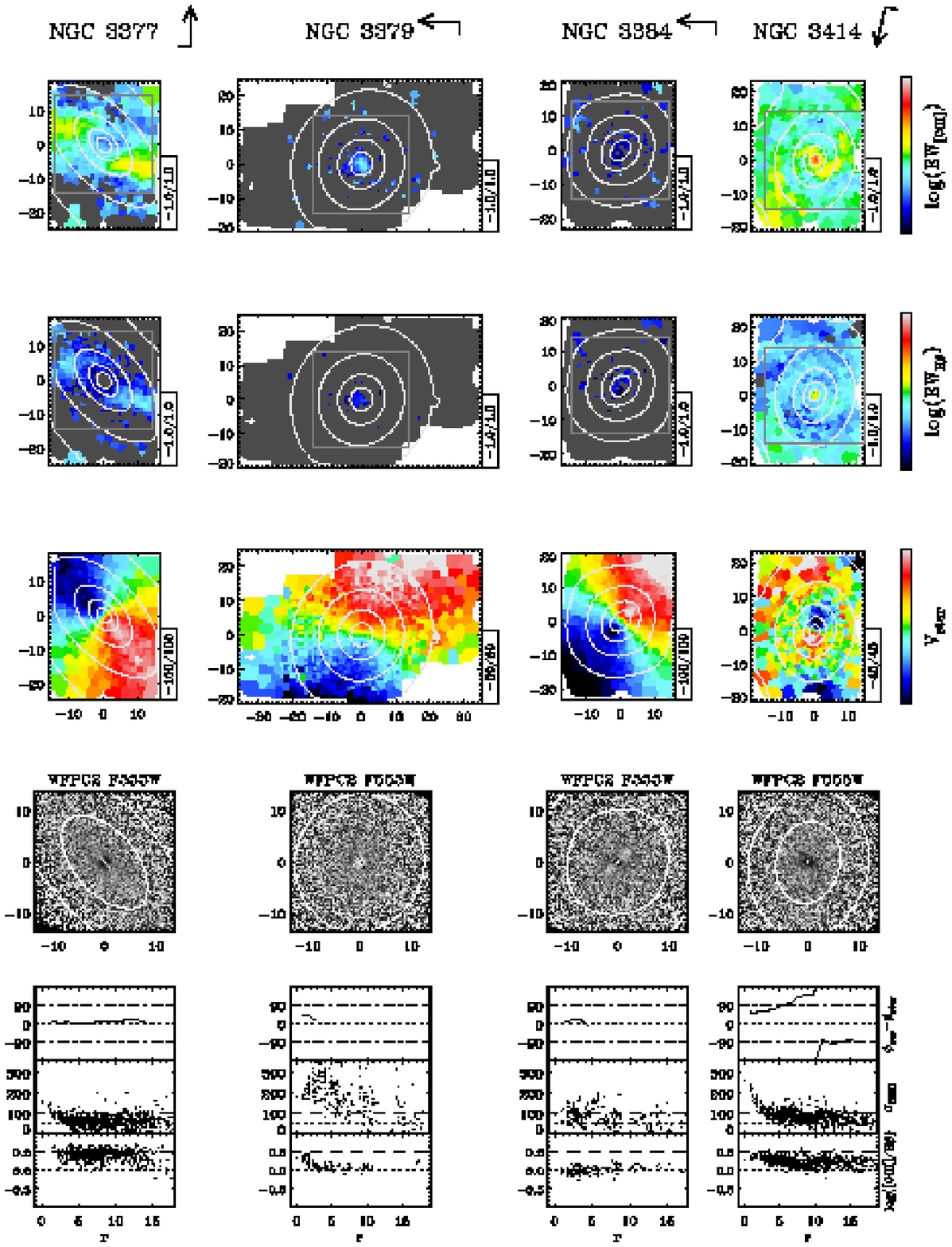}
\end{center}
\caption[]{Continue}
\end{figure*}

\clearpage
\addtocounter{figure}{-1}
\addtocounter{subfigure}{-1}
\begin{figure*}
\begin{center}
  \includegraphics[width=0.98\textwidth, trim=0cm 0cm 0cm 0cm]{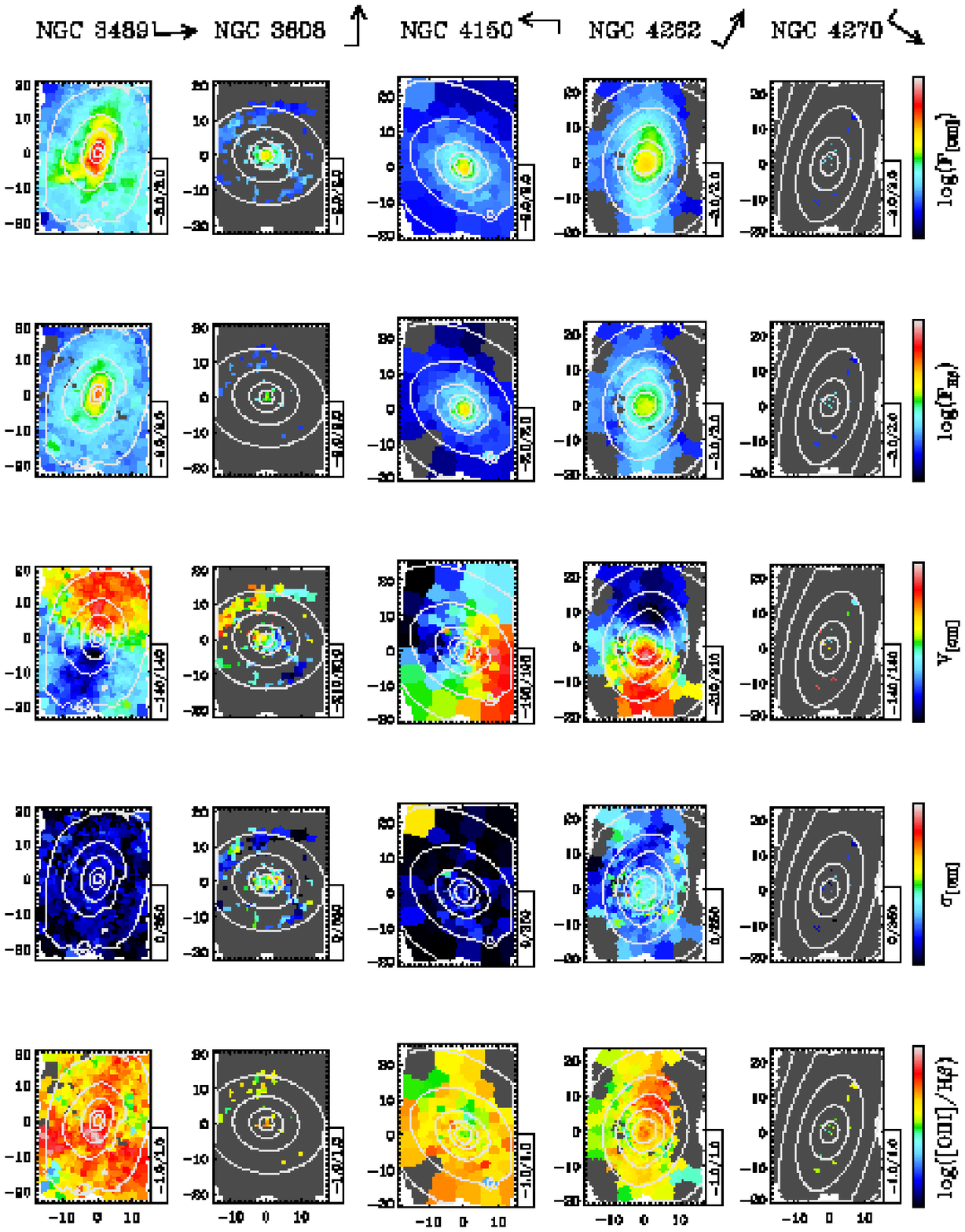}
\end{center}
\caption[]{Continue}
\end{figure*}
\addtocounter{figure}{-1}
\addtocounter{subfigure}{1}
\begin{figure*}
\begin{center}
  \includegraphics[width=0.98\textwidth, trim=0cm 0cm 0cm 0cm]{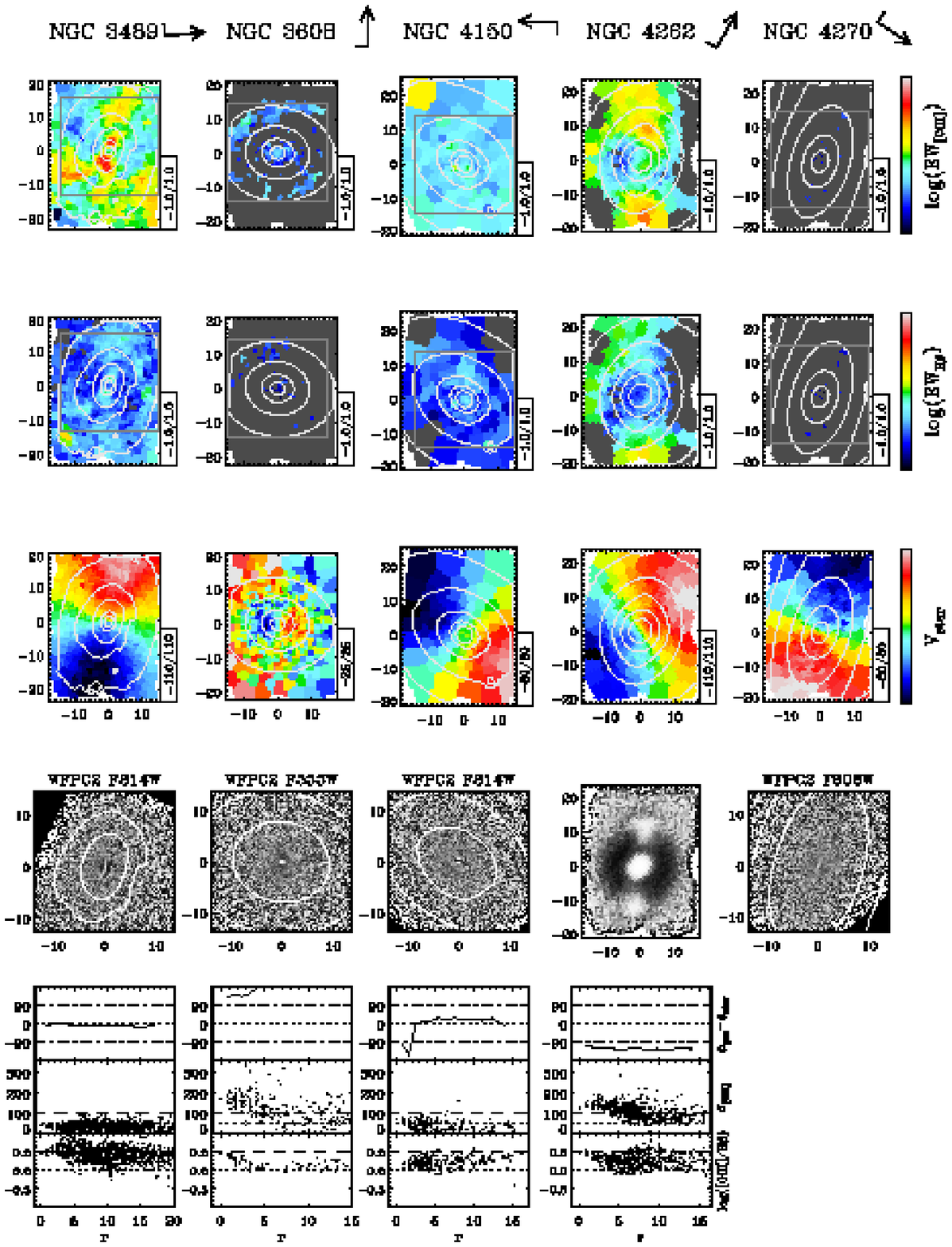}
\end{center}
\caption[]{Continue}
\end{figure*}

\clearpage
\addtocounter{figure}{-1}
\addtocounter{subfigure}{-1}
\begin{figure*}
\begin{center}
  \includegraphics[width=0.98\textwidth, trim=0cm 0cm 0cm 0cm]{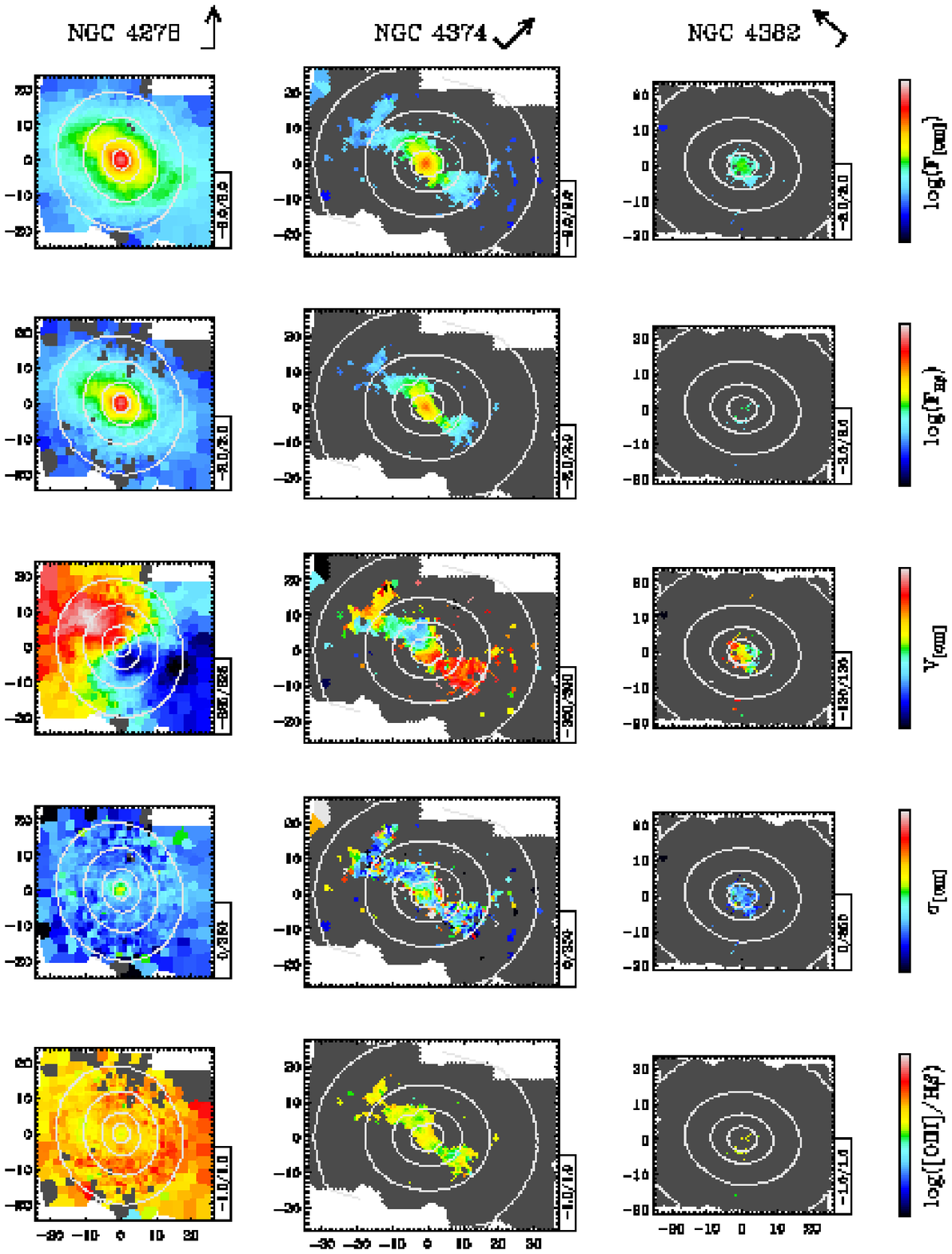}
\end{center}
\caption[]{Continue}
\end{figure*}
\addtocounter{figure}{-1}
\addtocounter{subfigure}{1}
\begin{figure*}
\begin{center}
  \includegraphics[width=0.98\textwidth, trim=0cm 0cm 0cm 0cm]{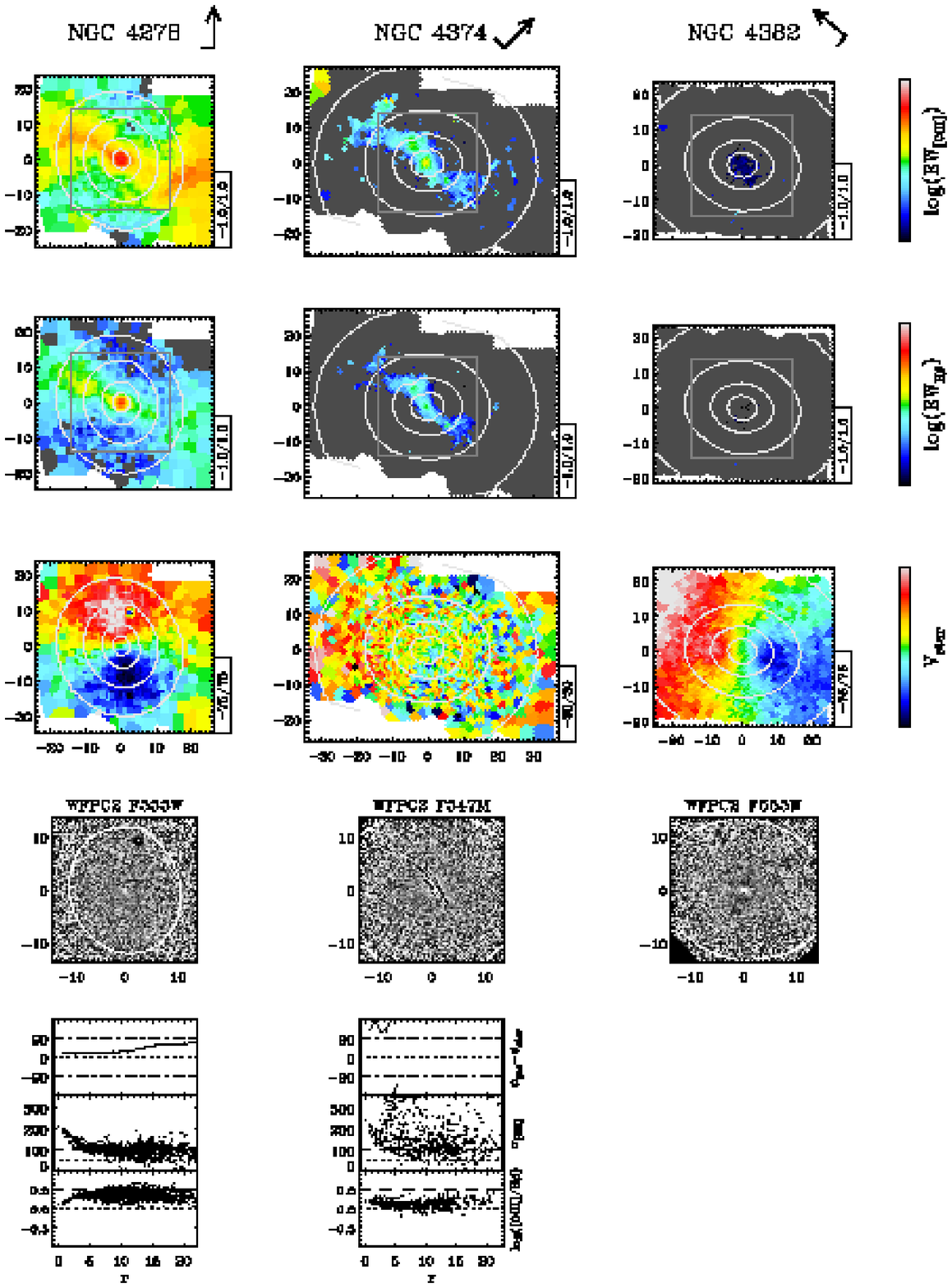}
\end{center}
\caption[]{Continue}
\end{figure*}

\clearpage
\addtocounter{figure}{-1}
\addtocounter{subfigure}{-1}
\begin{figure*}
\begin{center}
  \includegraphics[width=0.98\textwidth, trim=0cm 0cm 0cm 0cm]{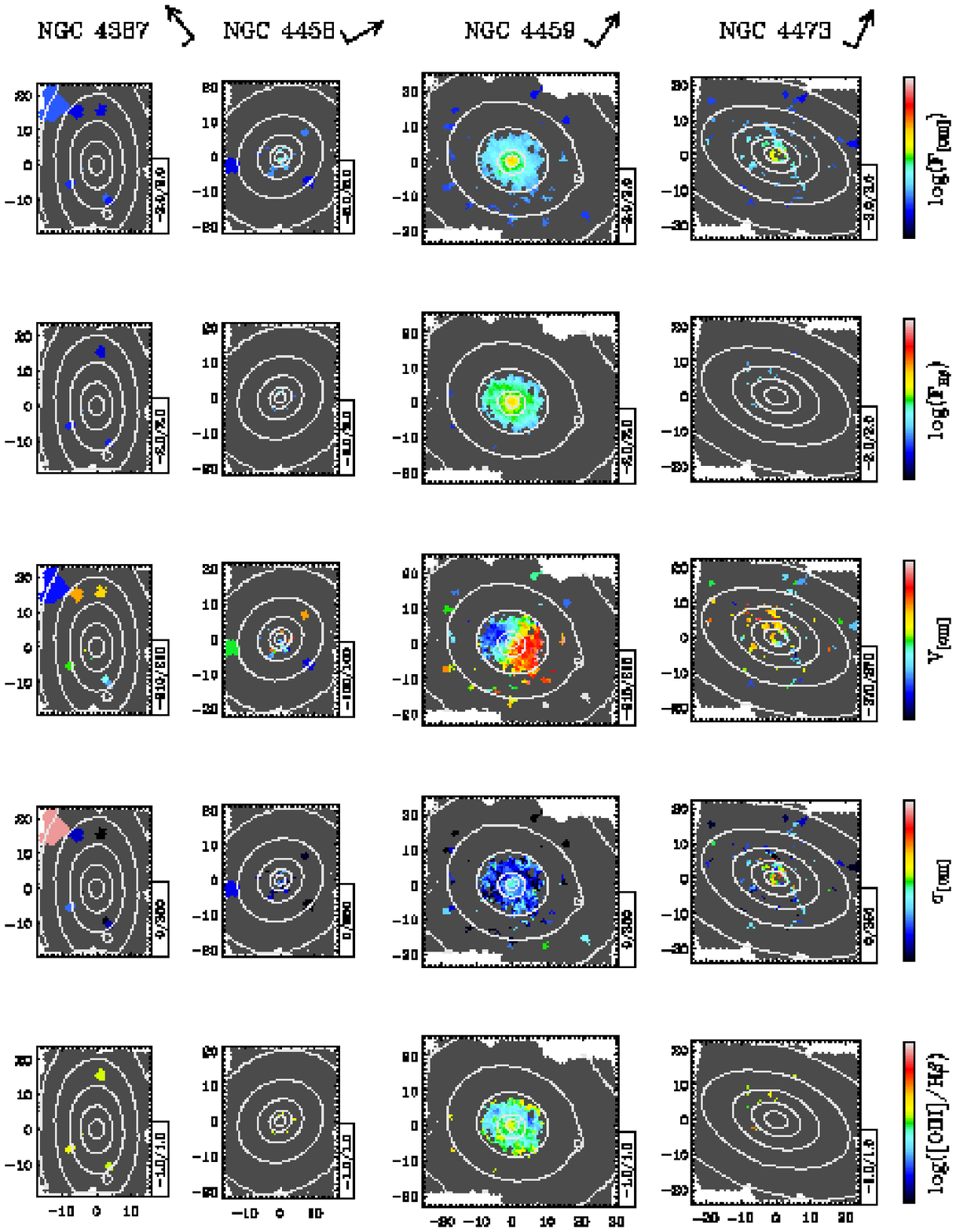}
\end{center}
\caption[]{Continue}
\end{figure*}
\addtocounter{figure}{-1}
\addtocounter{subfigure}{1}
\begin{figure*}
\begin{center}
  \includegraphics[width=0.98\textwidth, trim=0cm 0cm 0cm 0cm]{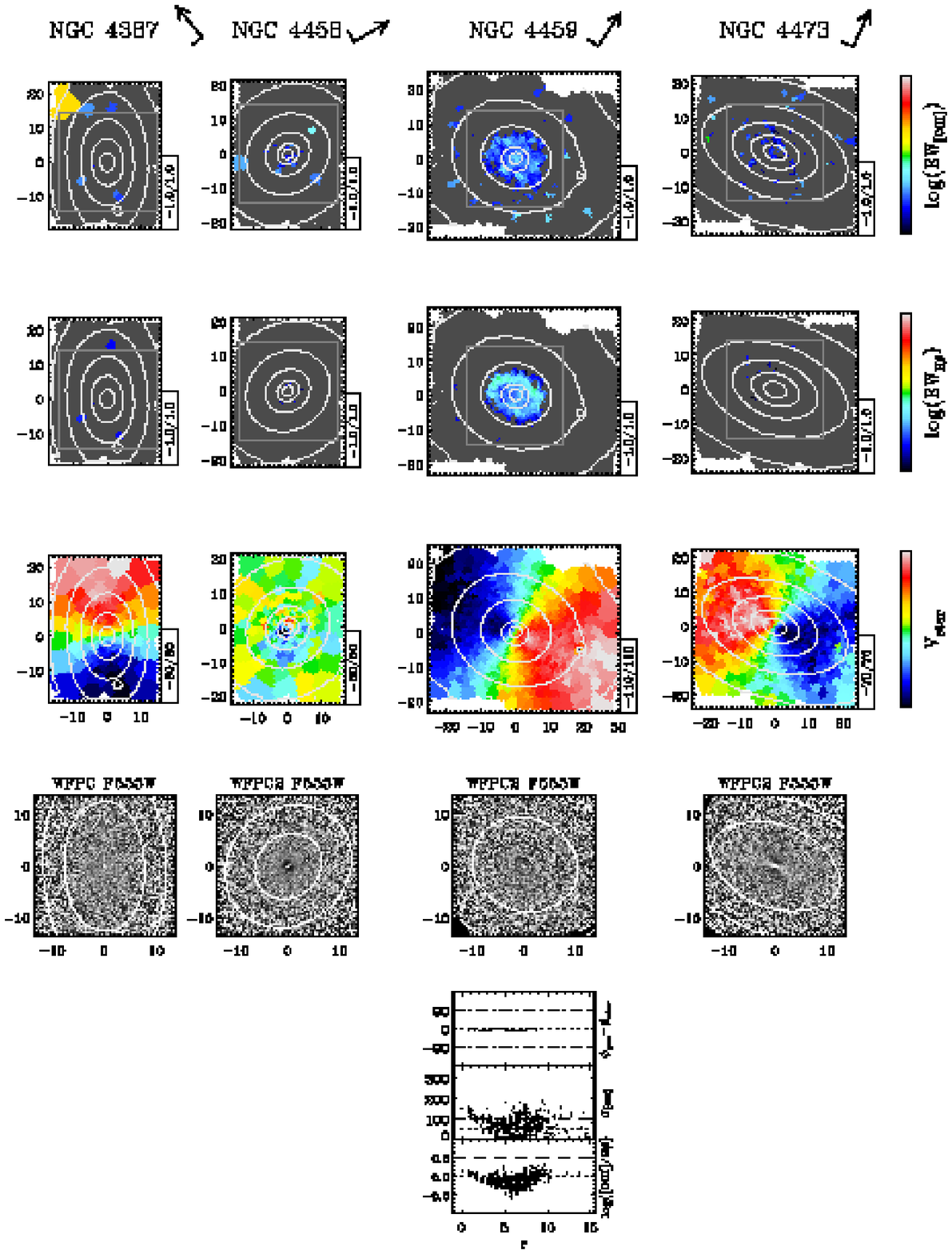}
\end{center}
\caption[]{Continue}
\end{figure*}

\clearpage
\addtocounter{figure}{-1}
\addtocounter{subfigure}{-1}
\begin{figure*}
\begin{center}
  \includegraphics[width=0.98\textwidth, trim=0cm 0cm 0cm 0cm]{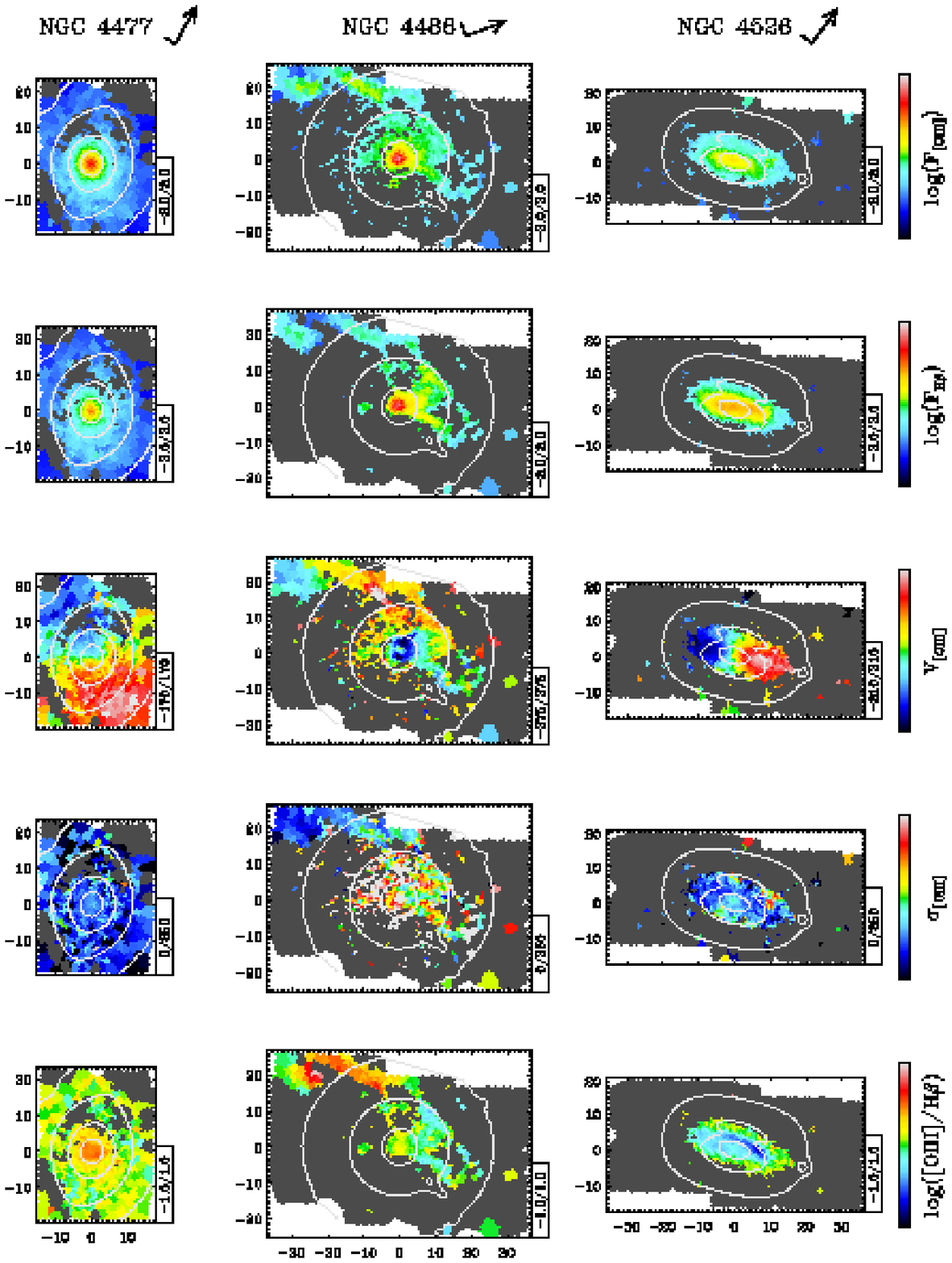}
\end{center}
\caption[]{Continue}
\end{figure*}
\addtocounter{figure}{-1}
\addtocounter{subfigure}{1}
\begin{figure*}
\begin{center}
  \includegraphics[width=0.98\textwidth, trim=0cm 0cm 0cm 0cm]{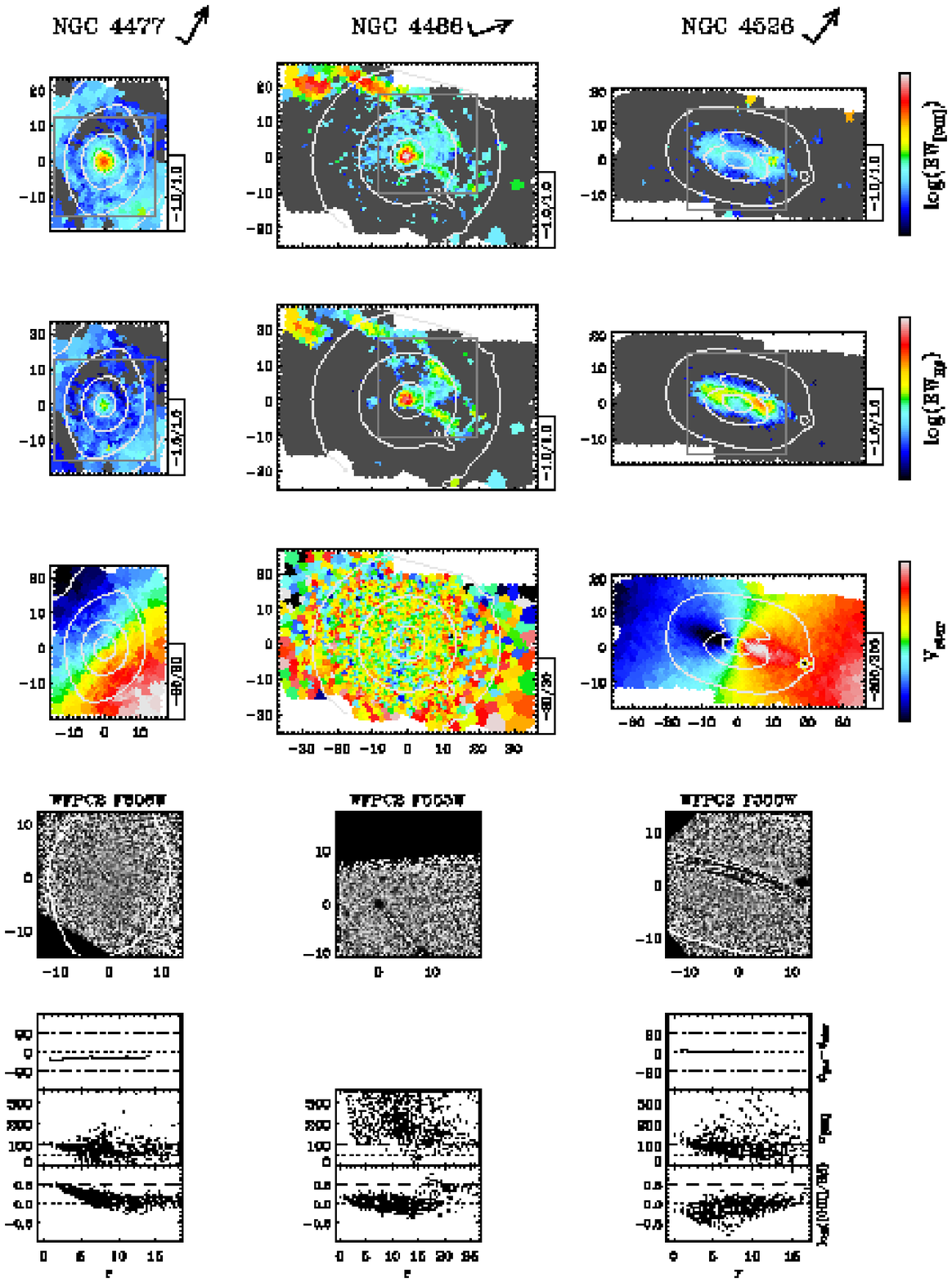}
\end{center}
\caption[]{Continue}
\end{figure*}

\clearpage
\addtocounter{figure}{-1}
\addtocounter{subfigure}{-1}
\begin{figure*}
\begin{center}
  \includegraphics[width=0.98\textwidth, trim=0cm 0cm 0cm 0cm]{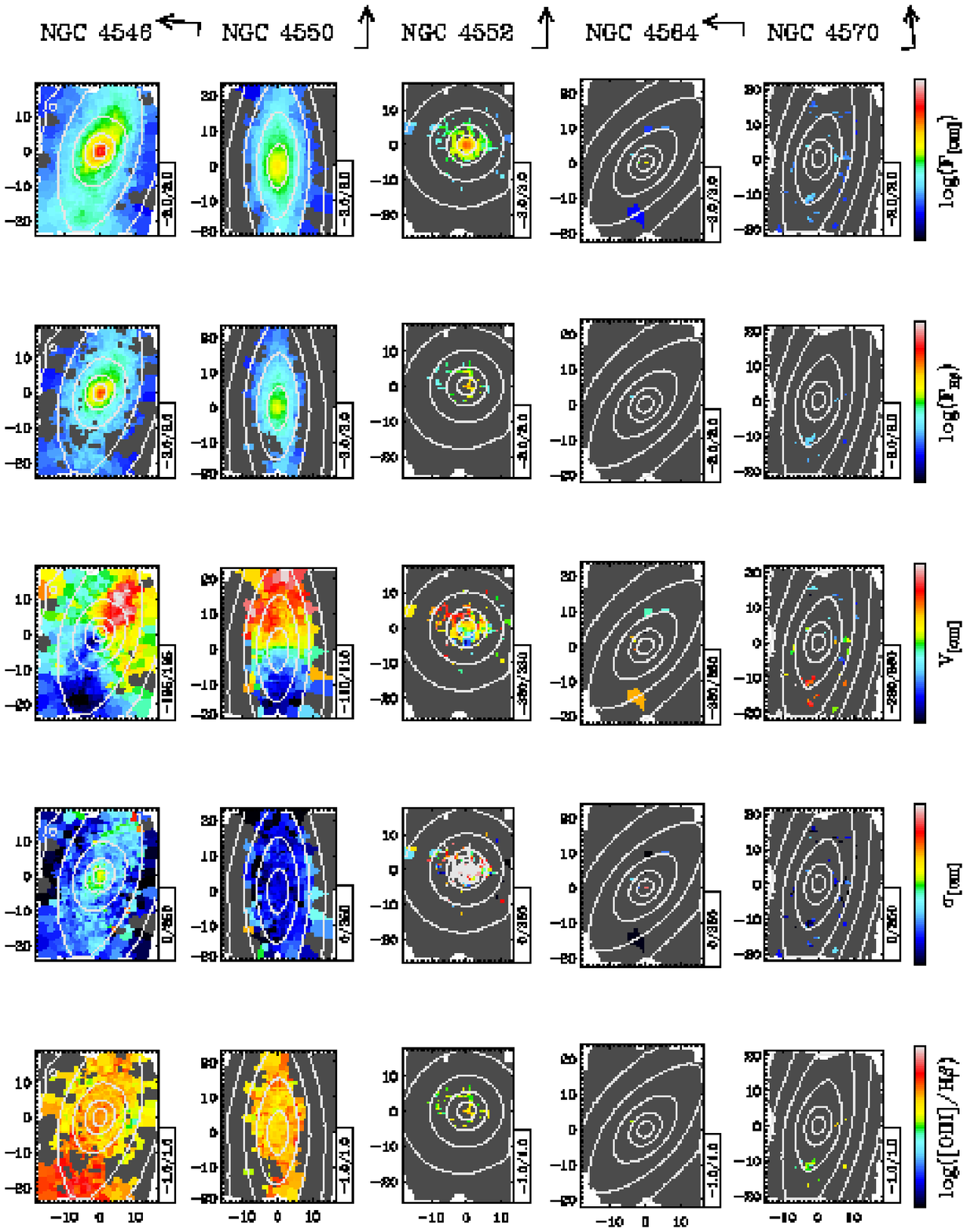}
\end{center}
\caption[]{Continue}
\end{figure*}
\addtocounter{figure}{-1}
\addtocounter{subfigure}{1}
\begin{figure*}
\begin{center}
  \includegraphics[width=0.98\textwidth, trim=0cm 0cm 0cm 0cm]{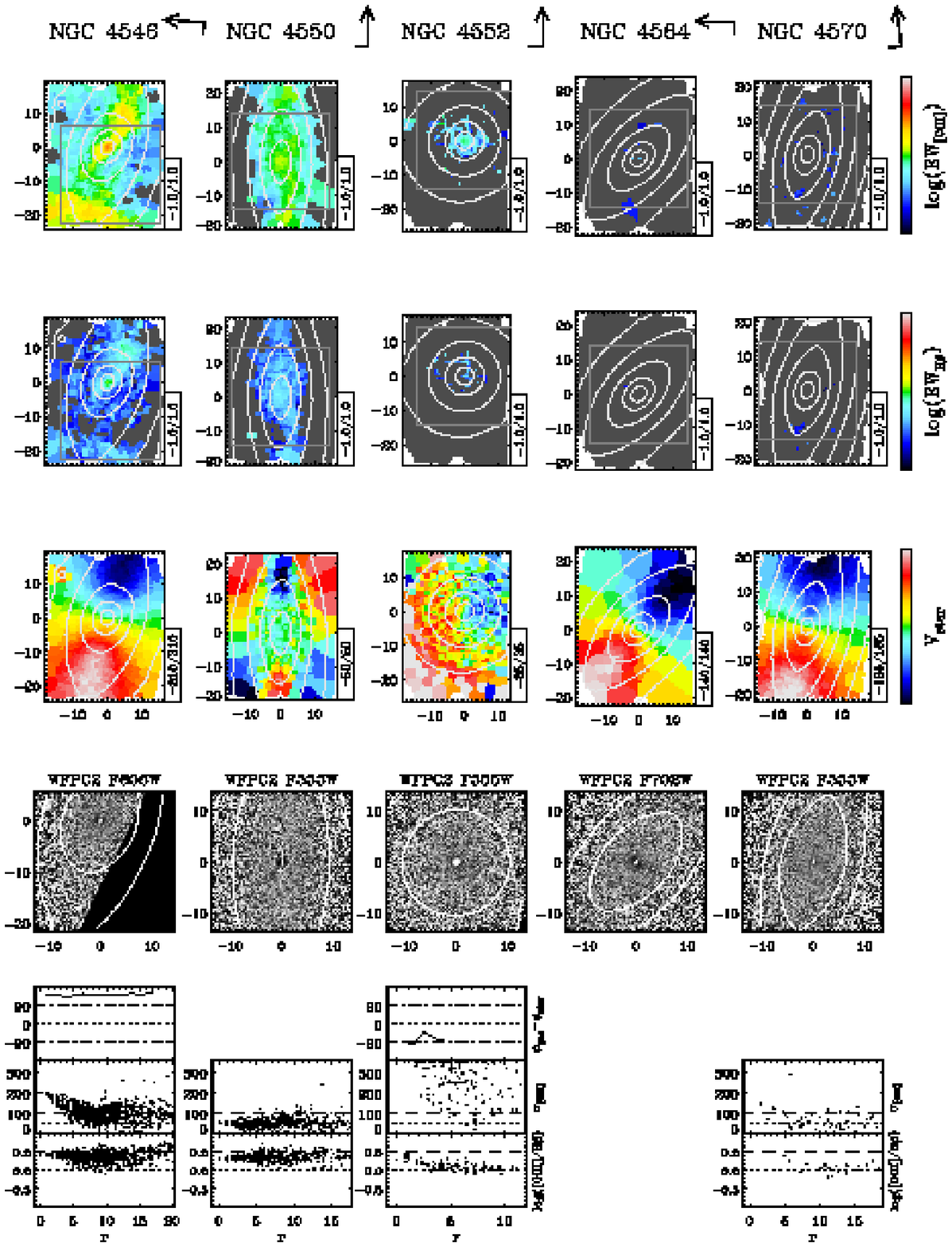}
\end{center}
\caption[]{Continue}
\end{figure*}

\clearpage
\addtocounter{figure}{-1}
\addtocounter{subfigure}{-1}
\begin{figure*}
\begin{center}
  \includegraphics[width=0.98\textwidth, trim=0cm 0cm 0cm 0cm]{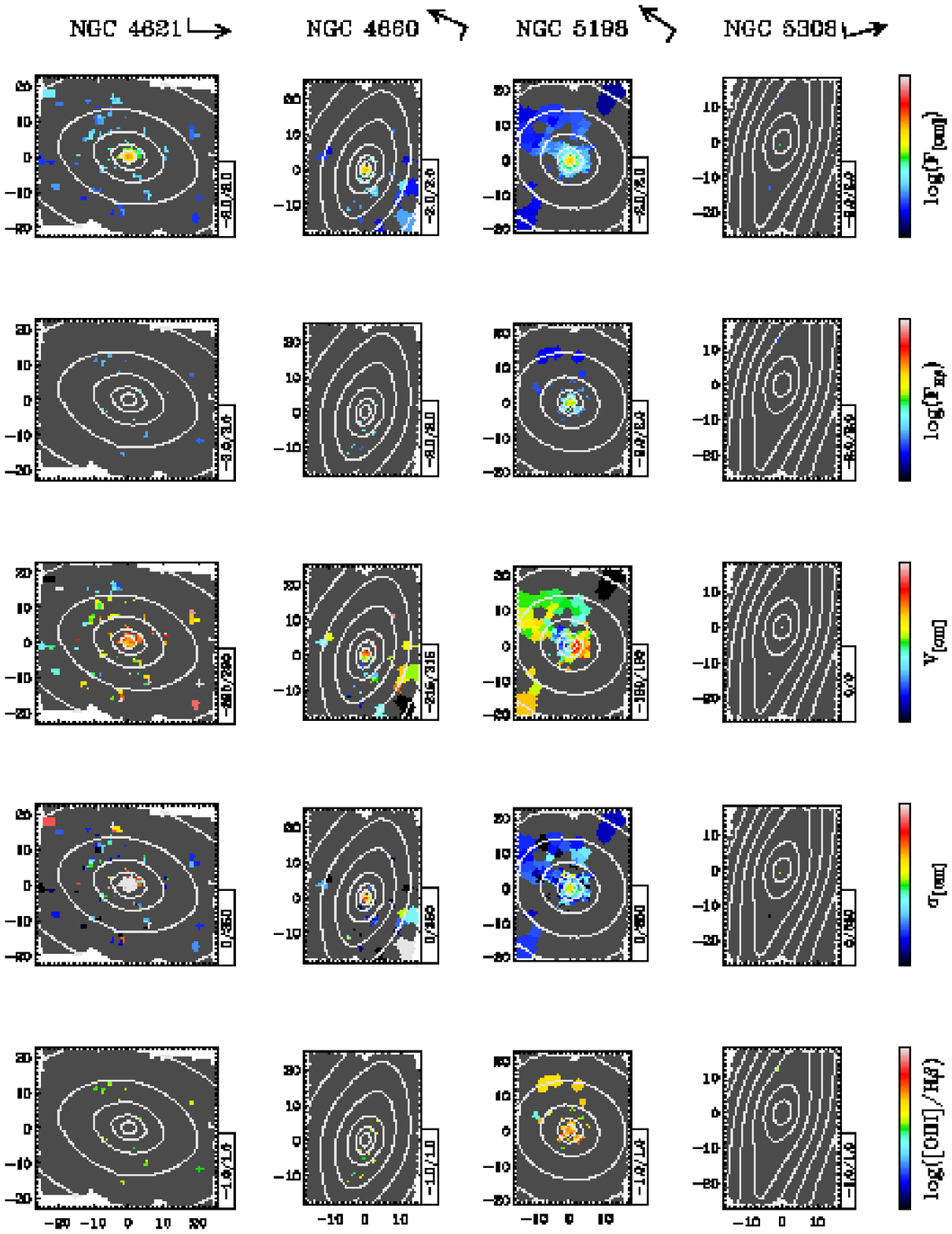}
\end{center}
\caption[]{Continue}
\end{figure*}
\addtocounter{figure}{-1}
\addtocounter{subfigure}{1}
\begin{figure*}
\begin{center}
  \includegraphics[width=0.98\textwidth, trim=0cm 0cm 0cm 0cm]{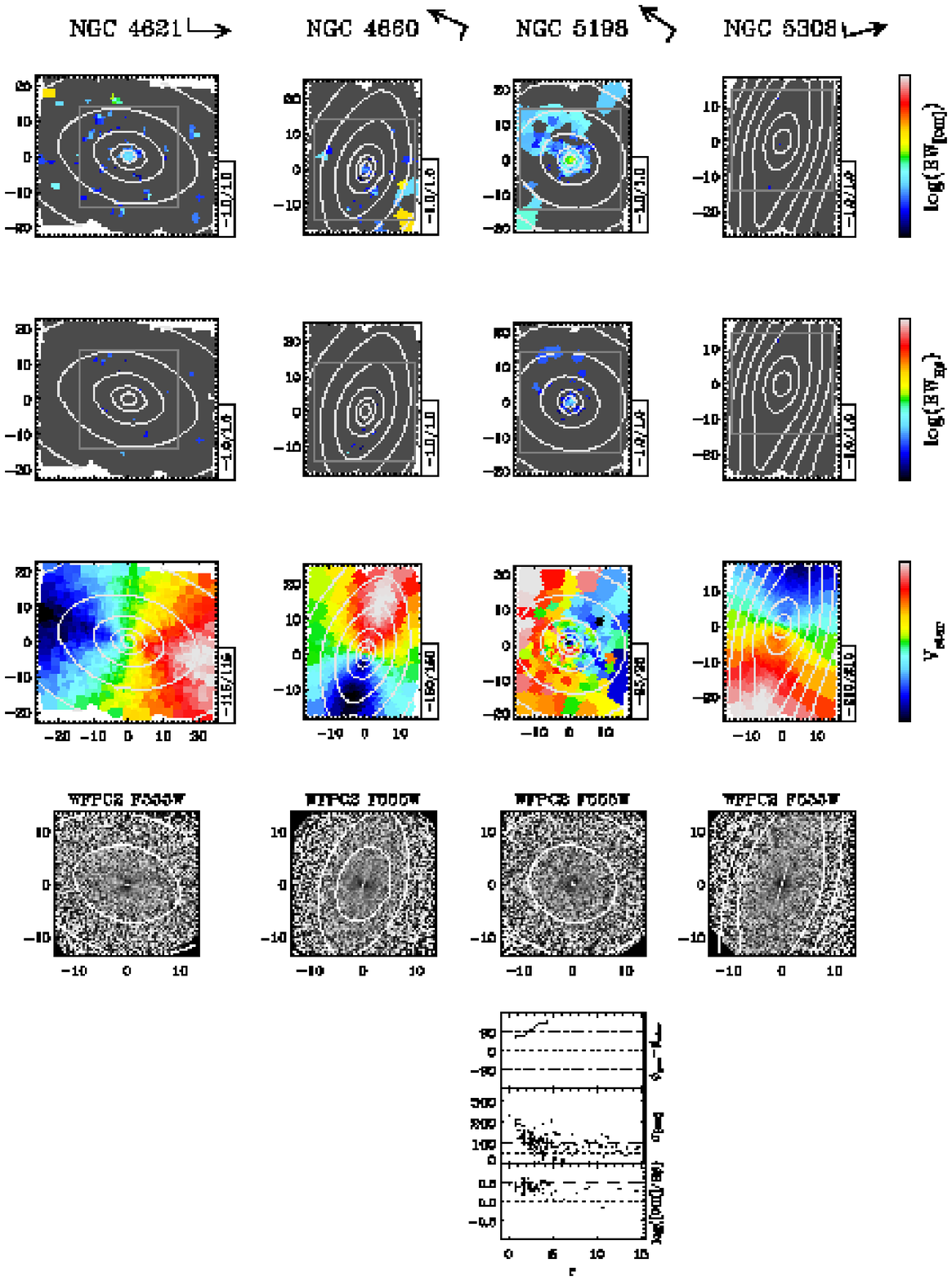}
\end{center}
\caption[]{Continue}
\end{figure*}

\clearpage
\addtocounter{figure}{-1}
\addtocounter{subfigure}{-1}
\begin{figure*}
\begin{center}
  \includegraphics[width=0.98\textwidth, trim=0cm 0cm 0cm 0cm]{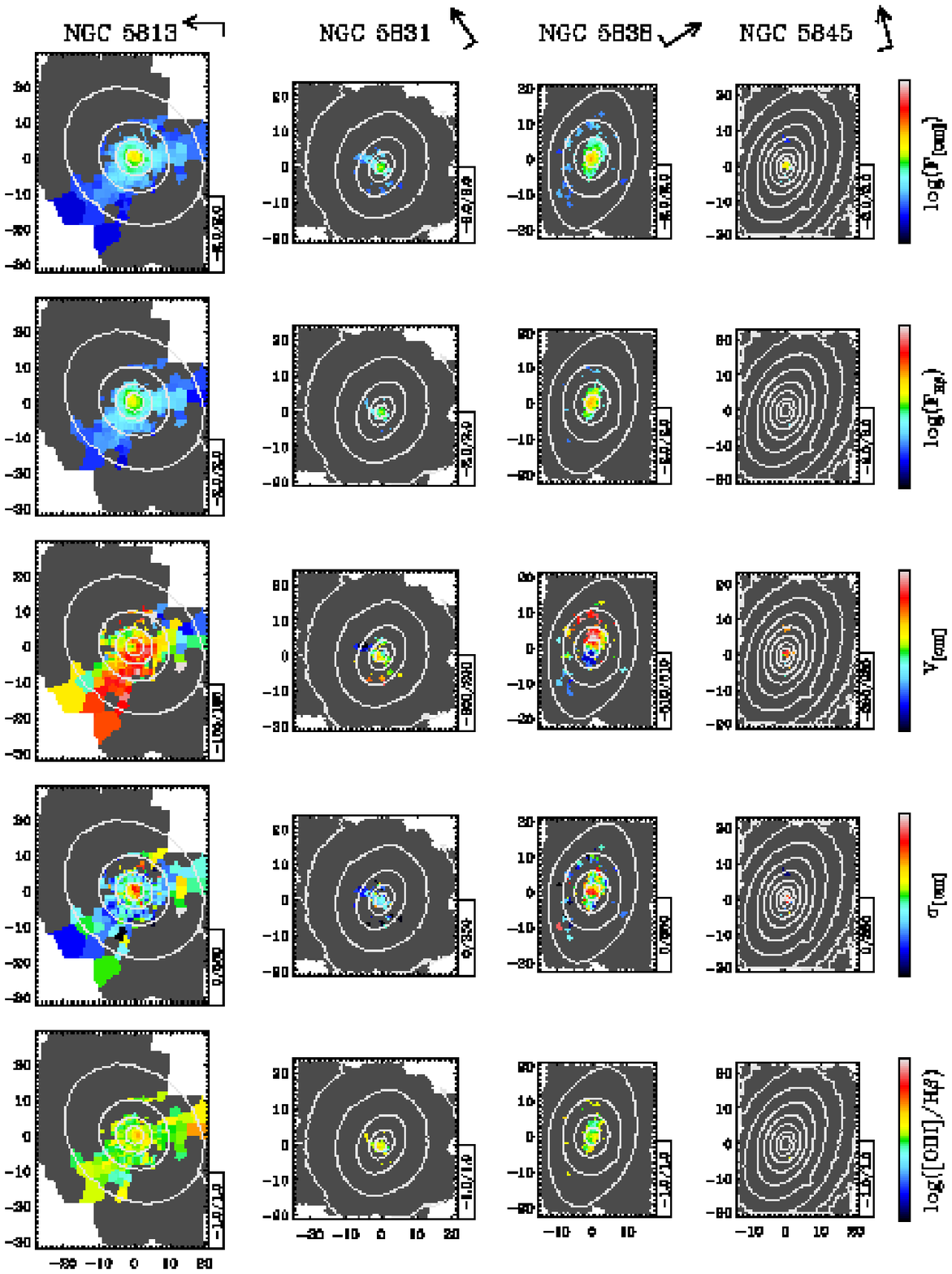}
\end{center}
\caption[]{Continue}
\end{figure*}
\addtocounter{figure}{-1}
\addtocounter{subfigure}{1}
\begin{figure*}
\begin{center}
  \includegraphics[width=0.98\textwidth, trim=0cm 0cm 0cm 0cm]{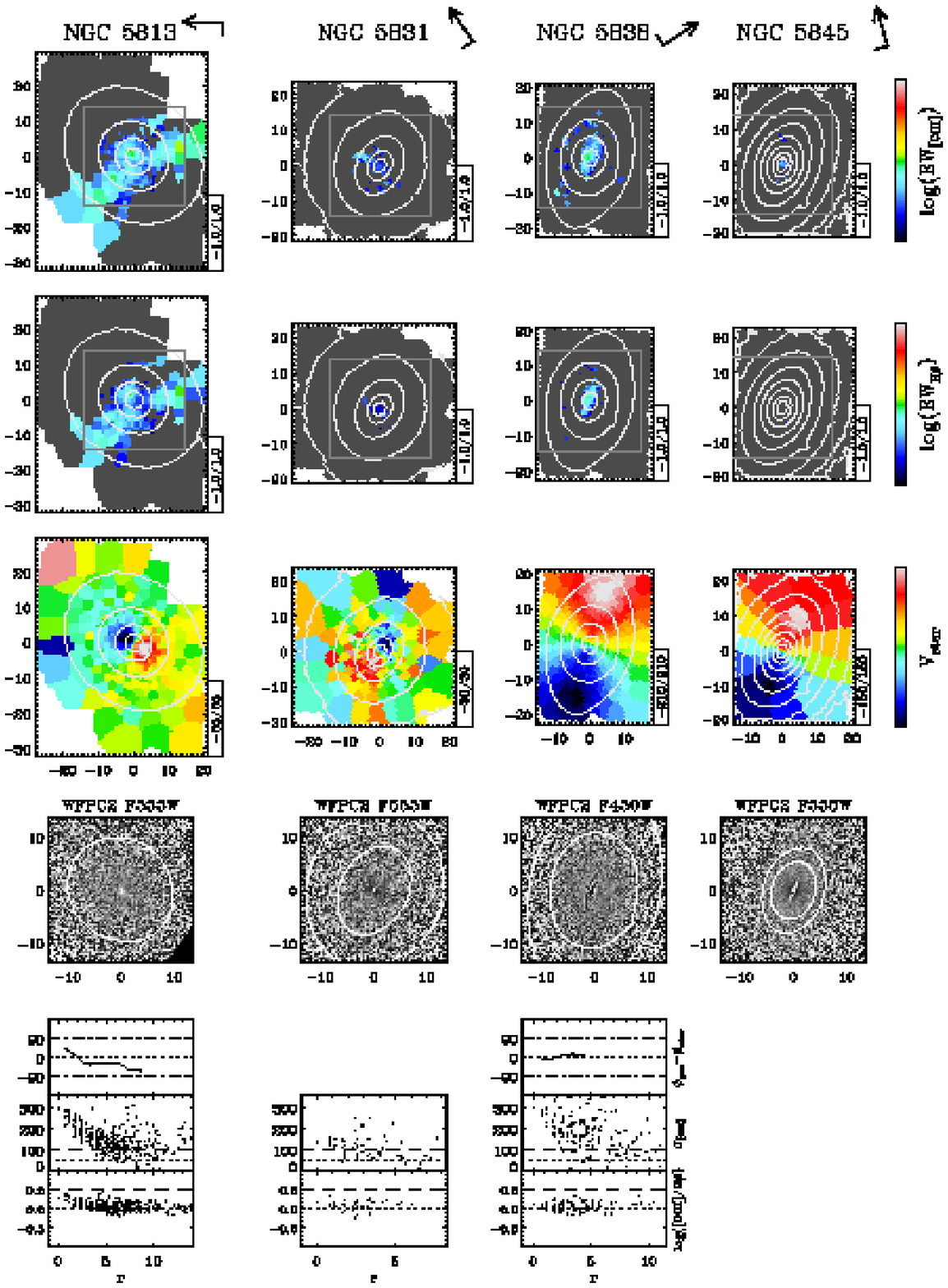}
\end{center}
\caption[]{Continue}
\end{figure*}

\clearpage
\addtocounter{figure}{-1}
\addtocounter{subfigure}{-1}
\begin{figure*}
\begin{center}
  \includegraphics[width=0.98\textwidth, trim=0cm 0cm 0cm 0cm]{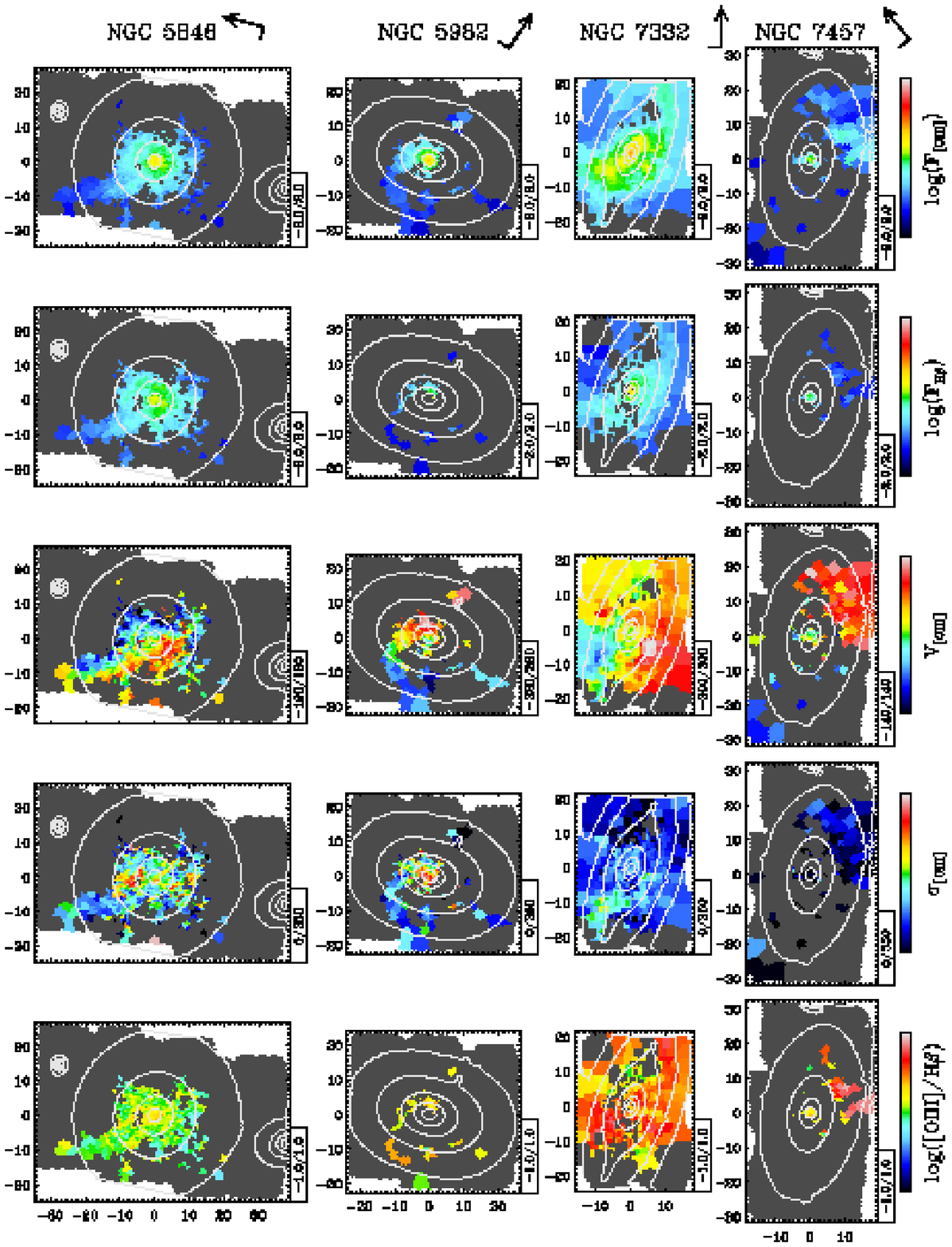}
\end{center}
\caption[]{Continue}
\end{figure*}
\addtocounter{figure}{-1}
\addtocounter{subfigure}{1}
\begin{figure*}
\begin{center}
  \includegraphics[width=0.98\textwidth, trim=0cm 0cm 0cm 0cm]{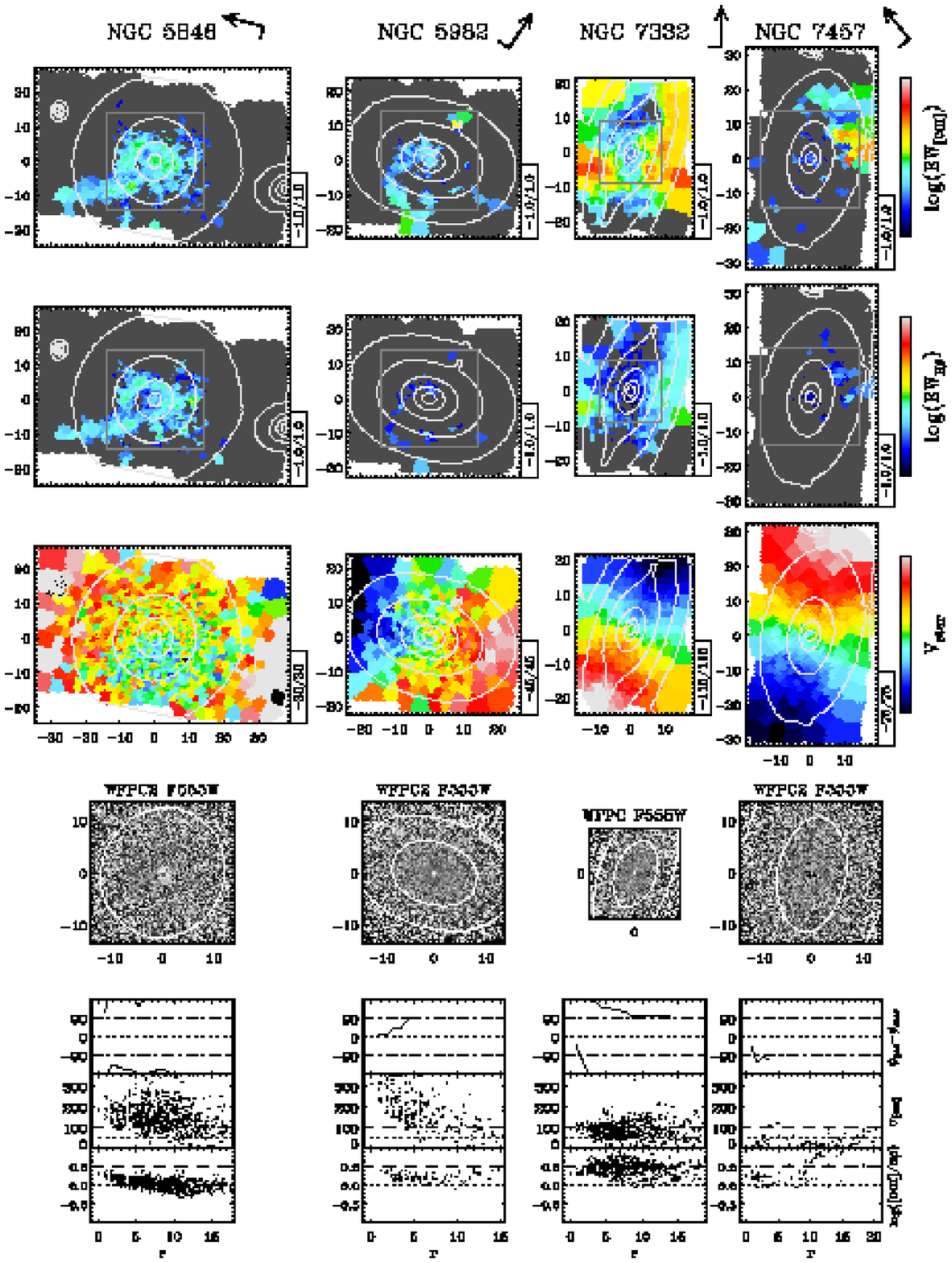}
\end{center}
\caption[]{Continue}
\end{figure*}

\clearpage
\renewcommand{\thefigure}{\arabic{figure}}
}

\newcommand{\placetabone}{
\begin{table*}
\caption{Ionised-gas emission the 48 E/S0 \sauron\ galaxies.}
\label{tab:allgal}
\begin{center}
\begin{tabular}{llcccccccccc}
\hline
NGC & Type & Environment & $\Delta m$ & $M_B$  & $\epsilon_{\rm 25}$  & Emission & \Ni\ & Dust & $\log{F(\rm H\beta)}$ & $\log{L(\rm H\alpha)}$ & $\log{\rm M_{H_{II}}}$ \\
x2 ~(1) &~(2) & (3) & (4) & (5) & (6) & (7) & (8) & (9) & (10) & (11) & (12)\\
\hline
\noalign{\smallskip}
\phantom{0}474  &S0$^0$(s)       & Field	 & 32.50 & -20.42 & 0.19 & yes	        &   no  & no  & -14.09 & 39.44 &  4.821 \\
\phantom{0}524  &S0$^+$(rs)      & Field	 & 32.58 & -21.40 & 0.01 & yes	        &   no  & yes & -14.57 & 39.00 &  4.373 \\
\phantom{0}821  &E6?             & Field	 & 31.86 & -20.44 & 0.32 &  no	        &   no  & no  & --     & --    &  --    \\
1023            &SB0$^-$(rs)     & Field	 & 30.06 & -20.42 & 0.56 & yes	        &   no  & no  & -14.10 & 38.46 &  3.835 \\
2549            &S0$^0$(r)sp     & Field	 & 31.12 & -19.36 & 0.68 & yes	        &   no  & no  & -14.18 & 38.80 &  4.179 \\
2685            &(R)SB0$^+$pec   & Field	 & 30.79 & -19.05 & 0.51 & yes	        &  yes  & yes & -13.48 & 39.37 &  4.747 \\
2695            &SAB0$^0$(s)     & Field	 & 31.83 & -19.38 & 0.27 &  no	        &   no  & --  & --     & --    &  --    \\
2699            &E:              & Field	 & 31.83 & -18.85 & 0.06 & yes	        &   no  & yes & -15.25 & 38.02 &  3.393 \\
2768            &E6:             & Field	 & 31.66 & -21.15 & 0.42 & yes	        &  yes  & yes & -13.55 & 39.65 &  5.025 \\
2974            &E4              & Field	 & 31.93 & -20.32 & 0.39 & yes	        &  yes  & yes & -13.35 & 39.96 &  5.333 \\
3032            &SAB0$^0$(r)     & Field	 & 31.68 & -18.77 & 0.11 & yes	        &  yes  & yes & -13.79 & 39.42 &  4.793 \\
3156            &S0:             & Field	 & 30.90 & -18.08 & 0.38 & yes	        &   no  & yes & -13.71 & 39.18 &  4.561 \\
3377            &E5-6            & Leo I group   & 30.14 & -19.24 & 0.39 & yes	        &   no  & yes & -13.63 & 38.96 &  4.337 \\
3379            &E1              & Leo I group   & 30.14 & -20.16 & 0.08 & yes	        &   no  & yes & -14.41 & 38.18 &  3.557 \\
3384            &SB0$^-$(s):     & Leo I group   & 30.14 & -19.56 & 0.49 & yes	        &   no  & yes & -14.67 & 37.92 &  3.297 \\
3414            &S0~pec          & Field	 & 31.52 & -19.78 & 0.17 & yes	        &  yes  & yes & -13.42 & 39.72 &  5.099 \\
3489            &SAB0$^+$(rs)    & Leo I group   & 30.14 & -19.32 & 0.38 & yes	        &   no  & yes & -12.95 & 39.64 &  5.017 \\
3608            &E2              & Field	 & 30.96 & -19.54 & 0.21 & yes	        &   no  & no  & -14.77 & 38.15 &  3.525 \\
4150            &S0$^0$(r)?      & Coma I cloud	 & 30.68 & -18.48 & 0.30 & yes	        &   no  & yes & -13.90 & 38.91 &  4.283 \\
4262            &SB0$^-$(s)      & Virgo cluster & 31.06 & -18.88 & 0.09 & yes	        &   no  & --  & -13.55 & 39.41 &  4.785 \\
4270            &S0              & Virgo cluster & 31.06 & -18.28 & 0.53 & traces       &   no  & no  & --     & --    &  --    \\
4278            &E1-2            & Coma I cloud	 & 30.68 & -19.93 & 0.06 & yes	        &  yes  & yes & -12.86 & 39.95 &  5.323 \\
4374            &E1              & Virgo cluster & 31.06 & -21.23 & 0.12 & yes	        &  yes  & yes & -13.66 & 39.30 &  4.675 \\
4382            &S0$^+$(s)pec    & Virgo cluster & 31.06 & -21.28 & 0.22 & weak [OIII]	&   no  & no  & --     & --    &  --    \\
4387            &E               & Virgo cluster & 31.06 & -18.34 & 0.34 & no	        &   no  & no  & --     & --    &  --    \\
4458            &E0-1            & Virgo cluster & 31.06 & -18.42 & 0.06 & traces       &   no  & no  & --     & --    &  --    \\
4459            &S0$^+$(r)       & Virgo cluster & 31.06 & -19.99 & 0.23 & yes	        &   no  & yes & -14.07 & 38.89 &  4.265 \\
4473            &E5              & Virgo cluster & 31.06 & -20.26 & 0.38 & traces       &   no  & no  & --     & --    &  --    \\
4477            &SB0(s):?        & Virgo cluster & 31.06 & -19.96 & 0.09 & yes	        &   no  & yes & -13.54 & 39.42 &  4.795 \\
4486            &E0-1$^+$pec     & Virgo cluster & 31.06 & -21.79 & 0.30 & yes	        &  yes  & yes & -13.26 & 39.70 &  5.075 \\ 
4526            &SAB0$^0$(s)     & Virgo cluster & 31.06 & -20.68 & 0.63 & yes	        &  yes  & yes & -13.74 & 39.22 &  4.595 \\ 
4546            &SB0$^-$(s):     & Virgo cluster & 31.06 & -19.98 & 0.50 & yes	        &  yes  & yes & -13.22 & 39.74 &  5.115 \\
4550            &SB0$^0$:sp      & Virgo cluster & 31.06 & -18.83 & 0.71 & yes	        &   no  & yes & -13.67 & 39.29 &  4.665 \\
4552            &E0-1            & Virgo cluster & 31.06 & -20.58 & 0.09 & yes	        &   no  & yes & -14.27 & 38.69 &  4.065 \\
4564            &E               & Virgo cluster & 31.06 & -19.39 & 0.45 &  no	        &   no  & no  & --     & --    &  --    \\
4570            &S0~sp           & Virgo cluster & 31.06 & -19.54 & 0.68 & yes	        &   no  & no  & -16.02 & 36.94 &  2.315 \\
4621            &E5              & Virgo cluster & 31.06 & -20.64 & 0.24 & traces       &   no  & no  & --     & --    &  --    \\
4660            &E               & Virgo cluster & 31.06 & -19.22 & 0.21 & traces       &   no  & no  & --     & --    &  --    \\
5198            &E1-2:           & Field	 & 32.80 & -20.38 & 0.14 & yes	        &   no  & no  & -14.52 & 39.13 &  4.511 \\
5308            &S0$^-$~sp       & Field	 & 32.26 & -20.27 & 0.82 &  no	        &   no  & no  & --     & --    &  --    \\
5813            &E1-2            & Field	 & 32.10 & -20.99 & 0.24 & yes	        &  yes  & yes & -14.01 & 39.36 &  4.741 \\
5831            &E3              & Field	 & 31.79 & -19.73 & 0.13 & yes	        &   no  & no  & -15.06 & 38.19 &  3.567 \\
5838            &S0$^-$          & Field	 & 31.36 & -19.87 & 0.59 & yes	        &  yes  & yes & -14.29 & 38.79 &  4.165 \\
5845            &E:              & Field	 & 31.69 & -18.58 & 0.32 & weak [OIII]  &   no  & yes & --     & --    &  --    \\
5846            &E0-1            & Field	 & 31.98 & -21.24 & 0.06 & yes	        &  yes  & yes & -13.85 & 39.48 &  4.853 \\
5982            &E3              & Field	 & 33.11 & -21.46 & 0.30 & yes	        &   no  & no  & -14.79 & 38.99 &  4.365 \\
7332            &S0~pec~sp       & Field	 & 31.42 & -19.93 & 0.73 & yes	        &   no  & yes & -13.43 & 39.67 &  5.049 \\
7457            &S0$^-$(rs)?     & Field	 & 30.46 & -18.81 & 0.41 & yes	        &   no  & no  & -14.49 & 38.23 &  3.605 \\

\hline
\end{tabular}
\end{center}
Notes: (1)~NGC number.  (2)~Hubble type (RC3: \citeauthor{RC3}
\citeyear{RC3}).  (3)--(5)~Galactic Environment, distance modulus $\Delta m$ in 
mag, and absolute $B$-band magnitude, from Paper~II. (6)~Ellipticity
$\epsilon_{\rm 25}$ of the 25 $B$-band mag arcsec$^{-2}$ isophote
(LEDA). (7)~Presence of \Hb\ or \Oiii$\lambda\lambda$4959,5007
emission. (8)~Presence of \Ni$\lambda\lambda$5198,5200 lines.
(9)~Presence of dust features in the \HST\ images.  (10) --
(12)~Estimated (see text) total \Hb\ flux, \Ha\ luminosity and mass of
the ionised gas in $\rm erg\,s^{-1}cm^{-2}$, $\rm erg\,s^{-1}$, and
$M_\odot$, respectively, for objects with clearly detected
emission.
\end{table*}
}

\newcommand{\placetabtwo}{
\begin{table}
\caption{Kinematic misalignment between gas and stars}
\label{tab:misal}
\begin{center}
\begin{tabular}{lccc}
\hline
NGC & $\phi_{\rm gas}-\phi_{\rm star}$ & $\Delta(\phi_{\rm gas}-\phi_{\rm star})$& $r_{\rm max}$ \\
 ~(1) &~(2) & (3) & (4) \\
\hline
\noalign{\smallskip}
\phantom{0}474  &   74 & 16 & 3.5 \\ 
\phantom{0}524  &    1 & 22 & 20  \\ 
1023            &  -31 & 10 & 18  \\ 
2549            &   11 & 15 & 3.5 \\ 
2685            &   73 & 14 & 20  \\ 
2768            &  -95 &  7 & 20  \\ 
2974            &    2 &  7 & 20  \\ 
3032            & -151 & 22 & 11  \\ 
3156            &  -11 &  9 & 15  \\ 
3377            &   11 &  6 & 15  \\ 
3379            &   43 & 11 &  3  \\ 
3384            &   13 & 11 &  5  \\ 
3414            &   75 & 32 &  9  \\ 
3489            &   -6 &  6 & 18  \\ 
3608            &  148 & 19 &  5  \\ 
4150            &   21 & 12 & 2-15\\ 
4262            & -119 &  6 & 15  \\ 
4278            &   29 & 19 & 22  \\ 
4374            & -179 & 44 &  6  \\ 
4459            &   -1 &  2 &  9  \\ 
4477            &  -28 &  5 & 15  \\ 
4526            &    6 &  5 & 10  \\ 
4546            &  144 &  5 & 17  \\ 
4550            &    0 & -- & --  \\ 
4552            &  -78 & 22 &  5  \\ 
4570            &    0 & -- & --  \\ 
5198            &   75 & 16 &  3  \\ 
5813            &   28 & 33 &  9  \\ 
5838            &   11 & 11 &  5  \\ 
5846            & -156 & 30 & 10  \\ 
5982            &   13 & 17 &  3  \\ 
7332            &  132 & 68 & 15  \\ 
7457            &  -95 & 29 &  4  \\ 
     
\hline
\end{tabular}
\end{center}
Notes: 
(1)~NGC number. 
(2)~Median kinematic misalignment in degrees.
(3)~Standard deviation for the kinematic misalignment in degrees.
(4)~Maximum distance from the center used to derived the values in
Columns (2) and (3). For NGC~4150 we excluded also the central 2\arcsec.
\end{table}
}    

\newcommand{\placefigApponea}{
\renewcommand{\thefigure}{A\arabic{figure}\alph{subfigure}}
\setcounter{subfigure}{1}
\begin{figure*}
\begin{center}
 \includegraphics[height=\textwidth, angle=90, trim=0cm 0cm 0cm 0cm]{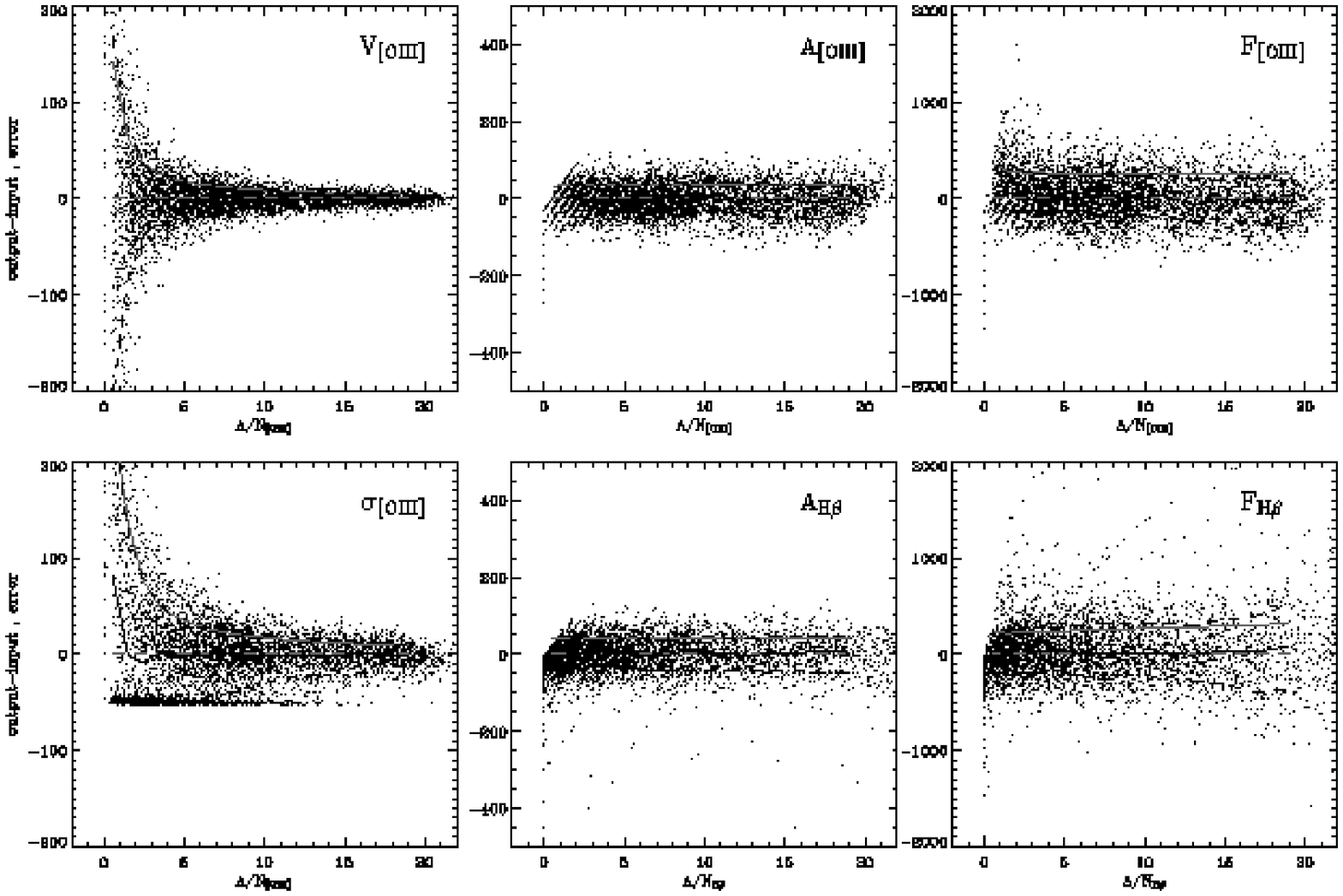}
\end{center}

\caption[]{
Recovery accuracy of the emission-line parameters as a function of the
measured $A/N$ for simulations with a statistical $S/N=60$.
{\it Left panels:} Difference between the output and input gas
velocity ({\it top\/}) and intrinsic velocity dispersion ({\it
bottom\/}). The {\it solid\/} and {\it dashed lines\/} indicate the
median values for the output-input difference and the 68\% confidence
region around them, respectively. The {\it grey lines\/} show the
median values of the formal uncertainties on the measured parameters.
{\it Middle panels:} Same as left, but now for the amplitudes of the
\Oiii\ ({\it top\/}) and \Hb\ ({\it bottom\/}) lines.
{\it Right panels:} Same as left, but now for the fluxes of the
\Oiii\ ({\it top\/}) and \Hb\ ({\it bottom\/}) lines.
The accuracy in estimating \Vg\ and \Sg\ is a strong function of
$A/N$, while the measurement of the amplitudes is not affected by
it. The same applies for the fluxes of the \Oiii\ lines, but for
strong \Hb\ lines these can be subject to larger errors if the width
of the lines have been estimated from quite weak \Oiii\ lines. In these
simulations the main limit to the recovery accuracy of the amplitudes
is set by the level of statistical noise in the spectra.}
\label{fig:appendix1a}
\end{figure*}
}

\newcommand{\placefigApponeb}{
\addtocounter{figure}{-1}
\addtocounter{subfigure}{1}

\begin{figure*}
\begin{center}
 \includegraphics[height=\textwidth, angle=90, trim=0cm 0cm 0cm 0cm]{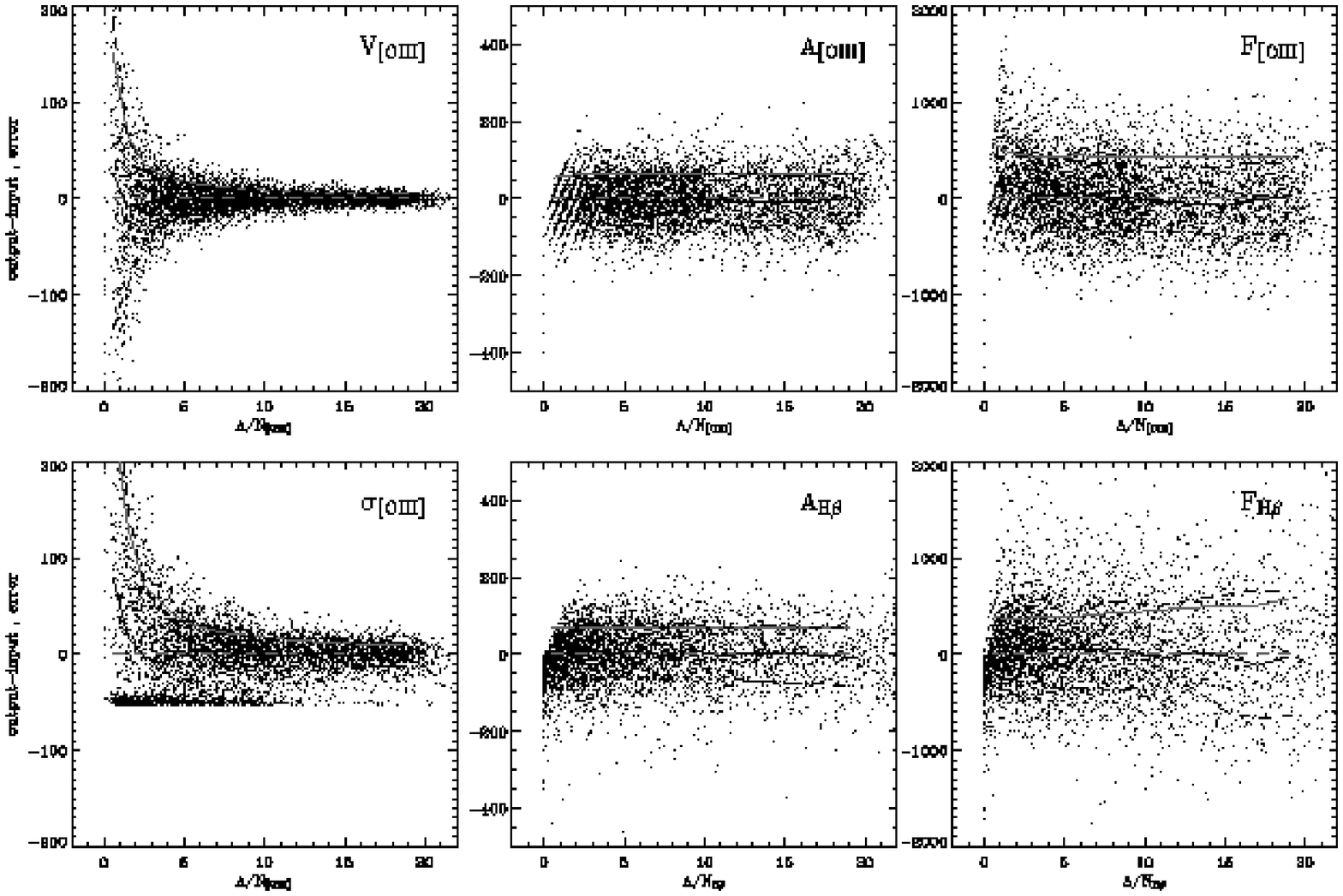}
\end{center}
\caption[]{
Same as Figure~\ref{fig:appendix1a} but for $S/N=100$. Notice how the
amplitude and flux of the lines are less accurately estimated.}
\label{fig:appendix1b}
\end{figure*}
}

\newcommand{\placefigApponec}{
\addtocounter{figure}{-1}
\addtocounter{subfigure}{1}

\begin{figure*}
\begin{center}
 \includegraphics[height=\textwidth, angle=90, trim=0cm 0cm 0cm 0cm]{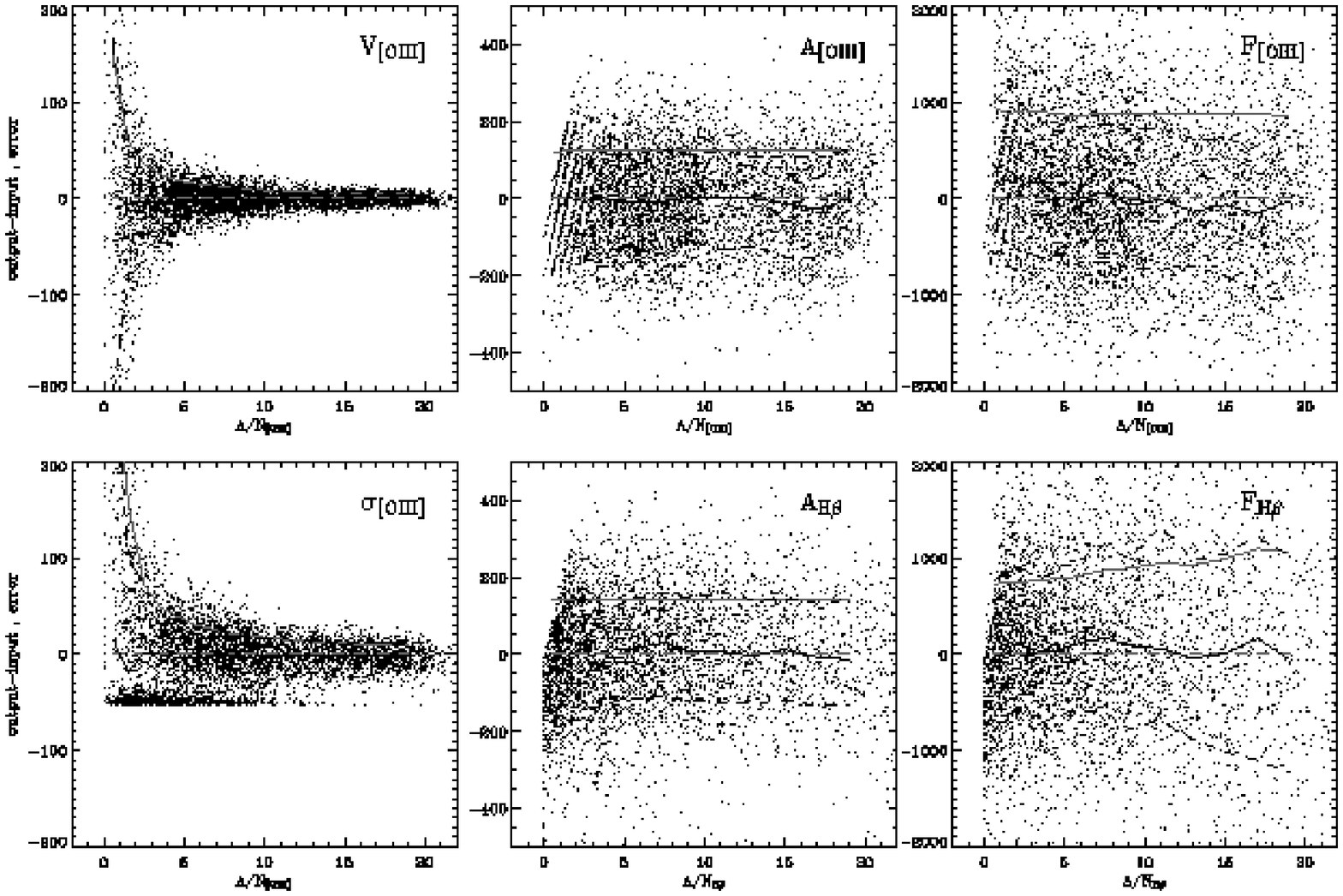}
\end{center}
\caption[]{Same as Figure~\ref{fig:appendix1a} but for $S/N=200$}
\label{fig:appendix1c}
\end{figure*}
}

\newcommand{\placefigApptwo}{
\renewcommand{\thefigure}{A\arabic{figure}}
\begin{figure*}
\begin{center}
 \includegraphics[height=\textwidth, angle=90, trim=0cm 0cm 0cm 0cm]{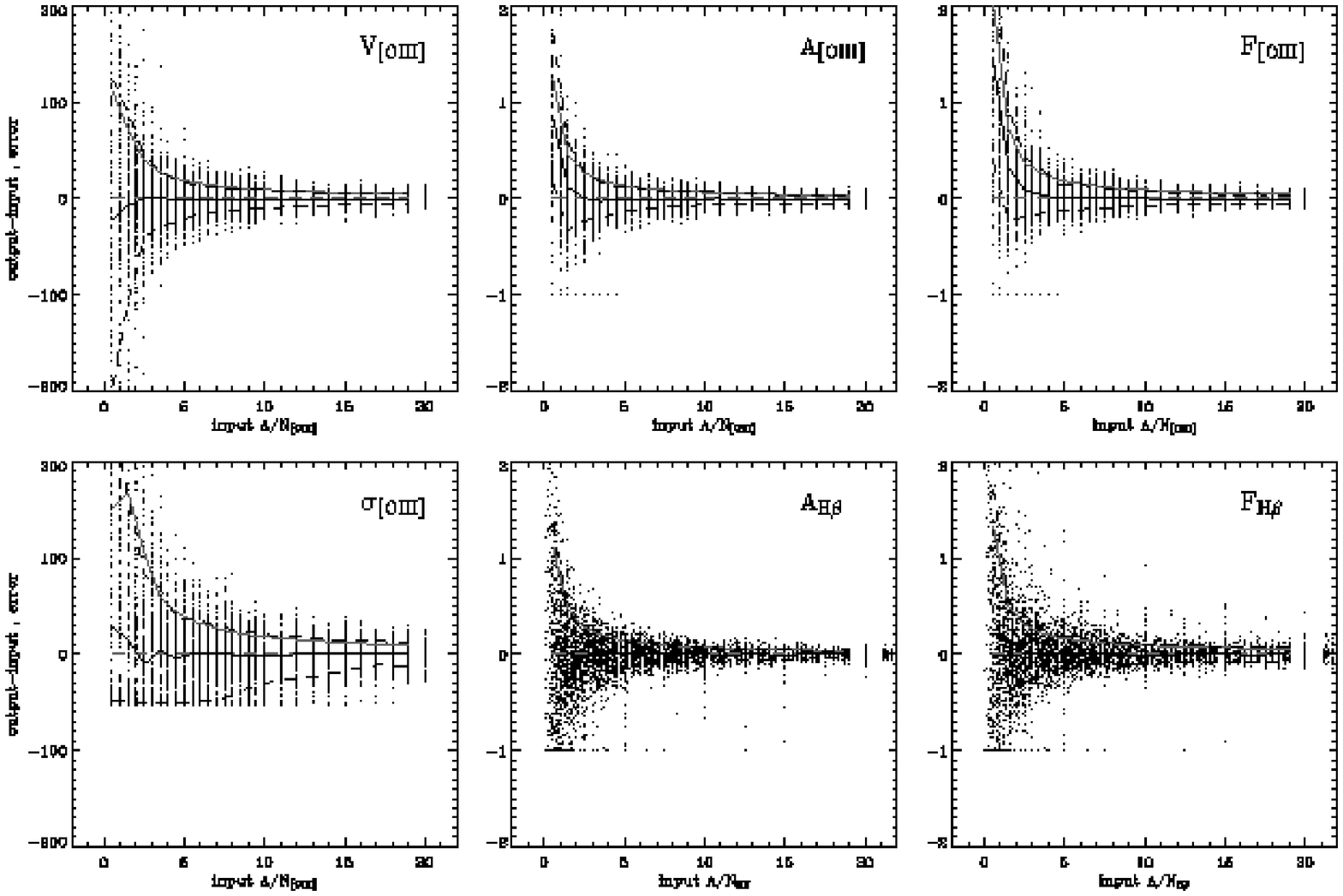}
\end{center}
\caption[]
{Same as Figure~\ref{fig:appendix1a} but for now showing the recovery
accuracy against the input $A/N$ values. Furthermore for the
amplitudes and the fluxes of the lines ({\it middle and left panels}),
both the input-output deviations and the formal uncertainties are
now relative to the input values.}
\label{fig:appendix2}
\end{figure*}
}

\newcommand{\placefigAppthree}{
\begin{figure*}
\begin{center}
 \includegraphics[height=\textwidth, angle=90, trim=0cm 0cm 0cm 0cm]{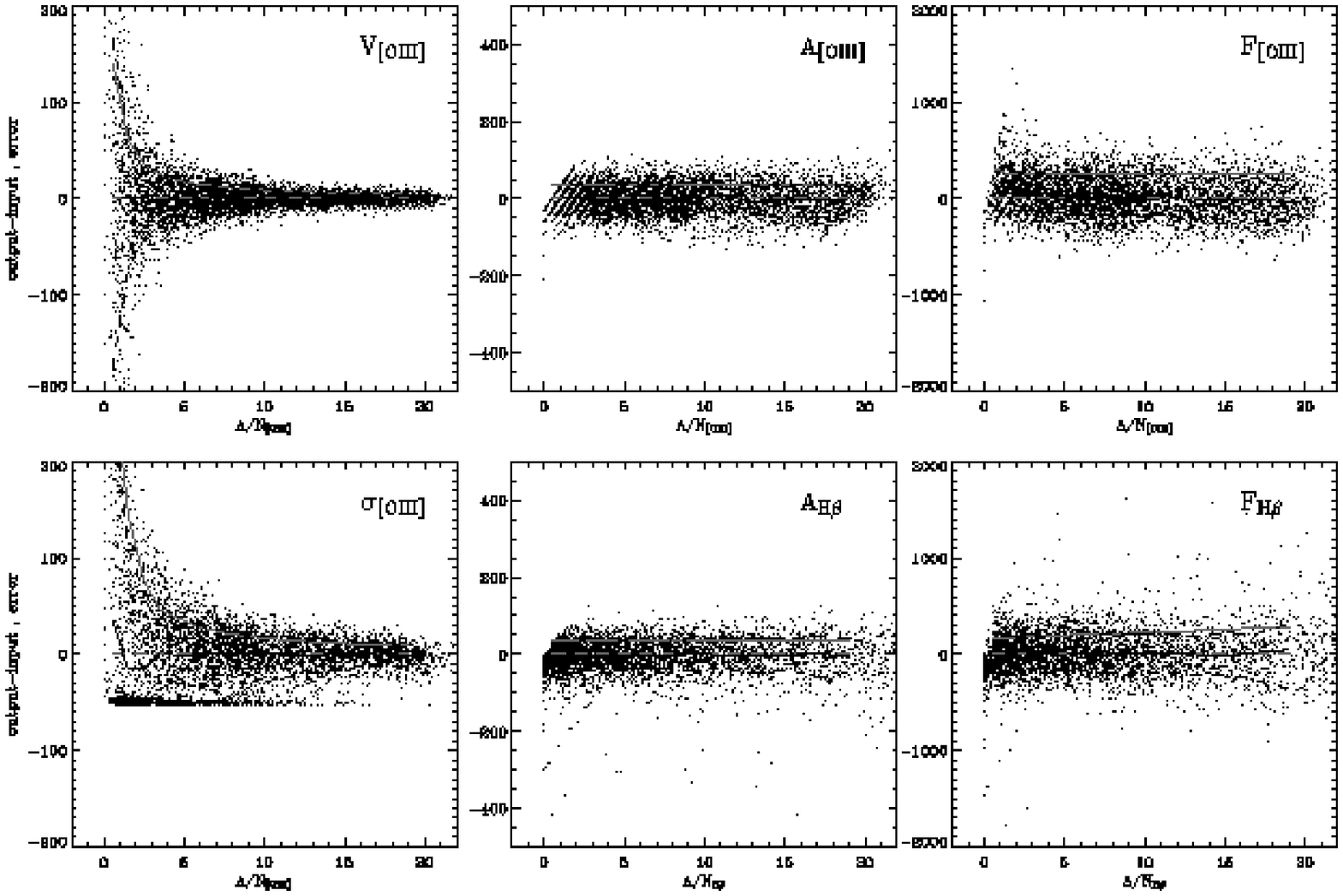}
\end{center}
\caption[]
{Same as Figure~\ref{fig:appendix1a} but for models where only the right,
input, stellar template have been used to match the stellar continuum.
Notice how the accuracy in any of derived parameters for the \Oiii\
lines has not substantially increased with respect to those shown in
Figure~\ref{fig:appendix1a}, while the improvement is clearly visible
for the \Hb\ amplitudes and fluxes.}
\label{fig:appendix3}
\end{figure*}
}

\newcommand{\placefigAppOIIIind}{
\begin{figure*}
\begin{center}
 \includegraphics[height=0.98\textwidth, angle=-90, trim=0cm 0cm 0cm 0cm]{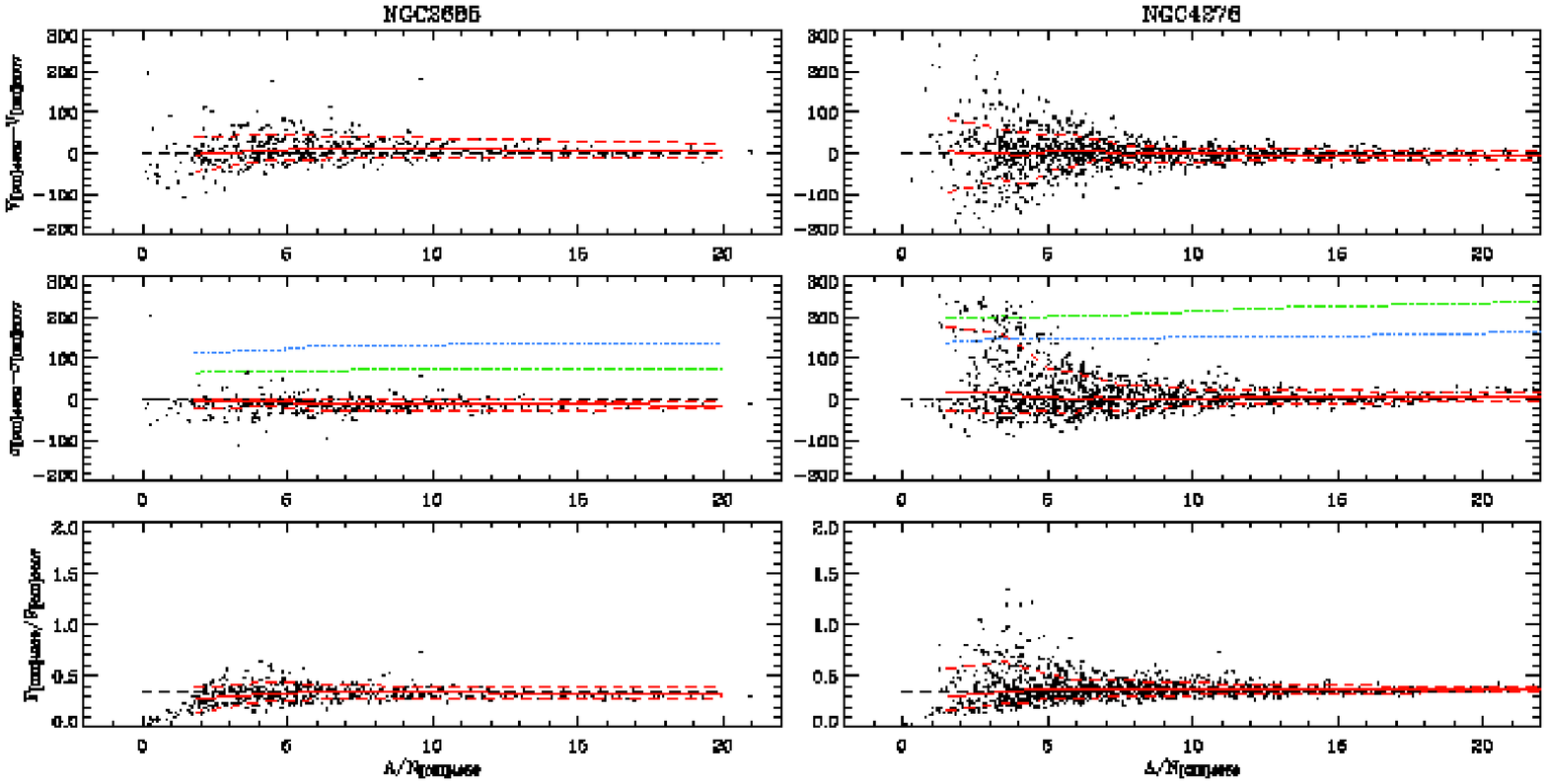}
\end{center}
\caption[]
{Comparison between emission-line parameters measured independently
from the \Oiii$\lambda4959$ and \Oiii$\lambda5007$ lines, as a
function of the $A/N$ of \Oiii$\lambda4959$ and for the case of
NGC~2865 ({\it left}) and NGC~4278 ({\it right}).
The {\it top\/} and {\it middle} panels show the difference between
the measured gas velocities and {\it observed\/} velocity dispersion.
The {\it lower} panel shows the ratio between the fluxes of the
\Oiii$\lambda4959$ and \Oiii$\lambda5007$ lines.
In each panel the {\it horizontal dashed line\/} indicates our
expectations (i.e. no difference between the velocities and velocity
dispersions and a flux ratio of 0.33), while the {\it solid\/} and
{\it dashed red lines\/} show the median values for the plotted
differences or ratios and the 68\% confidence region around them,
respectively.
In addition, in the {\it middle} panels the {\it blue dotted line\/}
indicates the gas velocity dispersion as traced by the median width of
the \Oiii$\lambda5007$ lines, while the {\it green dashed line} shows the
median stellar velocity dispersion at the location of the
emission-line fits.}
\label{fig:appendix4}
\end{figure*}
}

\newcommand{\placefigAppfoura}{
\renewcommand{\thefigure}{C\arabic{figure}\alph{subfigure}}
\setcounter{subfigure}{1}

\begin{figure*}
\begin{center}
 \includegraphics[width=0.98\textwidth,trim = 0cm 0cm 0cm 0cm]{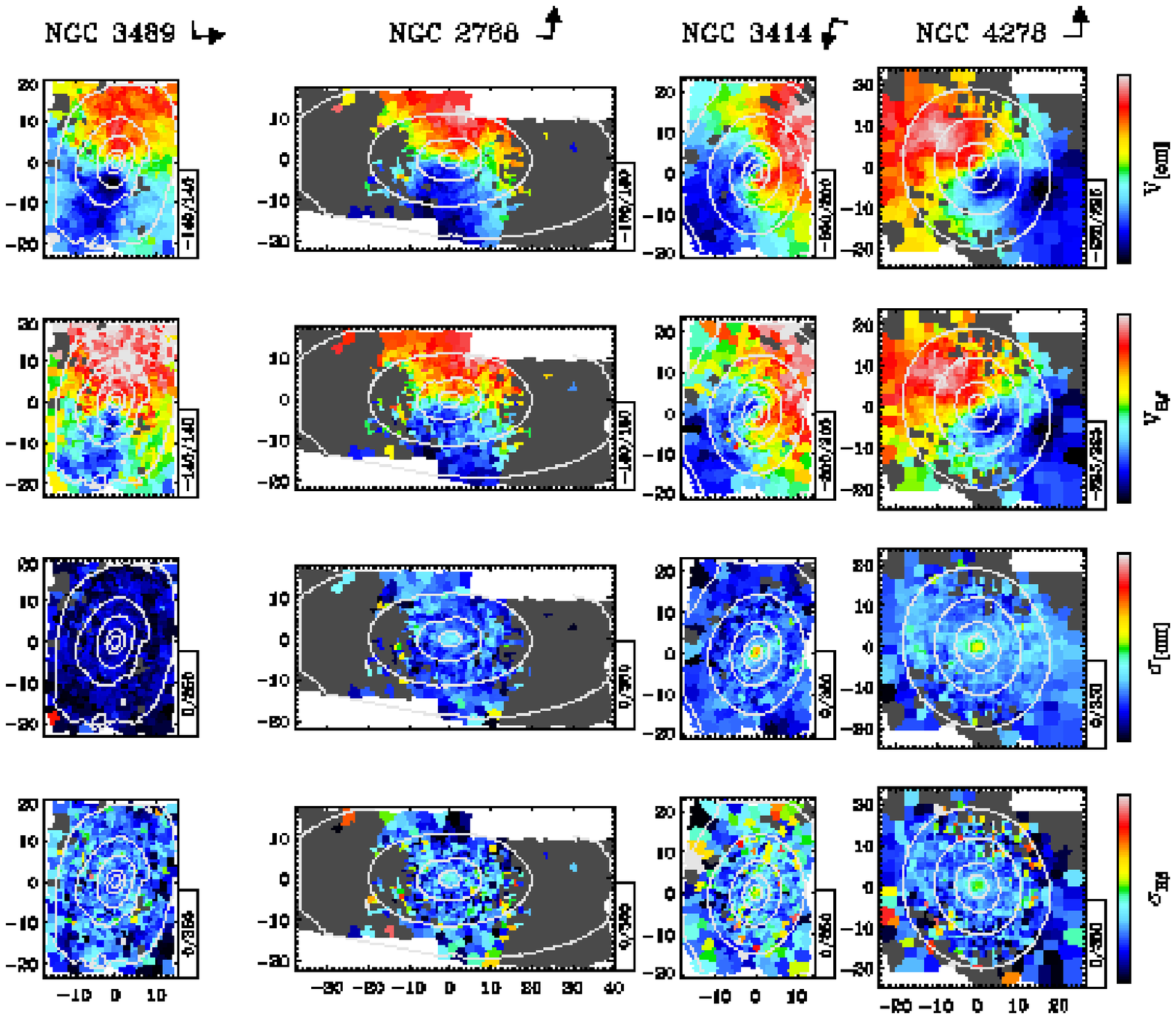}
\end{center}

\caption[]{Maps for independently derived \Oiii\ and \Hb\ kinematics.  
From top to bottom: i) and ii) mean velocity of the \Oiii\ and \Hb\ lines
iii) and iv) intrinsic velocity dispersion of the \Oiii\ and \Hb\
lines. The maps show only regions where both \Oiii\ and \Hb\ lines
were detected following our standard approach.
Left: the case of NGC~3489 is shown again to highlight the main
characteristics of an unreliable \Hb\ kinematics. In addition to the
velocity bias discussed in \S\ref{subsec:GasPaper_method}, the
spurious \Hb\ lines are considerably broader (by $\sim50$\kms) than
the \Oiii\ lines.
Right: three galaxies with reliable independently derived \Hb\
kinematics over most of the regions where emission is observed. In the
central regions of all three galaxies the \Oiii\ lines trace faster
rotation velocities than the \Hb\ line. In NGC~3414 and NGC~4278 the
central \Sg\ increase of the \Hb\ lines appears also to be much less
dramatic than the \Sg\ increase of the \Oiii\ lines. For NGC~3414,
template-mismatch is affecting the \Hb\ kinematics measured in the
outer parts of the maps.}
\label{fig:appendix5a}
\end{figure*}
}

\newcommand{\placefigAppfourb}{
\addtocounter{figure}{-1}
\addtocounter{subfigure}{1}

\begin{figure*}
\begin{center}
 \includegraphics[width=0.98\textwidth,trim = 0cm 0cm 0cm 0cm]{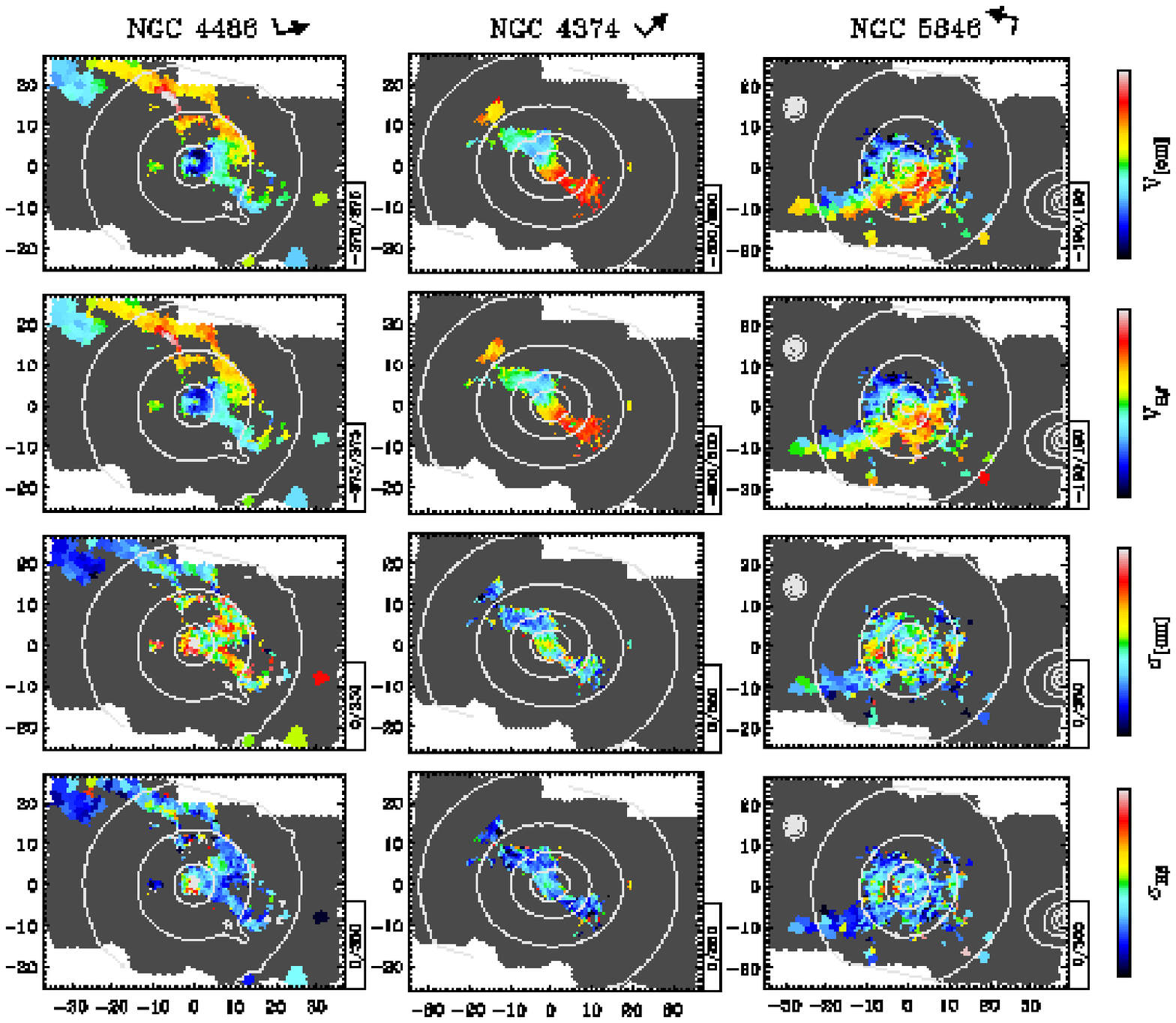}
\end{center}
\caption[]{Same as Figure \ref{fig:appendix5a}, but now showing giant elliptical galaxies. 
Notice the narrower width of the \Hb\ lines at all radii.}
\label{fig:appendix5b}
\end{figure*}
}

\newcommand{\placefigAppfourc}{
\addtocounter{figure}{-1}
\addtocounter{subfigure}{1}

\begin{figure*}
\begin{center}
 \includegraphics[width=0.98\textwidth,trim = 0cm 0cm 0cm 0cm]{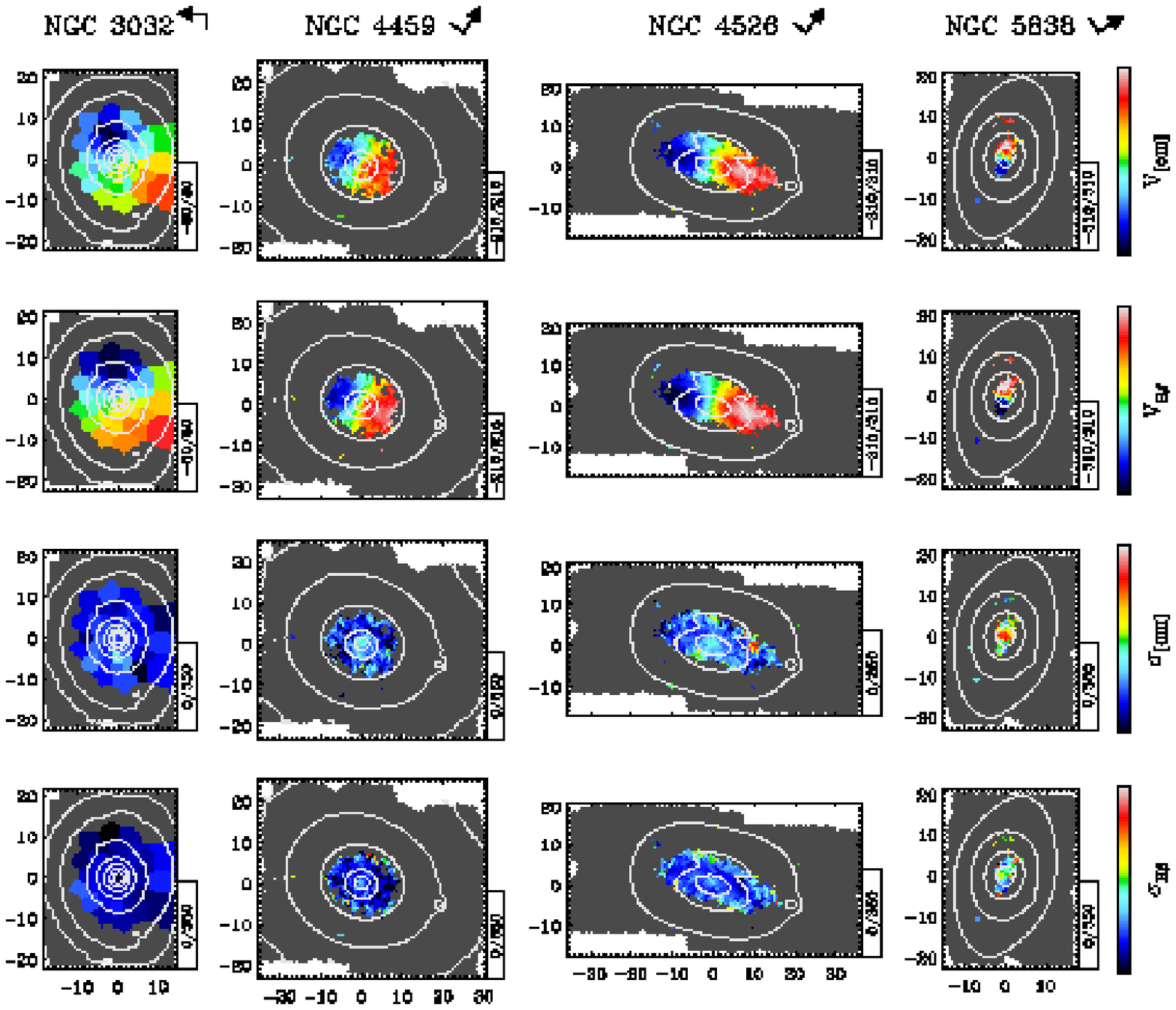}
\end{center}
\caption[]{Same as Figure \ref{fig:appendix5a}, but now showing galaxies with 
circularly symmetric dust lanes. In all cases the \Hb\ kinematics show
faster rotation and smaller velocity dispersions than the \Oiii\
kinematics.}
\label{fig:appendix5c}
\end{figure*}
}

\title[The \sauron\ project -~V]
{The SAURON project - V. Integral-field emission-line kinematics of 48
elliptical and lenticular galaxies}
\author[Sarzi et al.]  {Marc Sarzi,$^1$\thanks{sarzi@astro.ox.ac.uk} 
Jes\'us Falc\'on-Barroso,$^2$ Roger L.\ Davies,$^1$ 
Roland Bacon,$^3$ Martin Bureau,$^1$ \newauthor Michele
Cappellari,$^2$ P.~Tim de Zeeuw,$^2$ Eric Emsellem,$^3$ Kambiz
Fathi,$^{4,6}$ \newauthor Davor Krajnovi\'c,$^1$ Harald
Kuntschner,$^5$ Richard M.\ McDermid,$^2$ Reynier F.\ Peletier$^4$\\
$^1$Denys Wilkinson Building, University of Oxford, Keble Road,
Oxford, United Kingdom \\
$^2$Leiden Observatory, Postbus 9513, 2300 RA Leiden, The
Netherlands\\
$^3$Centre de Recherche Astronomique de Lyon, 9~Avenue Charles
Andr\'e, 69230 Saint Genis Laval, France\\
$^4$Kapteyn Astronomical Institute, Postbus 800, 9700 AV Groningen,
The Netherlands\\
$^5$Space Telescope European Coordinating Facility, European Southern
Observatory, Karl-Schwarzschild-Str~2, 85748 Garching, Germany\\
$^6$Rochester Institute of Technology, Rochester, New York 14623, USA}

\pagerange{\pageref{firstpage}--\pageref{lastpage}} \pubyear{2005}

\begin{document}
\date{Submitted on July 19, 2005. Accepted on November 8, 2005}
\label{firstpage}
\maketitle

%
%
\begin{abstract}
We present the emission-line fluxes and kinematics of $48$
representative elliptical and lenticular galaxies obtained with our
custom-built integral-field spectrograph \sauron\ operating on the
William Herschel Telescope.
H$\beta$, \Oiii$\lambda\lambda$4959,5007, and [{\sc
N$\,$i}]$\lambda\lambda$5198,5200 emission lines were measured using a
new procedure that simultaneously fits both the stellar spectrum and
the emission lines.
Using this technique we can detect emission lines down to an
equivalent width of 0.1\AA\ set by the current limitations in
describing galaxy spectra with synthetic and real stellar templates,
rather than by the quality of our spectra.
Gas velocities and velocity dispersions are typically accurate to
within 14~\kms\ and 20~\kms, respectively, and at worse to within
25~\kms\ and 40~\kms. The errors on the flux of the \Oiii\ and \Hb\
lines are on average 10\% and 20\%, respectively, and never exceed
30\%.
Emission is clearly detected in 75\% of our sample galaxies, and comes
in a variety of resolved spatial distributions and kinematic
behaviours.
A mild dependence on the Hubble type and galactic environment is
observed, with higher detection rates in lenticular galaxies and field
objects. More significant is that only 55\% of the galaxies in the
Virgo cluster exhibit clearly detected emission.
The ionised-gas kinematics is rarely consistent with simple coplanar
circular motions. 
However, the gas almost never displays completely irregular
kinematics, generally showing coherent motions with smooth variations
in angular momentum.
In the majority of the cases the gas kinematics is decoupled from the
stellar kinematics, and in half of the objects this decoupling implies
a recent acquisition of gaseous material.
Over the entire sample however, the distribution of the mean
misalignment values between stellar and gaseous angular momenta is
inconsistent with a purely external origin.
The distribution of kinematic misalignment values is found to be
strongly dependent on the apparent flattening and the level of
rotational support of galaxies, with flatter, fast rotating objects
hosting preferentially co-rotating gaseous and stellar systems. 
In a third of the cases the distribution and kinematics of the gas
underscores the presence of non-axisymmetric perturbations of the
gravitational potential.
Consistent with previous studies, the presence of dust features is
always accompanied by gas emission while the converse is not always
true.
A considerable range of values for the \Oiii/\Hb\ ratio is found both
across the sample and within single galaxies. Despite the limitations
of this ratio as an emission-line diagnostic, this finding suggests
either that a variety of mechanisms is responsible for the gas
excitation in E and S0 galaxies or that the metallicity of the
interstellar material is quite heterogeneous.
\end{abstract}

\begin{keywords}
galaxies: bulges -- galaxies: elliptical and lenticular, cD --
galaxies: evolution -- galaxies: formation -- galaxies: kinematics and
dynamics -- galaxies: structure -- galaxies: ISM
\end{keywords}

%
%
\section{Introduction}
\label{sec:intro}
Early-type galaxies were once considered uniform stellar systems with
little gas, dust, and nuclear activity. A number of imaging and
spectroscopic studies both from the ground and space have changed this
view \citep[see][for a review]{Gou99}.
Based on these surveys, we now know that early-type galaxies commonly
contain dust in either organised or complex structures, which is
almost always associated with optical nebular emission
\citep[\eg][]{Sad85,vDo95,Tra01}.
Early-type galaxies also show nuclear emission in 60\% of the cases
\citep{Ho97c}.

Still, a number of issues remain open.
What is the origin of the interstellar material in E/S0 galaxies? Is
it material lost by stars during their evolution or does it have an
external origin?
It has long been demonstrated \citep{Fab76} that during their life
stars would reinject more than enough material in the interstellar
medium to explain the observed gas emission. Yet, the finding that the
angular momentum of the gas or the orientation of the dust is very
often decoupled from that of the stars \citep[\eg][]{Ber92,vDo95}
suggests an external origin.
And what is its fate? Does it cool down to form stars or does it
become hot, X-ray emitting gas?
With the advent of new mm-wave detectors and the Chandra space
telescope, both molecular gas and X-ray detections have become more
common in early-type galaxies \citep[\eg][]{You02,Fabb03}.
And finally, what powers the observed nebular emission?  Is it a
central AGN? Is the warm ($\sim10^4$K) gas ionised by the hot
($\sim10^7$K) gas, through thermal conduction \citep[the ``evaporation
flow'' scenario, \eg][]{Spa89,dJo90}? Is the gas ionised by stars,
either young \citep{Shi92} or old \citep[\eg\ post-AGB
stars,][]{dSe90,Bin94}?  Or is the gas excited by shocks, as also
proposed for low-ionisation nuclear emission-line regions
\citep[LINERs,][]{Dop95,Dop96}.
If the extended emission observed in many early-type galaxies is
unlikely to be powered by nuclear activity \citep{Gou99}, all
other ionising mechanisms are plausible.

So far, the kinematics and ionisation of the gas in early-type
galaxies have been studied mostly through long-slit observations
\citep[\eg][]{Zei96,Cao00,Phi86,Ho97b}, while imaging surveys have
investigated the distribution of the ionised gas and dust
\citep[\eg][]{Bus93,Gou94,Mac96,Tra01}.
Integral-field spectroscopic (IFS) data can combine both spatial and
spectroscopic information, mapping the flux and kinematics of the
ionised gas across large sections of nearby galaxies. Using
two-dimensional measurements it is possible to accurately measure
and compare the projected angular momentum of gas and stars, assess
the regularity of the gas-velocity fields, and investigate the
possible sources of ionisation for the gas.

In this paper we present maps for the ionised-gas kinematics and
distribution within the effective radius of 48 representative E and S0
galaxies in both ``cluster'' and ``field'' environments, which were
obtained with the \sauron\ integral-field spectrograph \citep[][Paper
I]{Bac01}. These galaxies were observed in the course of the \sauron\
survey, a study of the structure of 72 representative nearby
early-type galaxies and bulges. The objectives of the survey along
with the definition and properties of the sample are described in
\citet[][Paper~II]{dZe02}. The stellar kinematics for the 48 E and S0
galaxies in the survey are presented in \citet[][Paper~III]{Ems04}.

This paper is organised as follows.
In \S 2 we describe the extraction of the ionised-gas kinematics and
fluxes, compare the \sauron\ measurements with published data, and set
detection thresholds for the gas emission and the sensitivity of our
survey.
In \S 3 we present the maps for the ionised-gas distribution and
kinematics, discuss the incidence of emission, and describe the main
features of the gas distribution and kinematics.
In the same section we also describe the maps for the \Oiii/\Hb\ line
ratio and review systematic differences between the kinematics of the
\Oiii\ and \Hb\ lines. 
\S \ref{sec:DustandGas} is devoted to the relation between gas
and dust, while in \S\ref{sec:starandgas} we compare the kinematics
of gas and stars in order to discuss the origin of the gas.
We further discuss the observed gas phenomenology and draw our
conclusions in \S6.

\section{Measuring the Ionised-Gas Kinematics and Flux Distribution}
\label{sec:GasPaper_method}

\placefigMasknoMask

Within the limited wavelength range of the \sauron\ observations
(4830-5330~\AA) there are three well-known emission lines we can expect
to detect, the \Hb$\lambda$4861 Balmer line and the
\Oiii$\lambda\lambda$4959,5007 and [{\sc
N$\,$i}]$\lambda\lambda$5198,5200 forbidden-line doublets.
The starting point for our emission-line measurements is described in
Paper~II. It consists of measuring the emission lines on residual
spectra obtained by subtracting from the \sauron\ spectra a detailed
description for the stellar spectrum, itself constructed for measuring
the stellar kinematics.
As described in Paper~III and in Cappellari \& Emsellem (2004), the
latter process requires an optimal combination of templates
representative of the galaxy stellar population, excluding spectral
regions that could be contaminated by emission lines.

Further testing revealed that this procedure does not work well for
all emission lines if there is insufficient information in the
emission-free part of the spectrum to adequately constrain the stellar
population content.
Indeed when the wavelength range is limited it is possible that some
of the emission lines we wish to measure lie very close to, or are
coincident with, the most age- or metallicity-sensitive absorption
features.
By masking the regions potentially affected by emission, these
important absorption features will be partially or even entirely
excluded from the template-fitting process.
This can lead to substantial biases in the resulting combination of
templates, evident in the residual spectrum as spurious features in
the masked regions that in turn contaminate the measurement of the
emission lines.
This is exactly the case for the \sauron\ observations and
particularly for the \Hb\ and \Ni\ doublet emission lines.
The \Oiii\ lines are less affected by this problem.

\subsection{The Method}
\label{subsec:GasPaper_method}

In this paper we extend the idea of Paper~II, draw from the software
of Cappellari \& Emsellem (2004), and exploit the results of Paper~III
to develop a more refined procedure to measure the gas kinematics and
fluxes without any spectral masking.
The key ingredient is to treat the emission lines as additional
Gaussian templates and, while iteratively searching for their best
velocities and velocity dispersions, to solve linearly at each step
for their amplitudes and the optimal combination of the stellar
templates
\footnote{The composition of the template library was improved with
  respect to the one used in Paper~III. Three stars from the
  \citet{Jon97} library were exchanged with more suitable ones that
  allowed a better match to the spectra of large early-type galaxies,
  thus improving the emission-line measurements. The impact on the
  published stellar kinematics is negligible.}
, which are convolved by the best stellar line-of-sight velocity
distribution (LOSVD).
In this way both the stellar continuum and the emission lines are
fitted {\it simultaneously\/}.

Here we adopt the same stellar kinematics and spatial binning scheme
of Paper~III \citep[from][]{Cap03} and hence solve for the gas
kinematics in exactly the same spectra.
Furthermore for each galaxy we adjust the continuum shape of the
stellar templates using a multiplicative Legendre polynomial of the
same order (typically 6) as needed in Paper~III.
We adopt a multiplicative polynomial adjustment to ensure that
no extra dilution of the absorption features is introduced.
As mentioned before, we assume a Gaussian LOSVD for the gas clouds.
In the case of doublets, each component has the same mean velocity and
width, and their relative strength is fixed by the ratio of the
corresponding transition probabilities.
Figure~\ref{fig:MasknoMask} shows for two specific cases the advantages
of this new method, when all emission-lines are fitted with the same
kinematics. 
Across our sample, template-mismatch features in the masked regions
lead to overestimated \Hb\ fluxes in average by 10\% and by up to 
$~40\%$ and, if the \Oiii\ lines are weak, also to systematically
overestimated line-widths by $\sim10\%$.

In principle, we could have used the new method imposing the same
kinematics on lines emitted from different atomic species or while
fitting each of them independently.
In practise, however, neither the \Hb\ nor the \Ni\ lines could be
always measured confidently without first constraining their
kinematics, as contamination due to template mismatch can still be
important even using our new approach.

The main difficulty with the \Hb\ measurement is the presence of a
number of metal features, mainly from chromium and iron, around
$4870$\AA.
If we allow the position and width of the \Hb\ emission to vary, when
the stellar templates cannot match the strength of this absorption
feature, the overall spectral shape between $4855$\AA\ and $4875$\AA\
can be described using metal-poor templates and a spurious \Hb\ line
placed roughly half-way between the \Hb\ and Cr-Fe absorption
features.
Figure~\ref{fig:HbProblem} shows an example of this problem.
The strength of this emission artifact can be significant and produce
a spurious detection. Fortunately, because the spurious \Hb\ emission
always falls at the same wavelength region in the rest frame, the mean
velocity of this line appears shifted with respect to the stellar
velocity by an approximately constant amount. Hence it is possible to
recognise this problem across the field by comparing \Hb\ velocity
maps to the stellar velocity maps (Figure~\ref{fig:HbProblem}).

\placefigHbProblem

The lines of the \Ni\ doublet are normally quite weak so
that it is almost never possible to constrain their kinematics
independently.
To complicate matters further, the \Ni\ doublet sits close to the
continuum region that is generally the worst matched by the templates,
because of an enhancement in the magnesium over iron ratio that is
neither included in the \citet{Vaz99} models (which form the bulk of
our template library) nor observed in stars in our Solar
neighbourhood. The impact of template-mismatch on the measurement of
the [{\sc N$\,$i}] lines is therefore more difficult to estimate than
in the case of the \Hb\ emission.

To measure the fluxes of the \Hb\ and \Ni\ lines we therefore
constrained their kinematics to those of the \Oiii\ lines, which was
obtained first.

Using the spatial binning, spectral library, and stellar kinematics of
Paper~III, we obtain the emission-line fluxes and kinematics in each
(binned) spectrum of our sample galaxies by following these exact
steps:

\begin{itemize}
   \item[{\bf a)}] Mask all spectral regions within $\pm300$\kms\ of
   the location of the \Hb\ and \Ni\ lines at the galactic
   systemic velocity $V_{\rm sys}$, taken from Paper~III. This covers
   the typical range of emission-line velocities and widths found in
   a preliminary analysis of our sample.

   \item[{\bf b)}] Convolve all stellar templates in our library with
   the corresponding best stellar LOSVD from Paper~III.

   \item[{\bf c)}] Solve for the best amplitude $A_{\rm gas}$, mean
   velocity \Vg\ and intrinsic velocity dispersion \Sg\ of the \Oiii\
   lines, while also optimising the continuum shape of the templates
   using a multiplicative polynomial adjustment. $V_{\rm sys}$ is used
   as initial velocity guess for \Vg.

   \item[{\bf d)}] Remove the mask and find the best $A_{\rm gas}$
   for all the lines, while imposing on them the \Oiii\ kinematics.

\end{itemize}

More specifically, at step {\bf c)} the best \Vg\ and \Sg\ and the
best coefficients of the multiplicative polynomial correction are
found through a Levenberg-Marquardt least-squares minimisation.
At each iteration we construct a Gaussian template for each emission
line with the current position and width (accounting for the \sauron\
spectral resolution), and with unit amplitude. For the \Oiii\ doublet,
the template is formed by two Gaussians, with amplitude of 0.33 for
the \Oiii$\lambda$4959 line \citep{Sto00}.  After multiplying the
convolved stellar templates by the current polynomial correction, we
fit for the best linear combination of both stellar and emission-line
templates (with positive weights), excluding the regions potentially
affected by \Hb\ and [{\sc N$\,$i}] emission using the mask built in
{\bf a)}. The weights assigned to the emission-line templates provide
the best emission-line amplitudes $A_{\rm gas}$ at each iteration, and
eventually, the final best values.

Step {\bf d)} is similar to {\bf c)}, although only the coefficients
of the polynomial adjustment are solved for non-linearly, while the
contribution of the stellar and emission-line templates is still
optimised at each iteration. Gaussian templates are constructed also
for the \Hb\ and \Ni\ lines, and the entire spectrum is used in the
fit.

In order to follow the most general method described at the beginning
of this section, only steps {\bf b)} and {\bf c)} are needed, without
masking any spectral region and by searching for the gas kinematics of
all lines at the same time.

\subsection{Constructing the Emission-Line Maps}
\label{subsec:GasPaper_Maps_Construction}

In our sample galaxies the ionised-gas emission is neither uniformly
distributed nor always strong enough to be detectable.
It is therefore crucial to understand the level to which we are
confident of detecting gas emission.
In Appendix A we present a number of experiments specifically designed
to address this issue, and here we summarise the results obtained
there.

The accuracy with which the position and width of an emission-line can
be recovered depends on how much the line protrudes above the noise in
the stellar spectrum. We call this quantity the line
amplitude-to-noise ratio, hereafter $A/N$.
The accuracy in recovering the amplitude of a line, on the other hand,
scales only with the noise level in the spectrum.
The ability to estimate the amplitude of the lines is also the
dominant factor in the error budget of the line fluxes.
Hence, in the limit of purely statistical fluctuations the line fluxes
are subject to larger errors in spectra of higher quality, although
the equivalent width of the lines is better estimated at these
regimes.
Better spectra also allow a more accurate description of the stellar
continuum, which is equally crucial to the emission-line
measurements.
 
Indeed, since emission lines are measured while simultaneously fitting
the stellar spectrum using a template library, any systematic mismatch
between the templates and the galaxy stellar population will
constitute a further source of error.
It is therefore important to include such deviations when estimating
the noise level against which the emission-line amplitudes are
compared.
We use a robust biweight estimator \citep{Hoa83} to measure the
scatter in the residuals of the fit to the integrated stellar
spectrum, as an estimate of both statistical fluctuations and
systematic deviations. We will refer to this as to the ``residual
noise''.

The simulations of Appendix A also show that the formal uncertainties
returned by our emission-line fitting procedure correctly estimate the
accuracy with which the input parameters are recovered. At low $A/N$,
however, our measurements become dominated by systematic, rather than
random, effects and the formal uncertainties cannot account for the
observed biases.
The minimum $A/N$ values below which these problems arise naturally
sets our detection threshold.

According to our emission-line measurement scheme, we must first
assess the presence of the \Oiii\ emission, without which no other
emission line can be detected.
The simulations show that our ability to estimate the velocity
dispersion of the \Oiii\ lines steadily deteriorates for $A/N \le
2.5$, while the line fluxes start to be overestimated by more than
10\% for $A/N \le 5$.
As these experiments cannot fully account for the limitations of our
template library, we conservatively settled on an overall $A/N$
threshold of 4.
This limit was chosen also considering that in galaxies with larger
velocity dispersions the impact of template-mismatch can be more
important and cause the line widths to be overestimated, as shown by
our experiment based on independent fits to the two lines in the
\Oiii\ doublet (\S\ref{app:simu_tm_o3ind}).
For an intrinsic $\sigma_{\rm gas}\,=50$\kms\ and $A/N=4$ the
simulations show that the errors on \Vg\ and \Sg\ are typically
25~\kms\ and 40~\kms, respectively, while errors on fluxes are 30\%.
The large error on the intrinsic \Sg\ is not due to a poor fit but
arises because the instrumental resolution is much larger than the
adopted value of 50~\kms. In fact, the observed width of the lines is
matched to within 17~\kms. Measurements for intrinsically broader
lines will be more accurate for a given $A/N$.

When \Oiii\ lines are detected, \Hb\ emission is also likely to be
present.
The possibility that the \Hb\ emission could be dominated by gas with
different kinematics than that contributing to the \Oiii\ emission
will not change our ability to estimate the \Hb\ fluxes.
Indeed, the accuracy in recovering fluxes is not dramatically affected
by how well the position and width of the lines are measured.
This is not only observed in our simulations, but also in the few
galaxies where the \Oiii\ and \Hb\ lines could be fitted independently
(\S\ref{subsec:Hbfaster}).
When we impose the \Oiii\ kinematics on the \Hb\ line, our simulations
show that, for $A/N \ge 2.5$, the estimates of the \Hb\ flux are
unbiased. At this level, they are accurate to within $\sim$30\%.
Unbiased \Hb\ fluxes can be measured with lower $A/N$ because the \Hb\
fit involves a lower number of parameters (\Vg\ and \Sg\ are fixed).
Considering that template-mismatch affects the measured
\Hb\ fluxes even in the framework of ideal simulation, we
conservatively settled on a $A/N$ threshold of 3 for this line.

Finally, we come to the harder problem of assessing the presence of
the \Ni\ doublet.
These lines are the most affected by our limited ability to describe
the stellar populations of our sample galaxies, a problem that we
cannot properly simulate.
The most likely source of template mismatch in this spectral region is
the absence of templates with super-solar abundance ratios.  A more
robust detection threshold can be established by comparing the [{\sc
N$\,$i}]$\lambda$5200 amplitude to the residual-noise level measured
more specifically in the Mg region, $N_{\rm Mg}$.
Furthermore, we do not expect to observe \Ni\ lines without
strong \Hb\ and \Oiii\ emission.
Hence, we consider the detection of the \Ni\ doublet to be
reliable only if both the \Hb\ and \Oiii\ lines have already been
detected, and if the \Ni\ lines satisfy the quite
conservative detection requirement of $A/N_{\rm Mg} \ge 4$.

For our survey we will therefore show maps for the flux, equivalent
widths, velocity, and velocity dispersion of the \Oiii\ emission-lines
for which $A/N \ge 4$. Maps for the flux and equivalent width
of the \Hb\ lines and for the \Oiii/\Hb\ ratio will show regions where
additionally $A/N \ge 3$ for the \Hb\ line.
We will not show maps for the \Ni\ lines, as these are detected only
in the central regions of 13 objects.
Instead, the galaxies with \Ni\ emission are listed in
Table~\ref{tab:allgal}.

To conclude, we stress that the detection thresholds adopted here will
never perfectly exclude all spurious measurements or guarantee that
all regions with real emission appear the maps.
For completeness, the public data release will contain all
emission-line measurements with associated errors, for all spectra
corresponding to single lenses or larger spatial bins shown in the
maps, with a flag marking the measurements we deem unreliable.

\subsection{Sensitivity Limits}
\label{subsec:GasPaper_DetectionLimits}

The detection thresholds adopted in the previous section set the
sensitivity of our survey. The equivalent width ($EW$) of the weakest
\Oiii\ and \Hb\ lines that we detect are around 0.1\AA\ and 0.07\AA\
for the \Oiii$\lambda$5007 and \Hb\ lines, respectively.
These limits can be understood as follows. If $S$ is the continuum
level in the spectra, $F$ the flux in the lines, and $\sigma$ their
typical observed widths, then an $A/N$ threshold can be translated
into a limiting $EW$ considering that for a Gaussian line:
\begin{equation}
EW = \frac{F}{S} = \frac{A\times\sqrt{2\pi}\sigma}{S} \;\;{\rm or}\;\;
EW = \frac{A/N\times \sqrt{2\pi}\sigma}{S/N}
\label{eq:one}
\end{equation}
Hence, for emission lines with a typical intrinsic broadening of
$\sigma_{\rm gas}=\,$50~\kms, which due to instrumental broadening
appears as $\sigma=\,$120~\kms, or 2\AA\ at 5007\AA, the $EW$ of a
barely detected \Oiii$\lambda$5007 line ($A/N=4$) would be
0.2\AA\ for $S/N=100$.

In the nuclear region of our sample galaxies the \sauron\ spectra are
of extremely high quality, with nominal $S/N$ up to 500 per pixel, so
that very weak lines should be detected. However, what matters for the
detection of emission lines is the ``residual noise'', not just the
statistical fluctuations in the stellar spectrum.
When the ``residual noise'' is compared to the continuum level, the
corresponding $S/N$ ratios reach values only up to $\sim 200$,
dominated by template mismatch.
With this upper limit for $S/N$ the $EW$ limits become 0.1\AA\ for
\Oiii$\lambda$5007 and 0.07\AA\ for \Hb, as found.
At a lower surface brightness level, however, the statistical
fluctuations still dominate the ``residual noise'', and the
signal-to-noise ratio is close to nominal. In Paper~III we adopted a
binning scheme with $S/N \ge 60$.  Therefore our sensitivity is never
worse than 0.3\AA\ and 0.2\AA\ for \Oiii$\lambda$5007 and \Hb,
respectively.

\placefigSauronVsPalomar

\subsection{Comparison with Published Emission-Line Measurements}
\label{subsec:GasPaper_Palomar_Comparison}

In Paper~II we showed for NGC~5813 that the \sauron\ ionised-gas
kinematics is consistent with published data. An ideal source for a
more general comparison is the Palomar spectroscopic survey of Ho,
Filippenko, \& Sargent (1995, 1997a), who observed 37 of our 48 sample
galaxies and also carefully subtracted the stellar light prior to the
emission-line measurements.
We compare the \Oiii/\Hb\ emission-line ratios as well as the width of
the forbidden emission, in this case relating the \sauron\ \Oiii\
lines width to the \Nii$\lambda$6583 width of \citet{Ho97a}.
For consistency with the Palomar data, we analysed nuclear spectra
extracted from our \sauron\ cubes within central apertures that match
the size ($2\arcsec \times 4\arcsec$) and orientation of the Palomar
long-slit observations.
For these spectra we first derived the stellar kinematics as in Paper
III, using the penalised pixel fitting algorithm of \citet{Cap04}, and
then followed our procedure to measure the gas emission.

The result of this comparison is shown in
Figure~\ref{fig:SauronVsPalomar} for the 21 objects with measured
emission-lines in the Palomar survey. Considering the number of
possible systematic factors that could enter this exercise (\eg\
different observing conditions, data quality and reduction, starlight
subtraction), the agreement between the \sauron\ and Palomar
measurements is satisfactory, as these are on average consistent to
within 20\% (see Figure~\ref{fig:SauronVsPalomar}).
In particular, for the \Oiii\ over \Hb\ ratios, the quality of the
Palomar blue spectra is significantly worse than that of the \sauron\
data, with nominal $S/N\sim 20-30$ per pixel as opposed to at least
500 for our nuclear extractions.
On the other hand, the higher quality of the Palomar red spectra
($S/N\sim 30-80$) is consistent with the better match between the
widths of the \Oiii\ and \Nii\ lines.

Forbidden lines can have different widths depending on their
respective critical densities \citep{Fil84,dRo86}. The fact that in
the majority of the \sauron\ galaxies the width of the \Nii\ lines is
comparable to that of the \Oiii\ lines could be explained if emission
predominantly originates in low-density reservoirs, as suggested by
\citet{Ho97b} in the case of \Nii\ and [{\sc S$\,$ii}], which also
often display similar widths.
On the other hand, the origin of the four outliers in
Figure~\ref{fig:SauronVsPalomar} with significantly broader \Oiii\
lines can be explained considering that, according to the positive
correlation between line width and critical density, \Oiii\ should be
broader than \Nii.

Of the 16 objects that Ho et al. identified as emission-line free, 9
show clear emission in the \sauron\ central apertures. Apparently
low-luminosity nuclear activity in early-type galaxies is even more
common than already established.

\section{Ionised-Gas Distribution and Kinematics}
\label{sec:maps}

Figure~\ref{fig:Mapsa}-\ref{fig:Mapsb} shows maps for the flux and
equivalent widths of the \Oiii$\lambda5007$ and \Hb\ emission lines,
for the velocity and intrinsic velocity dispersion of the \Oiii\
lines, and for the \Oiii$\lambda5007$/\Hb\ ratio of all 48 E and S0
galaxies in our sample. The maps were constructed according to the detection
thresholds set in \S\ref{subsec:GasPaper_Maps_Construction}.
The \sauron\ observations were not always carried out in photometric
conditions. Therefore the derived fluxes should be considered only as
approximate values.

Appendix~\ref{app:list} contains a description of the gas distribution
and kinematics in each of our sample galaxies, along with comments on
the dust distribution, the connection to the stellar kinematics, the
\Oiii/\Hb\ ratios, and references to previous narrow-band imaging and
long-slit spectroscopic work.
In the following we quantify the incidence of ionised-gas emission and
describe the general properties of the gas distribution, the
morphology of the gas velocity and velocity dispersion field, and the
range of \Oiii/\Hb\ ratios observed in
Figure~\ref{fig:Mapsa}-\ref{fig:Mapsb} for galaxies with clear
emission-line detections.
We will also briefly investigate the presence of systematic
differences between the kinematics of the \Oiii\ and \Hb\ lines.

\subsection{Incidence of Ionised-Gas Emission}
\label{subsec:gasincidence}

The maps reveal that ionised-gas emission is very common in early-type
galaxies, and that it comes with a variety of spatial distributions,
degrees of regularity in the observed kinematics, and large variations
in the \Oiii$\lambda5007$/\Hb\ ratio.
\Oiii\ and \Hb\ emission is clearly detected in 36/48 galaxies in our
sample (75\%), while in a further 7 objects we found either weak
central \Oiii\ lines only (NGC~4382, NGC~5845) or patchy traces of
emission (NGC~4270, NGC~4458, NGC~4473, NGC~4621, NGC~4660).
Whether strong, weak or patchy, such emission is always spatially
resolved.
Just 5 galaxies do not show any significant emission (NGC~821,
NGC~2695, NGC~4387, NGC~4564, NGC~5308).
\Ni\ lines are found in 13 galaxies.
Table \ref{tab:allgal} lists the basic properties of the E/S0 \sauron\
sample and identifies galaxies in which \Hb\ and \Oiii\ emission was
detected, and where \Ni\ lines were found.

Among the clear detections, the incidence of emission lines is higher in
lenticular galaxies, where emission is found in 20/24 objects (83\%),
compared to 16/24 (66\%) for ellipticals.
The dependence on the galactic environment appears similarly marginal,
with 20/24 field galaxies showing emission compared to 16/24 in
clusters, where the definition of ``field'' and ``cluster'' is as in
Paper~II.
However, the fraction of galaxies with clearly detected emission in
the Virgo cluster drops to only 55\% (10/18), with just 3/9
ellipticals showing the presence of ionised gas. 
These 3 galaxies are also the brightest ellipticals that we observed
in Virgo. More generally the incidence of emission does not seem to
depend on the galaxy luminosity.
The incidence of gas emission is the same for barred and unbarred S0
galaxies.

Table \ref{tab:allgal} also lists the total flux of the \Hb\ emission
in each galaxy with clearly detected emission. This includes the
contribution of regions where only the \Oiii\ lines were detected,
assuming a constant line ratio of \Oiii$\lambda 5007/$\Hb\ $=3$.
Assuming Case B recombination, a temperature of $T=10^4\,{\rm K}$, and
an electron density of $n_e=10^2\,{\rm cm^{-3}}$, we also report total
\Ha\ luminosities and ionised-gas masses \citep[following
][]{Kim89}. In computing the \Ha\ luminosity we adopted the distance
moduli of Paper~II, and used the theoretical value of 2.86 for the
flux ratio of the \Ha\ and \Hb\ lines \citep{Ost89}, thus ignoring the
impact of dust absorption which would lead to higher \Ha\ luminosities
and larger gas masses.
We caution against using such flux and mass measurements for
quantitative applications.

\subsection{Ionised-Gas Distribution}
\label{subsec:gasdistrib}

The distribution of the ionised-gas is traced by the flux maps in
Figure~\ref{fig:Mapsa}. Figure~\ref{fig:Mapsb} also includes maps for
the $EW$ of the lines, which highlight structures in the emission-line
distribution that are otherwise hidden in the flux maps (\eg\
NGC~2974, NGC~3414).
This is because the flux distribution of the lines mostly follows that
of the stellar continuum, with a wide dynamic range.
Small fluctuations are therefore harder to recognise in the flux maps,
whereas the $EW$-maps have the variation of starlight divided out.
Although $EW$ structures are not directly related to the flux
distribution of the lines, they show specific emission regions such as
rings or spiral arms.
The $EW$-maps also provide a picture of the robustness of our
emission-line measurements and illustrate how close they come to the
detection limit.
The $EW$ is indeed closely related to the $A/N$ ratio
(Eq.~\ref{eq:one}), which sets our detection thresholds and defines
the accuracy of the kinematic measurements.
The $EW$-maps should be used only as supplementary tools to the flux maps
when investigating the gas distribution.
Overall, by complementing the flux maps with the $EW$-maps, a better
picture of the ionised-gas distribution emerges.

The simplest case is that of an extended distribution consistent with
a disk of gas.
NGC~524, NGC~4459, NGC~4526, NGC~5838 belong to this category, as do
the polar-ring galaxies NGC~2685 and NGC~2768 where the gas
distribution is perpendicular to the stellar body.
Additional spiral or ring features are found in NGC~2974, NGC~3414,
NGC~4550, and possibly in NGC~3608.
In some other galaxies (NGC~5198, NGC~5846, NGC~5982) an elongated
structure extends from a central disk.
A very distinct class of objects is formed by galaxies where a
loosely-wound spiral feature, similar to an integral sign, is
superimposed on a more diffuse component (\eg\ NGC~474, NGC~3377,
NGC~4262, NGC~4278, NGC~4546).
Among more complex distributions are NGC~3156 and NGC~3489, which
exhibit a central ring and filamentary structures, and objects
characterised by outer emission-line regions that are strongly
misaligned with the main body of the galaxy (NGC~1023, NGC~2549,
NGC~7332, NGC~7457).
Finally, there are giant ellipticals where the gas distribution is
mainly confined to lanes across the galaxy (NGC~4374, NGC~5813) or in
filaments (NGC~4486).

We note that planetary nebulae (PNe) could be responsible for many of
the isolated patches of \Oiii\ emission observed in our sample
galaxies. For instance, the typical observed fluxes of the PNe in
NGC~3379 \citep[$\sim10^{-16}\,\rm erg\,s^{-1}cm^{-2}$ for a narrow-band
magnitude $m_{5007}=26$, ][]{Cia89} would correspond to equivalent
widths well above our sensitivity limit, except in the very central
region (within 5\arcsec) and for the largest outer bins. The detection
of PNe in NGC~3379 is discussed in Appendix~\ref{app:list}.

\subsection{Ionised-Gas Kinematics}
\label{subsec:gaskin}

\subsubsection{Velocities}
\label{subsubsec:Vgas}

A number of galaxies clearly show very regular gas velocity fields
consistent with circular gas motions in a disk (\eg\ NGC~524, NGC~4459,
NGC~4526).
Yet, the majority of the objects appear to deviate from this simple
situation, showing gradual twisting of the overall velocity field
(NGC~2974, NGC~3377, NGC~3414, NGC~4278), more complex leading or
trailing features (\eg\ NGC~2768, NGC~4550), and sometimes both
(NGC~3489, NGC~4374).
More mundanely, heavy spatial binning or weak emission leads to more
noisy velocity fields (\eg\ NGC~4150).
In some objects the radial variation of the gas angular momentum is
more abrupt. NGC~474 is the clearest example of this behaviour, but
also NGC~2549, NGC~7332, NGC~7457, and possibly NGC~1023, show
kinematics that are distinct in the inner and outer regions.
Overall across our sample the gas kinematics are never irregular, the
only exception being NGC~4486.

We note that regular motions occurs preferentially in objects with
a regular disk distribution, while a more complex kinematic behaviour
follows the presence of additional features in the gas distribution,
such as integral-sign and spiral structures. In particular, all
objects with relatively strong, misaligned emission in their outer
parts (\eg\ NGC~2549), also show strongly decoupled external and
internal gas motions, with the innermost system being more likely
settled onto the equatorial plane.

\subsubsection{Velocity Dispersions}
\label{subsubsec:Sgas}

The accuracy of the \Sg\ measurements is more sensitive to the impact
of weak emission than for the \Vg\ measurements.
This is exhibited as an increasing fluctuation in \Sg\ in regions with
barely detected \Oiii\ emission (\eg\ NGC~4374).
At low $A/N$ the width of the lines also tends to be overestimated,
due to template mismatch and correlation in the random fluctuations in
the spectra (see \S\ref{app:simu_lowaon} and \S\ref{app:simu_tm_o3ind}).

The \Sg-maps appear more uniform than the $EW$ and \Vg\ maps, and are
characterised by little variation of \Sg\ over large regions and by
the presence of a central gradient.
Both the typical width of the lines in the outer regions and the
magnitude of their central increase can vary considerably.
This is evident when the \Sg-values are plotted against their distance
from the centre (Figure~\ref{fig:Mapsb}).
In some cases the intrinsic dispersion of the lines remains as high as
100~\kms\ in the outer parts (\eg\ NGC~4278, NGC~4546) while in other
falls to 50~\kms\ (\eg\ NGC~3489, NGC~3156). Central values of \Sg\
reach $\sim$200~\kms\ (\eg\ NGC~2974, NGC~3414). In other cases \Sg\
remains constant (NGC~3489). The morphology of the central peaks can
vary, from sharp (\eg\ NGC~3377) to very extended (\eg\ NGC~2974),
elongated (\eg\ NGC~2768, NGC~3414), and even asymmetric (\eg\ NGC~4262,
NGC~4278, NGC~4546).
Finally, some objects show peculiar \Sg-features (\eg\ NGC~474 and
NGC~2685).

Except for asymmetric \Sg-peaks that occur in objects with
integral-sign $EW$-patterns, the most significant characteristics of
the \Sg-maps do not appear to correlate with the morphology of the gas
distribution and velocity field.
In the central regions, however, the observed velocities are important
to understand whether the \Sg-gradients are due to unresolved
rotation, an AGN, or a genuine increase of \Sg.
Detailed models are needed to address this issue, but we expect the
effects of purely unresolved rotation to be rather limited and
confined to the seeing-dominated regions.
The strongest and most extended $\sigma_{\rm gas}$-gradients therefore
suggest an intrinsic rise of \Sg\ \citep[\eg\ NGC~2974,][]{Kra05}.
On the other hand, the case for unresolved rotation is particularly
strong for sharp peaks that are elongated along the direction of
the zero-velocity curve (\eg\ NGC~2768, NGC~3414).

We note that \Sg\ always exceeds the expected value from thermal
broadening ($\sim10$\kms) and is generally smaller than the stellar
velocity dispersion $\sigma_{\ast}$. In some cases, however,
$\sigma_{\rm gas}\sim\sigma_{\ast}$, either only in the central
regions (\eg\ NGC~2974, NGC~3414, NGC~5813) or over most of the field
(\eg\ NGC~3156).
This suggests the presence of additional turbulence in the gas clouds
or that the latter do not follow perfectly circular orbits. A stronger
dynamical support for the gas motions \citep{Ber95} may be required to
explain cases where $\sigma_{\rm gas}\sim\sigma_{\ast}$.

\placefigMaps

\subsection{\Oiii/\Hb\ ratios}
\label{subsec:ratios}

The maps for the relative strength of the \Oiii$\lambda5007$ and \Hb\
lines in Figure~\ref{fig:Mapsa} can be used to identify regions where
emission could be powered by young stars and to trace variations of
the ionisation mechanism within single galaxies.
Low \Oiii/\Hb\ ratios tend to characterise star forming \Hii-regions,
whereas other mechanisms are in general responsible for emission with
high \Oiii/\Hb\ ratios. Only $\sim$20\% of the emission-line nuclei
with \Oiii/\Hb$>1$ are classified as \Hii\ nuclei, while this class of
objects represent $\sim$85\% of the nuclei with \Oiii/\Hb$\le1$
\citep{Ho97c}. On the other hand, \Oiii\ lines stronger than \Hb\ can
arise also in \Hii-regions if the metallicity of the gas is
sufficiently low \citep[\eg][]{Vei87}.
The metallicity of the interstellar medium can vary between different
objects (\eg\ if the gas has an external origin) but it is unlikely to
change abruptly across different regions of a galaxy.
Large fluctuations of the \Oiii/\Hb\ ratio within a galaxy suggest a
variation of the ionising mechanism rather than a change in the gas
metallicity alone.

\placetabone

The great variety of \Oiii/\Hb\ values found both across our sample
and within single objects (Figure~\ref{fig:Mapsa}) therefore suggests
either that different ionising sources could be at work in early-type
galaxies or that the metallicity of the interstellar medium is very
heterogeneous from galaxy to galaxy.
The radial profiles of Fig\ref{fig:Mapsb} show that more than half of
the objects have average \Oiii/\Hb\ values between 1 and 3.
Many galaxies (\eg\ NGC~3377, NGC~3489) display emission with
\Oiii/\Hb$> 3$ , but such a strong level of excitation
(typical of Seyfert nuclei) never dominates the entire field (except
for NGC~7332). On the other hand, \Oiii/\Hb\ ratios $\le 1$
characterise most of the emission observed in NGC~524, NGC~3032,
NGC~4459 and NGC~4526.
The radial profiles for the \Oiii/\Hb\ ratios also emphasise the
presence of central gradients. Towards the centre, with few
exceptions, \Oiii/\Hb\ always ranges between 1 and 3.

The ionisation structure we observe links to the gas distribution
and kinematics in two classes of objects in particular.
The first group includes the four galaxies with the lowest \Oiii/\Hb\
ratio in our sample: NGC~524, NGC~3032, NGC~4459 and NGC~4526. These
objects have clear disk-like gas distribution and kinematics, although
in NGC~3032 severe binning makes it hard to judge the regularity of
the velocity field. NGC~3032, NGC~4459, and NGC~4526 also share the
same radial trend for the \Oiii/\Hb\ ratio, with a minimum in a
circumnuclear region and stronger \Oiii/\Hb\ ratios toward the centre
{\it and\/} the edge of the gas disk. Emission in NGC~524 is too weak
to confirm this behaviour.
The second class is composed of NGC~4262, NGC~4278, and NGC~4546,
which show \Oiii/\Hb-maps that are not symmetric around their centre
despite the corresponding \Oiii\ and \Hb\ distributions being fairly
similar.
These objects also display an integral-sign feature in their gas
distribution. By contrast, NGC~3377, another object with this
signature, has a fairly symmetric \Oiii/\Hb\ distribution.

We note that many of the brightest elliptical galaxies in our sample
show rather uniform \Oiii/\Hb-maps for relatively intermediate values
of \Oiii/\Hb$\sim$1-2.

\subsection{\Hb\ and \Oiii\ kinematics}
\label{subsec:Hbfaster}

As clouds ionised by different sources need not share the same
kinematics, here we seek evidence for systematic differences between
the kinematics of clouds that emit more efficiently \Hb\ photons than
\Oiii.
We measured the \Hb\ and \Oiii\ kinematics independently in the
galaxies with the strongest emission, and checked if the \Hb\
measurements were subject to the systematics described in \S2.1 by
comparing the \Hb\ velocity field with the stellar one (see
Figure~\ref{fig:HbProblem}). The \Ni\ lines were still
fitted using the \Oiii\ kinematics.
Independent \Hb\ and \Oiii\ kinematics could be measured over most of
the region where emission is observed only in 10 galaxies.
The most significant result from this experiment, shown in full in
Appendix~\ref{app:o3vhb}, is that in objects with very regular gas
distribution and kinematics, such as NGC~4459 and NGC~4526, the \Hb\
velocities are higher than the \Oiii\ velocities, by up to 100~\kms
(Figure~\ref{fig:Hbfaster}).
This suggests that \Hb\ is a better tracer than \Oiii\ for the
circular velocity in the equatorial plane. 

In other cases the differences between the velocities are more limited
($\le 30$\kms) and show less clear patterns, except for the central
regions of NGC~3414 and NGC~4278 where a faster \Oiii\ component is
observed.
In general the width of the \Hb\ lines tend to be smaller than that of
the \Oiii\ lines, in particular towards the centre.

We note that even when the difference between the \Hb\ and \Oiii\
kinematics is extreme (100~\kms), the flux of the \Hb\ lines fitted
independently and of those measured with the \Oiii\ kinematics differ
by less than 20\%.
In terms of equivalent width the constrained fit can lead, in the most
extreme cases, to an overestimation of H$\beta$ by up to 0.2\AA.

\placefigHbfaster

\section{Dust and Gas}
\label{sec:DustandGas}

Dust in early-type galaxies is almost always associated with gas
emission \citep[\eg][]{Tom00,Tra01}. With \sauron\ we can correlate
the presence and spatial morphology of the dust not only with the
ionised-gas distribution but also with its two-dimensional kinematics.

To highlight the presence of dust we constructed unsharp-masked maps
using both the \sauron\ reconstructed intensity maps and archival
\HST\ images, the latter being available for 45/48 objects. This was
done by dividing each image by a smoothed version of itself, using a
Gaussian kernel with a FWHM of 0\farcs3 and 0\farcs8 for the \HST\ and
\sauron\ images, respectively.
Figure~\ref{fig:Mapsb} shows for each galaxy the \HST\ or
\sauron\ unsharp-masked image.
The unsharp-masked images reveal different dust morphologies, such as
perfect disks (NGC~524, NGC~3032, NGC~3379, NGC~4459, NGC~4526,
NGC~5838), less defined coplanar distributions, (\eg\ NGC~2974,
NGC~3489, NGC~4550), well defined lanes (\eg\ NGC~2685,
NGC~3377, NGC~4374) or more complex filamentary (e.g, NGC~4486,
NGC~5813, NGC~5846) and patchy (\eg\ NGC~4552) structures.
Whenever dust is present we also find emission, although very weak in
NGC~5845. The converse does not always hold, which is expected as dust
is harder to detect in close to face-on configurations.
Still, it is puzzling to see that in some cases (\eg\ NGC~5198,
NGC~5982) dust is not detected despite the observed high gas
velocities, which are hard to reconcile with motions in a face-on
disk.

Consistent with previous studies \citep[\eg][]{Gou94}, the dust
generally follows the ionised-gas distribution, even if it is not
always possible to connect features in the flux and $EW$-maps with
ones in the unsharp-masked images. In some objects, however, specific
\Hb-emitting regions have dusty counterparts.
Perfect examples are the circumnuclear \Hb-dominated regions of
NGC~4459 and NGC~4526 that correspond to the strongest absorption
features in the unsharp-masked images. 

Regular dust distributions trace well regular velocity fields,
consistent with \citet{Ho02}. All galaxies with perfect dusty disks
have indeed very regular gas kinematics, with NGC~3032 the only
possible exception. On the other hand, the absence of regular dust
lanes does not imply irregular kinematics, as very coherent gas
motions correspond also to less defined dust morphologies (\eg\
NGC~2974, NGC~4278), and even to quite complex ones (\eg\ NGC~4150,
NGC~5846).

The unsharp-masked maps also highlight the presence of separate
stellar components, such as nuclear disks (\eg\ NGC~821, NGC~5308),
peanut-shaped bulges (NGC~2549), and bilobate structures reminiscent of
a bar (NGC~2699, NGC~4262). Interestingly, nuclear disks are a common
feature in galaxies with no or only weak emission.

\section{Kinematic Misalignment Between Gas and Stars}
\label{sec:starandgas}

The distribution and kinematics of the ionised gas in early-type
galaxies have long been known to be often decoupled from that of the
stars \citep[see][for a review]{Ber99}.
Figure~\ref{fig:Mapsb} includes the stellar velocity fields from
Paper~III to facilitate the comparison between the \Oiii\ and stellar
kinematics, and shows that our sample galaxies are not an exception in
this respect.
The velocity maps show that the motion of gas is often decoupled from
that of the stars, and that the angular momenta of the gas and stars
have the same orientation in only a few cases.

A number of studies have used the distribution of the misalignment
between the spatial distributions \citep[\eg][]{Mar04} and angular
momenta \citep[\eg][]{Kan01} of gas and stars to constrain the origin of
the ionised gas in early-type galaxies. The orientation of the dust
relative to that of the stars has also served this purpose
\citep[\eg][]{vDo95,Tra01}.
With 48 representative E and S0 galaxies in both cluster and field
environments, the \sauron\ sample constitutes a good basis to test
simple hypotheses on the origin of the gas.

We used the velocity maps of Figure~\ref{fig:Mapsa}-~\ref{fig:Mapsb}
to accurately measure the direction of the stellar and gaseous
rotation, and compared the distribution for the kinematic
misalignments between gas and stars with what is expected if the gas
has either an internal or external origin.
We trace the direction of maximum rotation as a function of radius,
using the harmonic-expansion method for analysing two-dimensional kinematics maps
of \citet{Kra04a}.
We divided the velocity field in concentric circular annuli and fitted
the first four terms of a Fourier series to the velocity angular
profiles in each of the annuli. The angular phase of the first order
term corresponds to the direction of maximum rotation,
$\phi_{\rm gas}$ or $\phi_{\rm star}$.  The width of the annuli
increases geometrically to follow the increase in the \sauron\ bin
size. The fit automatically stops toward the edge of the field or of
the gas distribution \citep[see][]{Kra04a}.
In the lower panels of Figure~\ref{fig:Mapsb} we plot the difference
between $\phi_{\rm gas}$ and $\phi_{\rm star}$ as a function of
radius. The gas-star kinematic misalignment is always shown between
$\pm$180\degr, and is derived only in objects with clear gas detection
and where stellar rotation is observed. We further excluded galaxies
without sufficiently extended emission (NGC~2699 and NGC~5831). In
NGC~4570 and the external regions of NGC~7457 the emission is too
fragmented for the harmonic fit to converge. The complex stellar
kinematics in NGC~4550 are also hard to describe with this method.

\placetabtwo
From the misalignment profiles of Figure~\ref{fig:Mapsb} we derived
for each object a median value and the standard deviation, which are
tabulated in Table~\ref{tab:misal}.
We considered NGC~4550 and NGC~4570 as perfectly co-rotating systems
(see Appendix\ref{app:list}), and used only the central gaseous
systems of NGC~474 and NGC~2549, as they are more likely to be
settled.
For the same reason we excluded the filamentary structures extending
from the central disks in NGC~5198 and NGC~5982. In NGC~3414 we
considered only the central regions because of a strong decoupling in
the stellar velocity field. Table~\ref{tab:misal} lists the adopted
radial ranges.
The values for the standard deviation of the kinematic misalignment
always account for the observed twists in the gas and stellar velocity
maps since the harmonic-expansion generally measures $\phi_{\rm
gas}$ and $\phi_{\rm star}$ to within a few degrees.

\placefigHistMisalESO

Figure~\ref{fig:HistMisalES0} presents the distribution of the average
values for the kinematic misalignment between gas and stars in our
sample, now shown between 0\degr\ and 180\degr.
Its principal feature is a pronounced excess of objects with gas in
prograde orbits with respect to objects with gas in retrograde orbits.
Figure~\ref{fig:HistMisalES0} shows also that elliptical and
lenticular galaxies have very similar distributions. Despite the
modest number of objects, a KS-test shows that the two distributions
are identical at a 1$\sigma$-level ($p=77\%$). The distributions of
field and cluster galaxies are also not significantly different ($p=43\%$).
The distribution of kinematic misalignments is also independent of the
galaxy luminosity.

To interpret Figure~\ref{fig:HistMisalES0}, let us assume that the
intrinsic shape of our sample galaxies is mildly triaxial
\citep{Fra91}.
In this situation, stable closed orbits for the gas are allowed only
in two planes: the plane containing the short and intermediate axis
and the plane containing the long and intermediate axis.
When gas is acquired from random directions, it will settle more often
in the long-intermediate plane and the chances of settling in the
short-intermediate plane will scale with the degree of triaxiality
\citep{Ste82}.

If the origin of the gas is external, and assuming that prograde and
retrograde gas settle in the same way, the distribution of the
gas-star kinematic misalignments will therefore display three peaks:
two of equal intensity corresponding to counter- and co-rotating gas
and stars, and a weaker peak for gas in orthogonal rotation, assuming
the stars rotate along the short axis.
In a triaxial galaxy, however, the stellar rotation axis can lie
anywhere in the plane containing the short and long axes. Since gas in
equilibrium can rotate only around these two axes, intermediate values
for the kinematic misalignment will also be observed.
Projection effects will further dilute the distribution of the
observed kinematic misalignments, but overall the resulting
distribution will be {\it symmetric\/} around 90\degr, with an equal
number of counter- and co-rotating gas and stellar systems.

If the origin of the gas is internal (\eg\ from stellar-mass loss) the
gas will rotate in the same sense as the stars, and therefore the
distribution for the kinematic misalignments will be {\it
asymmetric\/}, with values mostly between 0\degr\ and~90\degr.

The observed distribution in Figure~\ref{fig:HistMisalES0} is
inconsistent with the prediction of either of these simple scenarios.
Half of the objects show a kinematic decoupling that implies an
external origin for the gas, but the number of objects consistent with
co-rotating gas and stars exceeds by far the number of cases with
counter-rotating systems, suggesting that internal production of gas
has to be important.

We note that our objects form a representative, but incomplete, sample
of the local early-type galaxies population.
The degree of incompleteness is known as our targets were
drawn from a complete sample (Paper~II).
Since galaxies with prograde and retrograde gaseous systems form
subsamples that are similarly representative of the local galaxy
population, incompleteness does not have a significant impact on the
previous discussion. Incompleteness is not responsible for the
observed asymmetry in Figure~\ref{fig:HistMisalES0}.

\placefigHistMisalFlRo

\subsection{The dependency on galaxy flattening and rotational support}
\label{sec:starandgas_ell}

The distribution of the average values for the kinematic misalignment
between stars and gas does not depend on Hubble type, galactic
environment, or galaxy luminosity.  It does, however, strongly depend
on the apparent large-scale flattening of galaxies. 
Figure~\ref{fig:HistMisalFlRo} shows that the roundest objects in our
sample ($\epsilon_{25} \le 0.2$) present a more symmetric distribution of
kinematic misalignments than flatter galaxies, which instead host
predominantly co-rotating stellar and gaseous systems.
The two distributions are different at a 99\% confidence level and
there is a 53\% probability that the distribution of rounder object
was drawn from a uniform distribution.
The galaxies in the two subsamples have no significantly different
luminosities or Hubble types and do not live in different
environments. The two samples are also equally incomplete. The flat
objects showing co-rotating gas and stars are also not significantly
different than the rest of the galaxies in this subsample.

Since for random orientations fairly round galaxies are likely to be
almost spherical and hence supported by dynamical pressure, rather
than by rotation, the degree of rotational support could also be
important to explain the observed dependency on the galaxy flattening.
A first distinction between slowly-rotating, rounder galaxies and
fast-rotating flatter objects, can be made using the classical
$V/\sigma-\epsilon$ diagram recently revised by Binney (2005). This
diagram alone, however, would fail to separate galaxies that are
characterised by overall rotation from non-rotating objects with a fast
rotating core, because the derivation of $V/\sigma$ includes only a
luminosity weighting. In Emsellem et al. (in preparation) we assess
the level of rotation support in a more quantitative way, adopting a
quantity that is closely related to the specific angular momentum of a
galaxy, thus overcoming the limitations of the $V/\sigma-\epsilon$
diagram.
Figure~\ref{fig:HistMisalRnR} shows the distribution of kinematic
misalignments between gas and stars in fast and slowly rotating
galaxies according to the criterion of Emsellem et al. \citep[see
also][for an illustration of this two kind of objects]{McD05}.
Consistent with our expectations, the two distributions are remarkably
different (at a $2\sigma$ level, $p=95\%$), as in the case of flat and
round objects (Figure.~\ref{fig:HistMisalFlRo}).
In the context of our simple first-order assumptions these results
suggest that external accretion of gaseous material is less important
than internal production of gas in flat and fast rotating galaxies. 
On the other hand, the more uniform distribution of kinematic
misalignments in rounder and slowly rotating objects suggests that
these objects aquire their gas more often.
However, it must be bear in mind that the interpretation of the
misalignment distribution of rounder galaxies can be complicated by
projection effects, because such objects often host kinematically
decoupled core, and by the fact that gas is subject to only weak
gravitational torques in almost spherical objects.

\placefigHistMisalRnR

%
%
\section{Discussion and Conclusions}
\label{sec:conclusions}

We have measured the ionised-gas fluxes and kinematics in 48
elliptical and lenticular galaxies both in cluster and field
environments, using a novel technique to measure emission lines in
galactic spectra where the stellar and ionised-gas contributions to
the spectrum are {\it simultaneously\/} described.
Extensive simulations were performed to test this procedure and assess
the detection limits in measuring gas emission.

The excellent quality of the \sauron\ data and the ability of our new
method to exploit the entire spectral range allowed us to detect
emission lines down to an equivalent width of 0.1\AA, which is set by
the current limitations in describing the spectra of early-type
galaxies with synthetic and real stellar templates.
Due to these limitations neither the \Hb\ nor the \Nii\
lines could be always measured without imposing on them the kinematics
of the \Oiii\ doublet.
In the case of \Hb, independent fits lead to biased gas kinematics
that are easily recognised across the field of a galaxy. This allowed us
to identify a few galaxies where the \Hb\ and \Oiii\ kinematics could be
independently derived and compared.

Across our sample, \Vg\ and \Sg\ are on average accurate to within
14~\kms and 20~\kms, respectively. Errors on the flux of the \Oiii\
and \Hb\ lines are on average 10\% and 20\%, respectively.
Although the \Hb\ and \Oiii\ kinematics can be different, imposing the
\Oiii\ kinematics on the \Hb\ lines does not dramatically affect our
ability to measure the \Hb\ fluxes. This is observed both in the
simulations and in the objects where independently derived \Hb\ and
\Oiii\ kinematics could be compared.
On the other hand, relying on the detection of \Oiii\ emission before
measuring \Hb\ does limit our ability to detect weak emission from
\Hii-regions, where the \Oiii\ lines are dimmer than \Hb.

Emission is clearly detected in 36/48 of our sample galaxies (75\%)
and only 5 objects do not show any significant emission. The remaining
7 galaxies exhibit weak \Oiii\ lines only or fragmented traces of
emission. 
Among clear detections, a mild dependence on the Hubble type and
galactic environment is observed, with higher detection rates in
lenticular galaxies and field objects. 
Emission is found in 20/24 lenticular galaxies in our sample (83\%)
and in 16/24 (66\%) of the ellipticals. This is remarkably close to
the detections rates of the imaging survey of \citet{Mac96},
who found \Ha+\Nii\ emission in 85\% of S0 and 68\% of E.
The dependence on the galactic environment is similarly marginal,
although when only the Virgo cluster is considered the fraction of
galaxies with clearly detected emission drops to 55\% (10/18), with
just 3/9 ellipticals exhibiting emission lines. These 3 objects are
also the brightest that we observed in this cluster.
\citet{Lau05} also find a significantly lower incidence (33\%) of
galactic dust in Virgo than in the rest of the local elliptical galaxy
population (47\%). They also found dust only in the brightest objects
of this cluster.

The observed emission comes with a variety of resolved distributions,
kinematic behaviours, and \Oiii/\Hb\ line ratios. It is very often,
although not always, associated with dust.
Two interesting classes of objects can be recognised.

The first group show settled gaseous systems where star-formation is
almost certainly occurring, particularly in circumnuclear regions.
The defining properties of this class, which includes the S0 galaxies
NGC~524, NGC~3032, NGC~4459, NGC~4526 and NGC~5838, are a regular
disk-like gas distribution and kinematics, very regular and circularly
symmetric dust lanes, and the lowest \Oiii/\Hb\ ratios in our sample.
The kinematic signature of emission from \Hii-regions in circular
motion on the equatorial plane is observed in the independently
derived kinematics of the \Hb\ lines, which show faster rotation and
smaller velocity dispersions than the \Oiii\ kinematics.
The detection in three galaxies of CO emission from dense molecular
clouds \citep[in NGC~3032, NGC~4459 and NGC~4526 by][]{Sag89} further
suggests star-formation activity.
On the other hand, it is likely that gas clouds departing from simple
rotation end up getting shocked, increasing the gas ionisation and the
\Oiii\ emission.

The second group of galaxies, which includes NGC~474, NGC~3377,
NGC~4262, NGC~4278, and NGC~4546, is characterised by an integral-sign
pattern in the ionised-gas distribution and by noticeable twists in
the gas velocity maps. In addition, NGC~4262, NGC~4278, and NGC~4546
display peculiar asymmetries in their \Sg\ and \Oiii/\Hb\ maps.
Since all these objects show misaligned stellar photometric and
kinematic axes (although only mildly for NGC~3377, see Paper~I), the
observed twisting in the ionised gas distribution and kinematics is
more likely tracing regions where gas accumulates while flowing in a
non-axisymmetric potential rather than a warped configuration.

The presence of a triaxial structure also appears to be underscored by
finding in the outer parts of a galaxy gas emission that is misaligned
with respect to the main stellar body and that is kinematically
decoupled from the gas kinematics in the central region, as in
NGC~1023, NGC~2549, NGC~7332 and NGC~7457.
NGC~1023 is indeed a well-known barred galaxy \citep[\eg][]{Deb02},
NGC~2549 shows a peanut-shape structure in the \HST\ images, and the
presence of a bar in NGC~7332 was extensively discussed by
\citet{Fal04}.
Taken together with NGC~2974, where the gas distribution observed with
both \sauron\ and \HST\ is consistent with the presence of nested bars
(Krajnovi\'c et al. 2005; Emsellem et al. 2003), these patterns show
how dramatically the gas can respond to the presence of even a modest
non-axisymmetric perturbation of the gravitational potential.

Weak bars or unsettled configurations can contribute to explain why
the ionised-gas kinematics is rarely consistent with simple coplanar
circular motions. Yet, despite complex structures are often observed
in the velocity maps, the gas almost never displays irregular
kinematics and instead generally shows coherent motions with smooth
variations in angular momentum. In the majority of the cases the gas
kinematics is decoupled from the stellar kinematics.

We have measured the kinematic misalignment between stars and gas 
and derived a distribution of mean kinematic misalignment values to 
draw on the origin of the gas.
Although half of the objects show a kinematic decoupling that implies
an external origin for the gas, the distribution of the misalignment
values between stellar and gaseous angular momenta is inconsistent
with a purely external origin.
In particular, the number of objects with co-rotating gas and stars
exceeds by far the number of cases with counter-rotating systems,
suggesting that internal production of gas has to be important.
The distribution of the kinematic misalignment between stars and gas
does not depend on the Hubble type, galactic environment, or
luminosity of our sample galaxies. It does, however, strongly depend
on the apparent flattening of galaxies, and their level of rotational
support.

These results demonstrate that the origin of the gas in early-type
galaxies is not yet a ``solved problem'' -- more clues are needed.
Measuring the metallicity of the interstellar medium can provide the
needed insight. Indeed if the gas originated in stellar-mass loss its
metallicity should be related to that of the surrounding stars, in
contrast to what is expected if the gas has an external origin. Yet,
to measure the chemical composition of the gas it is first necessary
to understand the ionisation of the gas, i.e. what causes some atomic
species to emit more efficiently than others.

In this respect, the \sauron\ \Oiii/\Hb\ maps reveal a wide range a
values across the sample and considerable structures within single
galaxies. Despite the limitation of the \Oiii/\Hb\ ratio as an
emission line-diagnostic, this finding suggests either that a variety
of mechanisms is responsible for the gas excitation in E and S0
galaxies or that the metallicity of the interstellar material is quite
heterogeneous from galaxy to galaxy.
The \Oiii/\Hb\ maps always show central gradients, where the line
ratios tend to values always between 1-3. Since at these \Oiii/\Hb\
regimes LINERs and Transition objects are the most common class of
emission-line nuclei, our measurements may be consistent with the
finding that the majority of E and S0 emission-line nuclei belong to
these two spectroscopic classes \citep{Ho97c}.
In the outer parts of our sample galaxies, except for the objects
where star-formation is occurring, the \Oiii/\Hb\ is also typically
between 1-3, although with large scatter.
This could also be consistent with a LINER-like classification for the
extended emission of early-type galaxies, which has been observed in
many cases \citep[\eg][]{Phi86, Zei96, Gou99}.
More emission-line diagnostics such as the \Nii/\Ha\ ratio are needed
to confirm the spectral classification of the observed emission.

Excluding a central AGN as the ionising mechanism for the emission
observed at kilo-parsec scales, shocks, thermal conduction, and both
young and post-AGB stars are all potential sources of ionisation.
Shocks are known to occur while gas flows in barred potential
\citep[\eg][]{Rob79}, therefore their r\^ole has to be important in
galaxies where the presence of non-axisymmetric perturbation of the
potential is underscored by characteristic patterns in the gas
distribution and kinematics as those discussed before.
Conductive heating of the warm gas by hot electrons from the X-ray
emitting gas is a readily available source of energy in most early-type
galaxies, sufficient to power the observed nebular emission 
\citep[\eg][]{Mac96}. High spatial resolution X-ray data have also 
shown a striking coincidence between the spatial distribution
of the X-ray and ionised-gas emission \citep[\eg][]{Tri02,Spa04}, 
supporting a casual link between them.
Post-AGB stars could represent the most common source of ionisation.
They can provide enough ionising photons to power the observe emission
\citep{Mac96} and lead to LINER-like line ratios
\citep{Bin94,Gou99}. The most compelling piece of evidence supporting
this scenario is the finding that the emission-line flux correlates
very well with with host galaxy stellar luminosity within the
emission-line region \citep{Mac96}. This correlation suggests indeed
that the source of ionising photons are distributed in the same way as
the stellar population. The \sauron\ observations further support this
view, since as noted in \S\ref{subsec:gasdistrib} the flux
distribution of the lines closely follows that of the stellar
continuum. 
This is shown by the smooth appearance of many EW-maps, in
particular for the \Hb\ line.
Finally, the r\^ole of \Hii-regions cannot be ruled out only on the
basis of \Oiii/\Hb\ ratios $> 1$, and should be explored in more
details in light of the recent claims based on GALEX data of Yi et al
(2005), that a substantial fraction of nearby early-type galaxies
recently underwent star-formation activity.

Complementing this survey with integral-field spectroscopic data
in the H$\alpha$+\Nii\ wavelength region will mark an important step
to further understand what powers the nebular emission in early-type
galaxies, with bearings on the origin of the gas in these systems.

\section*{Acknowledgements}
MS is grateful to James Binney, Sadegh Khochfar, Johan Knapen, John
Magorrian, and Joe Shields for the helpful discussions. The \sauron\
project is made possible through grants 614.13.003, 781.74.203,
614.000.301 and 614.031.015 from NWO and financial contributions from
the Institut National des Sciences de l'Univers, the Universit\'e
Claude Bernard Lyon~I, the Universities of Durham, Leiden, and Oxford,
the British Council, PPARC grant `Extragalactic Astronomy \& Cosmology
at Durham 1998--2002', and the Netherlands Research School for
Astronomy NOVA.  RLD is grateful for the award of a PPARC Senior
Fellowship (PPA/Y/S/1999/00854) and postdoctoral support through PPARC
grant PPA/G/S/2000/00729. The PPARC Visitors grant
(PPA/V/S/2002/00553) to Oxford also supported this work.
MB acknowledges support from NASA through Hubble Fellowship grant
HST-HF-01136.01 awarded by Space Telescope Science Institute, which is
operated by the Association of Universities for Research in Astronomy,
Inc., for NASA, under contract NAS~5-26555 during part of this work.
MC acknowledges support from a VENI grant 639.041.203 awarded by the
Netherlands Organization for Scientific Research.
JFB acknowledges support from the Euro3D Research Training Network,
funded by the EC under contract HPRN-CT-2002-00305.
KF acknowledges support for proposal HST-GO-09782.01 provided by NASA
through a grant from the Space Telescope Science Institute.
This project made use of the HyperLeda and NED databases. Part of this
work is based on data obtained from the ESO/ST-ECF Science Archive
Facility.
Finally, we wish to thank the referee, Dr. Thomas, for his suggestions.

%
%

%
%
\appendix

\placefigApponea
\placefigApponeb
\placefigApponec
\placefigApptwo
\placefigAppthree
\placefigAppOIIIind

\section{Tests on the emission-line measurements}
\label{app:simu}

In this appendix, we attempt through a number of experiments to
establish how accurately the position, width, amplitude, and flux of
an emission line can be measured as a function of its relative
strength with respect to both the statistical and systematic
deviations from the fit to the stellar continuum.

\subsection{The synthetic data}
\label{app:simu_synthdata}

To simulate the typical situation in the \sauron\ spectra, we
constructed a large number of synthetic spectra, adding to an old
stellar population varying amounts of \Hb\ and \Oiii\ emission and
statistical noise. Specifically we used a 10-Gyr-old solar-metallicity
template from the Vazdekis library, to which we added \Oiii\ emission
lines with $A/N$ ranging from 0.5 to 20, and \Hb\ emission ranging
from 1/10 to 10 times the \Oiii\ emission. No \Ni\ emission was
added. We included Poisson noise to simulate spectra with a $S/N$ in
the stellar continuum of 200, 100, and 60, the latter value being
typical of the spectra in the outer parts of our sample galaxies.
The stellar spectra were all broadened by the same line-of-sight
velocity distribution, in this case a simple Gaussian with
$\sigma=150$~\kms centred at zero.
The velocity of the \Oiii\ lines ranged from -250 to 250~\kms, with a
constant intrinsic width of $\sigma_{\rm gas}=50$~\kms. Since the
stellar spectra are kept at their rest frame, the gas velocities are
effectively relative to the stellar ones.
We imposed the same kinematics on all emission lines. 

Overall, we created 15390 synthetic spectra, which were analysed
following exactly the same procedure as for the real data
(\S\ref{subsec:GasPaper_method}). As the stellar kinematics can be
measured very accurately at the considered $S/N$ and $\sigma$ values,
we always adopted the correct $v$ and $\sigma$ values while matching
the stellar continuum.

\subsection{General Results}
\label{app:simu_genres}

Figures~\ref{fig:appendix1a}-\ref{fig:appendix1c} show how the
measured position, width, amplitudes, and fluxes of the lines deviate
from their input values as a function of the measured $A/N$ and for
different $S/N$ values.

The recovery accuracy of the position and width of the lines is a
strong function of the measured $A/N$ while the amplitudes do not seem
to depend on it.
The quality of the data, as quantified by the $S/N$ values, does not
seem to affect the \Vg\ and \Sg\ measurements, while it is dominant in
determining how well the amplitudes are estimated. The accuracy in
measuring the amplitudes of the lines is indeed limited only by the
level of statistical noise associated to the spectra.

The precision in estimating the emission-line fluxes turned out to be
dominated by how well the amplitudes are measured, as it also scales
mostly with $S/N$ instead of $A/N$.
A dependency on $A/N$ is introduced through the impact of the
line-widths in the flux measurements, although only to a lower extent.
For the \Oiii\ lines this is visible as a weak modulation in the flux
recovery accuracy with $A/N$, which indeed does not
exactly parallel the small variations in the amplitude estimation. 
The effect on the \Hb\ line is more obvious and is illustrated by a
loss of accuracy in measuring the fluxes for stronger lines.
This is due to synthetic spectra where the \Hb\ line is much stronger
than quite weak \Oiii\ lines, so that the measured width of the lines
is subject to larger fluctuations.
As the \Oiii\ kinematics is imposed on the \Hb\ lines while fitting
them, this can result in an error on the flux measurements that gets
worse the stronger the \Hb\ emission, which is exactly what is
observed in the simulations.
Fortunately, this situation rarely occurs in our sample galaxy, since
when the \Oiii\ lines are weaker than \Hb\ generally both lines are
quite strong.
We note that for weak \Hb\ lines the accuracy with which the fluxes
are recovered becomes again dominated by the error in estimating the
line amplitudes.

Figures~\ref{fig:appendix1a}-\ref{fig:appendix1c} also show how the
formal uncertainties on the measured parameters closely parallel the
standard deviations of the measured values from the input ones,
demonstrating that our error estimates are reliable.
Only in the case of the line fluxes, our formal error estimates
systematically over and underestimate by $\sim10$\% the observed
fluctuations for the \Oiii\ and \Hb\ lines, respectively.
 
\subsection{Loss of reliability at low $A/N$ values}
\label{app:simu_lowaon}

Figures~\ref{fig:appendix1a}-\ref{fig:appendix1c} show that as the
emission-lines get weaker our measurements become less reliable and
also start to be dominated by systematic effects.

In the case of the line velocities, there is little evidence for a
systematic bias, whereas below $A/N\sim2-3$ the line widths start to
be overestimated.
Two opposite biases are at work in this case.
As the lines become weaker, their width can either be overestimated as
their wings are increasingly lost in the noise level, or
underestimated as it becomes more likely that the fit just converges
on pixel-scale random fluctuations. In the latter case the returned
intrinsic width will always be near zero, which explains the
horizontal stripe of points in the bottom left panels of
Figures~\ref{fig:appendix1a}-\ref{fig:appendix1c}.
This last behaviour is seldom observed in real data, as some
correlation between pixels is introduced during the data reduction. As
a result, the width of the lines tends to be overestimated at low
$A/N$ regimes. In \S\ref{app:simu_tm_o3ind} it will be shown that
template-mismatch can also work in this direction.

Finally, both amplitudes and fluxes tend to be overestimated at low
$A/N$ regimes.
However, except for the fluxes of the \Oiii\ lines that can be already
biased for $A/N \leq 5$, this problem is not always visible in
Figures~\ref{fig:appendix1a}-\ref{fig:appendix1c} since the measured
$A/N$ values are themselves overestimated at this level. This is shown
in the central panels of
Figures~\ref{fig:appendix1a}-\ref{fig:appendix1c} by the skewed
stripes of points at very low $A/N$, which correspond to synthetic
spectra with lines of the same amplitude.
To highlight this bias, in Figure~\ref{fig:appendix2} we show the
recovery accuracy of the line parameters against the input $A/N$
values, clearly not accessible quantities in the real data.
In this figure both the input-output deviations and the formal
uncertainties on the amplitudes and fluxes of the lines are shown
relative to the input values.
This illustrates how accurate these measurements can be in principle,
as even for an input $A/N=3$ and a $S/N=60$, the flux and amplitudes
of both lines are recovered to a 30\% level, and are biased by less
than 10\%. 
The \Hb\ fluxes are less biased than the \Oiii\ fluxes at
low $A/N$ regimes (Figure~\ref{fig:appendix2}), as effectively the fit
to the \Hb\ line involves a lower number of degrees of freedom.

When considering the flux and amplitude deviations in
\ref{fig:appendix1a}-\ref{fig:appendix1c} in relative terms, the
observed biases remain quite limited, within 10\% for $A/N \ge 2.5$.

\subsection{Assessing the sensitivity to template mismatch: simulations}
\label{app:simu_tm_simu}

In real spectra our ability to match the stellar continuum is more
limited, as the range in metallicities and abundances in our template
library do not match what is observed in early-type galaxies.
This problem can considerably affect our emission-line measurement,
and it is difficult to simulate.
Yet, we can understand which line is more sensitive to template
mismatch by running a second set of simulations. Here only the input
template is used to match the stellar continuum, as opposed to our
standard procedure, where the whole template library is used.
The comparison of these simulations with the previous set, will tell
us if systematics are already introduced by allowing the full freedom
in our library even in fitting spectra that can be matched by our
templates and to which noise and emission lines have been added.

Figure~\ref{fig:appendix3} shows the result of this experiment. 
As expected the fit to the \Oiii\ lines is not affected by using
either the whole library of templates or only the true one, while the
estimation of the \Hb\ line amplitudes and fluxes is more sensitive in
this respect.
The accuracy in estimating the fluxes of the \Hb\ lines is on average
30\% worse when allowing the full freedom in the template library,
with very little dependence on either $A/N$ or $S/N$, even though the
models could perfectly account for the stellar continuum,

When fitting real spectra we can expect that intrinsic limitations in
our templates will make things even worse, although it is likely that
the impact on the \Oiii\ measurements will remain more limited.

\subsection{Assessing the sensitivity to template mismatch: 
\Oiii$\lambda4959$ and \Oiii$\lambda5007$ independent fits}
\label{app:simu_tm_o3ind}

The \Oiii$\lambda\lambda4959,5007$ doublet provides a natural testbed
to assess the impact of template-mismatch on our emission-line
measurements using {\it real\/} data.
When fitted independently, the \Oiii\ lines must indeed return the
same velocity and velocity dispersion, and their relative fluxes
should be consistent with theoretical predictions
\citep[\eg][]{Sto00}.

The previous simulations have demonstrated that the accuracy with
which the position and the width of the lines are recovered scales as
$A/N^{-1}$.
The weaker \Oiii$\lambda4959$ line will therefore display much larger
deviations from the real gas velocity and velocity dispersion than the
\Oiii$\lambda5007$ line. The \Oiii$\lambda4959$ deviations will
dominate the observed scatter between the velocities and velocity
dispersions derived from each of the \Oiii\ lines.
Hence, as long as the \Oiii$\lambda5007$ line is sufficiently strong,
we can use it to trace the gas LOSVD parameters, and use the results
of the \Oiii$\lambda4959$ fit to study how these parameters are
recovered as a function of $A/N$ in this particular spectral region.

Figure A4 shows the results of this experiment for two representative
galaxies in our sample. Plotted against the $A/N$ of the
\Oiii$\lambda4959$ line are, from top to bottom, the differences
between the gas velocities and {\it observed\/} velocity dispersions
measured from the two lines in the \Oiii\ doublet, and their flux
ratio. These measurements correspond only to fits where the $A/N$ of
\Oiii$\lambda5007$ was greater than 5.
The most striking feature in Figure~\ref{fig:appendix4} is the
difference between the behaviours observed in the two galaxies.
Whereas for NGC~2685 the \Oiii\ lines show position, widths, and
fluxes that are consistent with our expectations even for very weak
\Oiii$\lambda4959$ lines, in NGC~4278 the scatter between the
independently measured LOSVD parameters strongly depends on the $A/N$
of the \Oiii$\lambda4959$ line. Furthermore, as emission becomes
weaker, an increasing fraction of the fits return systematically
broader \Oiii$\lambda4959$ widths, particularly for $A/N\leq4$.

The observed behaviour is due to template-mismatch.
As shown by the green lines in Figure~\ref{fig:appendix4}, NGC~4278
has much larger stellar velocity dispersions than NGC~2685.
On the other hand, the \Oiii$\lambda5007$ lines have similar widths in
both cases.
A larger stellar velocity dispersion $\sigma$ exacerbates the impact
of template mismatch on the emission-line measurements in two ways.
First, more massive galaxies have stellar populations with higher
metallicities and enhanced abundances of their $\alpha$-elements
\citep[\eg][]{Wor92,Gre97,Tra00,Mar03,Tho05}, which are more difficult
to match with the templates in our library. 
Second, for a given strength of the template-mismatch features
affecting the emission-line measurements, in galaxies with larger
$\sigma$ the mismatch features are themselves broader and therefore
induce a positive bias on the line-widths.

\subsection{Concluding Remarks}
\label{app:simu_concl}

The previous simulations have shown that the accuracy with which the
position and width of an emission line can be measured scales only
with the relative strength of the line with respect to the deviations
from the fit to the stellar continuum. On the other hand the accuracy
in measuring amplitudes and fluxes is dominated by the entity of this
same residual noise, which itself depends on the quality of the
spectra and the limitations of the template library.

The mild dependence of the flux estimates on the precision with which
the emission-line kinematics is measured is reassuring in light of
fact that in real galaxies the \Oiii\ and \Hb\ lines do not
necessarily share the same kinematics.
Fortunately, the loss of accuracy in measuring the flux of the
strongest \Hb\ lines due to a poorly estimated kinematics from weak
\Oiii\ lines is not an issue for our sample.

At low $A/N$ regimes the emission-line measurements become dominated
by systematics effects and should not be deemed reliable.
Even in the framework of ideal simulations the accuracy in recovering
the \Hb\ fluxes is hampered by template-mismatch systematics. 
By independently fitting the lines of the \Oiii\ doublet we have
further investigated in {\it real\/} spectra the impact of
template-mismatch in the \Oiii\ spectral region. Template-mismatch is
more important in the presence of larger stellar velocity dispersions
and can induce to overestimated line-widths.

For completeness we report that simulations carried out by measuring
independently the position and width of the \Hb\ and \Oiii\ lines
show, unsurprisingly, that the \Hb\ kinematics is subject to larger
biases than in the case of the \Oiii\ lines. For instance, to achieve
the same accuracy in recovering the \Oiii\ kinematics for $A/N=4$, an
$A/N=5$ for \Hb\ is needed.

\section{Description of individual galaxies}
\label{app:list}

Here we describe the main structures observed in the \sauron\ maps for
the ionised-gas distribution and kinematics of each of the galaxies in
the E/S0 sample presented in this paper. We also comment on the
observed dust distribution, the connection to the stellar kinematics,
the \Oiii/\Hb\ ratios, and refer to previous narrow-band imaging and
long-slit spectroscopic work. We implicitly refer to
Figure~\ref{fig:Mapsa}-\ref{fig:Mapsb} when describing these
structures.

\begin{description} 

\item[{\bf NGC 474:}] This well-known shell galaxy \citep{Tur99}
displays a peculiar ionised gas distribution and kinematics.
The \Oiii\ distribution follows a twisted pattern, along which the
\Oiii\ lines are stronger at the northern side of the galaxy between
5\arcsec\ and 7\arcsec\ whereas they become stronger at the southern
end at larger radii. Although the \Hb\ lines are much fainter than the
\Oiii\ lines, they seem to follow a similar spatial distribution and
indeed escape detection on the northern side beyond 7\arcsec.
The gas velocity field is characterised by a central and an outer
component with almost opposite angular momenta. In both cases the gas
kinematics is decoupled from the stellar one. The width of the lines
appear to peak in the transition region between the inner and outer
gas velocity components, which can be explained as a bias introduced
by fitting a single Gaussian to what in fact in some spectra looks
like a superposition of two separate components with different
velocities.
The absence of strong absorption features is surprising given the
considerable amount of gas emission.

\item[{\bf NGC 524:}] This galaxy displays a disk of gas in very
regular rotation, in the same sense of the stars, consistent with the
IFS observations of \citet{Sil00}.
The emission is close to our detection limit, however, which explains
the patchy appearance of the maps. A similarly fragmented distribution
is shown by the \Ha+\Nii\ narrow-band images of \citet{Mac96}.
The gas distribution is also matched by an extremely regular
distribution of the dust, organised in concentric circular lanes.
The strength of the \Hb\ and the \Oiii\ emission are comparable,
although at the edge of the disk the \Oiii\ lines become stronger.

\item[{\bf NGC 821:}] 
No significant emission is detected in this close to edge-on galaxy.
The {\it HST\/} unsharp-masking image reveals a nuclear disk
structure but no dust, consistent with \citet{Rav01}.

\item[{\bf NGC 1023:}] This galaxy displays \Oiii\ emission that
remains strong in the outer regions, with a distribution that appears
to be skewed with respect the galaxy major axis.
Yet, the gas shows a coherent, though not perfectly regular, velocity
field with a reasonably defined zero-velocity curve, which indicates a
mild kinematic misalignment between gas and stars.
The approaching south-eastern arm indicates that the angular momentum
of the gas may vary in the outermost parts of the galaxy.
This would be consistent with the complex neutral hydrogen morphology
and kinematics found by \citet{San84}, suggesting an
interaction with three nearby companion galaxies.
No noticeable absorption feature is visible in the \HST\ image, which
also in this case reveals the presence of a nuclear stellar disk.

\item[{\bf NGC 2549:}] Another object with strong and misaligned
emission in its outer parts, where the gas kinematics is misaligned
with respect to the stellar kinematics.
Towards the centre, however, the angular momentum of the gas quickly
aligns itself behind the stellar angular momentum and along the galaxy
minor axis, suggesting the presence of a gaseous system well settled on
the galaxy equatorial plane.
Also in this case no dust is observed, although the unsharp-masked
image reveals the clear peanut-shape signature of a boxy bulge.

\item[{\bf NGC 2685:}] Also known as the Helix galaxy \citep{Bur59},
this famous peculiar object shows a gas distribution almost
perpendicular to the galaxy major axis. A spiral-like structure is
evident in the flux maps, in particular on the north-eastern side of
the galaxy. This feature could also be regarded as a warped
configuration where gas is settling on the galaxy equatorial plane
towards the central regions.
The gas kinematics is also strongly misaligned with respect to the
stellar kinematics, on average by 75\degr, and shows a complex
velocity field, with a zero-velocity curve that is skewed with respect
to the direction along which the most extreme velocities are observed.
Figure~\ref{fig:Mapsb} also shows that towards the centre the gas
angular momentum is aligning itself behind the stellar one. Near IR
narrow-band \HST\ images suggest the presence of a nuclear P$\alpha$
disk aligned with the galaxy major axis \citep{Boe99}, which would
confirm the presence of a settled central gaseous system.
The velocity dispersion map also shows peculiar features $\pm
5\arcsec$ away from the centre along the minor axis, where \Sg\
increases up to 150~\kms. No strong evidence for complex line profiles
is found.
The unsharp-masked image reveal a nuclear disk and almost
perpendicular dust lanes both across the centre and towards the
north-eastern side of the galaxy, where also the \Hb\ emission is
stronger than the \Oiii. Elsewhere the distribution of the \Oiii/\Hb\
ratio is complex.

\item[{\bf NGC 2695:}] No emission is detected in this galaxy, and no
\HST\ image is available. The \sauron\ unsharp-masked image does not
show peculiar features, besides highlighting a twisting of the inner
isophotes.

\item[{\bf NGC 2699:}] This object shows emission only in the central
regions, making it difficult to judge the gas kinematics.
The \HST\ image reveals a bilobate structure underscoring the presence
of a nuclear bar, and a very small nuclear dust disk that was noticed
also by \citet{Tra01}.

\item[{\bf NGC 2768:}] This galaxy is known for hosting a central dust
lane along the minor axis \citep{Kim89} and for showing a kinematic
decoupling between stars and gas \citep{Ber92} that was later
interpreted as the result of a gaseous polar-disk structure
\citep{Fri94}. The \sauron\ maps further reveal remarkably different
distributions for the \Oiii\ and \Hb\ emission. The \Oiii\ lines
follow a twisted pattern that is misaligned by $\sim$45\degr\ from the
direction of the dust lanes, while the \Hb\ distribution appears to be
rounder. \HST\ narrow-band images by \citet{Mar04} show that the
\Ha+\Nii\ emission in the central arcsecond also follows an
integral-sign pattern that is decoupled from the orientation of the
dust.
The \Oiii\ velocity field is well defined, showing peculiar
and almost symmetric twists at $\sim$10\arcsec\ from the centre.
Overall, the gas motions are almost perpendicular to the stellar ones.
The \Sg-map shows an elongated peak parallel to the zero-velocity curve,
consistent with the observations of \citet{Ber95}, and possibly also a
rise of \Sg\ corresponding to the peculiar twists in the velocity
field.
The crisscross appearance of the \Oiii/\Hb\ maps can be interpreted in
light of the different shape of \Oiii\ and \Hb\ distributions.

\item[{\bf NGC 2974:}] \sauron\ observations for this galaxy were
previously discussed in detail in \citet{Kra05}. Here we note how the
$EW$-maps highlights the presence of a nuclear bar surrounded by a
ring and of two spiral arms connecting to the outer parts of the field
where emission remains strong. The presence of nested bars within this
galaxy is discussed in the light of the \sauron\ and \HST\
observations by Krajnovi\'c et al. and \citet{Ems03}. The inner bar is
responsible in particular for the observed twist within the central
4\arcsec\ in the velocity field, which is otherwise very regular.
The velocity dispersion of the gas rises up to 250~\kms\ and remains
moderately high ($\sim$100~\kms) also in the rest of the map. The
possibility of a dynamical support for the gas motions was first
investigated by \citet{Cin94} and recently also by Krajnovi\'c et al.
Although not perfectly regular, the dust distribution appears to trace
the general orientation of the gaseous disk, following in particular
part of the $EW$-ring.

\item[{\bf NGC 3032:}] Although heavy binning limits our ability to
judge the regularity of the gas distribution and kinematics in this
galaxy, overall the gas appears to be consistent with a disk of gas in
circular rotation, in almost the opposite sense than the stars. The
presence of very regular dust-lanes and of a more regular \Hb\
kinematics (Figure~\ref{fig:appendix5c}) further suggest a disk
distribution for the gas.
This dusty galaxy features the strongest \Hb\ emission in our sample,
and the lowest values for the \Oiii/\Hb\ ratios. The relative strength
of the \Hb\ line compared to the \Oiii$\lambda5007$ line shows a clear
radial pattern: from twice as bright at the edge of the disk, to ten
times stronger at $\sim$4\arcsec, and finally giving way to the \Oiii\
lines very sharply in the central 2\arcsec. The detection of molecular
gas by \citet{Sag89} further suggest ongoing star formation activity.

\item[{\bf NGC 3156:}] This dusty galaxy shows a complex distribution
for the \Oiii\ emission, with filamentary structures extending
perpendicularly to the galaxy major axis. On the other hand, the \Hb\
emission follows more closely the stellar distribution and forms a
nuclear ring corresponding to the location of the central absorption
features visible in the \HST\ images.
Despite the peculiar features in the \Oiii\ distribution the gas
kinematics show rather coherent motions and very small line
widths. The \Oiii/\Hb\ map displays very large fluctuations, with high
\Oiii/\Hb\ ratios ($\ge 3$) corresponding to the main \Oiii\
structures and lower values corresponding to the nuclear \Hb\ ring.

\item[{\bf NGC 3377:}] \sauron\ observations for the ionised-gas
distribution and kinematics of NGC~3377 were previously presented in
Paper~I. Here we note how the ionised-gas distribution displays an
integral-sign shaped pattern that is parallelled by a similar twist in
the gas velocity field.
Along this pattern the emission-line fluxes display a much milder
decrease with galactocentric distance than the stellar continuum,
resulting in decreasing EW values towards the centre where the weaker
\Hb\ emission eventually escapes detection.
The single dust filaments visible on the northern side of the galaxy
was also recognised by \citet{Tra01}.

\item[{\bf NGC 3379:}] The largest member of the Leo Group, this
galaxy displays a central disk of gas and a number of isolated sources
of emission. The direction of maximum rotation of the central disk is
consistent with the orientation of the nuclear dusty ring visible in
the \HST\ images, and indicates a $\sim$45\degr\ kinematic
misalignment between gas and stars. The impact of template-mismatch in
biasing the \Sg\ measurements towards overestimated values
(\S\ref{subsubsec:Sgas}) is visible at the edge of the central disk,
where the strength of the emission lines approaches the detection
limit.
On the other hand, all isolated sources show very narrow \Oiii\ lines
and undetected \Hb\ emission, consistent with large \Oiii/\Hb\ ratios.
These characteristics suggest that these sources are in fact PNe.  
The two sources with the largest EW values $\sim$20\arcsec\ south and
south-west from the centre, have positions and velocities consistent
with the PNe \#4 and \#7 of \citet{Cia93}, respectively.

\item[{\bf NGC 3384:}] This galaxy shows weak \Oiii\ and \Hb\ emission
close to our detection limit. The unsharp-masked image suggests the
presence of two, nested, disk structures. \citet{Rav01} noticed the
same features in near-IR \HST\ images.

\item[{\bf NGC 3414:}] This peculiar S0 galaxy (Arp 162) shows a
spiral pattern in the distribution of the gas, which is more evident
in the EW maps and is accompanied by a 90\degr\ twist of the velocity
field.
In the central regions the gas rotates almost perpendicularly with
respect to the stars, as noticed by \citet{Sil04}.
The \Sg-map show a very sharp peak, which is elongated in the same
direction of the zero-velocity curve, as in the case of NGC~2768.
Weak filamentary absorption features are visible in the southern part
of this galaxy, consistent with \citet{Tra01}.

\item[{\bf NGC 3489:}] Similarly to NGC~3156, also this dusty galaxy
features a complex \Oiii\ distribution with filamentary structures
and, despite that, a fairly regular velocity field and very narrow
lines. It also shows a more uniform distribution for the \Hb\
emission, which explains the resemblance between the maps for the
\Oiii/\Hb\ ratio and the EW of the \Oiii$\lambda5007$ line.

\item[{\bf NGC 3608:}] As in NGC~3414, this galaxy shows a spiral
pattern in the ionised-gas distribution. In this case, however, the
emission lines are much weaker and no gas is detected between the
arms. Furthermore both NGC~3414 and NGC~3608 show little overall
stellar rotation except in the central regions, where the kinematics
of gas and stars are strongly decoupled. No clear absorption features
are visible in the \HST\ image, although \citet{Tom00} claim to detect
a dusty disk.

\item[{\bf NGC 4150:}] This object displays a extremely smooth
EW-maps, exemplifying how closely the gas emission can follow the
stellar continuum in early-type galaxies. Despite heavy binning the
\Vg- and \Sg-maps suggest fairly regular gas motions in a dynamically
cold disk.
A complex dust structure in the central 3\arcsec corresponds to the
region where the stellar kinematics reveal counter-rotating structure
(Paper~III) and to a minimum in the \Oiii/\Hb\ distribution.
Outside these central features, the gas and stellar kinematics are
mildly misaligned by 20\degr.

\item[{\bf NGC 4262:}] Like NGC~3377, this strongly barred galaxy
shows an integral-sign pattern in the gas distribution and a twisted
\Vg-field.  In addition, this object shows peculiar asymmetric
distributions for the central values of \Sg\ and for the \Oiii/\Hb\
ratio. The gas and stellar kinematics are strongly decoupled. The
unsharp-masked \sauron\ image shows a bilobated structure reminiscent
of that observed at smaller scales in NGC~2699.

\item[{\bf NGC 4270:}] This galaxy shows only fragmented traces of
emission and no evidence for significant absorption features.

\item[{\bf NGC 4278:}] This galaxy shows the strongest emission, in
terms of the EW of the lines, in our sample, allowing to appreciate in
detail the peculiar distribution and kinematics of the gas. As in
previous objects, the ionised gas displays an integral-sign pattern, a
twisted \Vg-field, and an asymmetric \Sg\ central peak and
distribution of the \Oiii/\Hb\ ratio.
Dust absorption features are visible only on the north-western side of
the galaxy. Their morphology suggests an orientation similar to that
of the integral-sign pattern.
The gaseous and stellar kinematics are misaligned by increasingly
wider angles, as the stellar and gaseous velocity fields twist in
opposite direction towards the outer parts of the field.
This object also hosts an outer \Hi\ ring \citep{Rai81} that may be
physically connected with the ionised gas given the similar
orientation of the two gaseous systems \citep{Gou94}.

\item[{\bf NGC 4374:}] Known for its low-luminosity radio jet
\citep[\eg][]{Lai87}, also this galaxy shows intense ionised-gas
emission. This is confined along a lane running across the galaxy from
east to west, which broadens up in the north-south direction at large
radii. The central dust lane also runs across the nucleus in the same
direction of the ionised-gas, and perpendicular to the radio jet.
The gas kinematics indicate a coherent rotation, except in the eastern
direction where the initially approaching gas gradually reverse its
angular momentum and eventually, beyond $\sim$15\arcsec\ is fully
receding.  Clear stellar rotation is observed only in the central
$\sim$3\arcsec, in almost opposite direction with respect to the
gas. The \Oiii/\Hb-map shows rather uniform values, between 1-2.
We note that Chandra observations reveal an H-shaped distribution for
the soft X-ray emission \citep{Fin01} that is broadly consistent with
the ionised-gas distribution.

\item[{\bf NGC 4382:}] Only weak \Oiii\ emission is detected in this
galaxy, mostly towards the centre. The \HST\ unsharp-masked image
suggests the presence of a dust lane running across the nucleus along
the minor axis direction. However, \citet{Lau05} show the absence of
dust reddening in the centre and find that NGC~4382 displays a central
minimum in its surface brightness distribution.

\item[{\bf NGC 4387:}] This galaxy shows only marginally detectable
emission in few spectra, corresponding to isolated sources, possibly
PNe.

\item[{\bf NGC 4458:}] This galaxy displays only fragmented traces of
emission, in particular close to the centre. The \HST\ unsharp-image
reveals a nuclear disk \citep{Mor04}, consistent with the presence of
a central fast-rotating component in the stellar velocity field.

\item[{\bf NGC 4459:}] This S0 galaxy shows an extremely well-defined
dust distribution and regular gas kinematics. The \Hb\ emission peaks
in a circumnuclear ring, corresponding to the region where the dust
distribution appears the most circularly symmetric and flocculent. On
the other hand, the \Oiii\ emission very closely follows the stellar
continuum. The values of \Oiii/\Hb\ ratio therefore show a very
similar radial pattern to NGC~3032, with a minimum in a circumnuclear
region and stronger \Oiii/\Hb\ ratios toward the centre and the edge
of the gas disk.
The independently derived \Hb\ kinematics show faster rotation and
smaller velocity dispersions than the \Oiii\ kinematics, suggesting
that the \Hb\ line is tracing better than the \Oiii\ doublet a
kinematically colder component of gas, which is currently forming
stars. The detection of molecular gas by \citet{Sag89} supports
this picture.

\item[{\bf NGC 4473:}] Also this galaxy displays scattered traces of
emission, mostly from the \Oiii\ doublet. \Nii+\Ha\ emission was found
by \citet{Mac96} but not by \citet{Gou94}. The \HST\ unsharp-masked
image highlights the presence of a nuclear disk, already noticed by
\citet{vdB94}. No dust is visible, contrary to the findings of
\citet{Fer99} but consistent with \citet{vDo95}.

\item[{\bf NGC 4477:}] Ionised-gas emission in this object is very
concentrated toward the centre whereas it decreases more gently than
the stellar continuum at outer radii. The gas kinematics is rather
regular, and misaligned with respect to the stellar kinematics by
$\sim$30\degr. \Hb\ and \Oiii\ emission have comparable strengths over
most of the field except in the central 5\arcsec\ where the \Oiii/\Hb\
ratio rises to values $\sim$3. Beside the evident central absorption
features, weak dust lanes are visible $\sim$7\arcsec\ to the south.

\item[{\bf NGC 4486:}] The central member of the Virgo cluster, M87
displays the most complicated gas structures in our sample.  The
ionised-gas distribution features several filamentary structures and a
very bright central component, consistent with previous narrow-band
observations \citep[\eg][]{Spa93}. In the central 2\arcsec\ the
emission lines have complex profiles that are poorly matched by single
Gaussians, which explains the large central values for \Sg. No \Oiii\
emission is detected along the radio jet, whereas we found \Hb\
emission when this line is fitted independently.
The gas velocities show extreme variations across the field,
particularly along the main filamentary structures to the east of the
nucleus. The ionised gas is outflowing from the centre to the north,
decelerating at larger radii. The main features of the \Vg-map are
remarkably consistent with the long-slit observations of
\citet{Spa93}.
Most of the observed emission comes with rather intermediate values
for the \Oiii/\Hb\ ratio, between 1 and 2, except in the south-eastern
filament and on the closer end of bright patch $\sim$30\arcsec
south-east from the centre.

\item[{\bf NGC 4526:}] Another prototypical example of an early-type
galaxy with a perfect dust disk, very regular gas kinematics, a
circumnuclear region where \Hb\ emission is particularly strong, and an
independently derived \Hb\ kinematics indicating faster rotation and
smaller velocity dispersions than the \Oiii\ kinematics. As in
NGC~3032 and NGC~4459 molecular gas was detected by \citet{Sag89}.

\item[{\bf NGC 4546:}] Another galaxy showing an integral-shape
pattern in ionised-gas distribution, along which the emission is
stronger on the western side of the galaxy $\sim$10\arcsec\ from the
centre, and on the eastern side beyond 15\arcsec. As in NGC~4262 and
NGC~4278 we observe a strong twisting of the gas kinematics, and an
asymmetric distribution for the central \Sg\ values and for the
\Oiii/\Hb\ ratios. The gaseous and stellar motions are strongly
decoupled, with almost opposite angular momenta \citep{Gal87}.
Absorption features are found only close to the centre.

\item[{\bf NGC 4550:}] Known for hosting two counter-rotating stellar
disks \citep{Rub92, Rix92}, the ionised-gas distribution and
kinematics of this peculiar object, together with the observed dust
morphology, is consistent with gas moving in a fairly inclined and
dynamically cold disk. A spiral arm extending to the south is evident
in the EW maps. \citet{Wik01} note a similarly lopsided distribution
for the molecular gas they detect. Consistent with the measurements of
Rubin et al. and Rix et al., the ionised-gas rotates against the stars
in the disk dominating the stellar kinematics along the major axis
beyond 10\arcsec. On the other hand, the gas rotates in the same sense
of the stars in the main body of the galaxy.
Since this implies that the gas does not necessarily have an external
origin, we decided to adopt a value of zero for the kinematic
misalignment between gas and stars in NGC~4550 for the purpose of the
analysis of \S\ref{sec:starandgas}. Independently derived maps for the
\Oiii\ distribution and kinematics were presented by \citet{Afa02}.

\item[{\bf NGC 4552:}] Ionised-gas emission, mostly from the \Oiii\
doublet, is clearly detected in the central regions of this object
although template-mismatch severely affects the measured values of
\Sg. In the central 4\arcsec\ the gas kinematics is strongly decoupled
from the stellar kinematics, confirming that the observed emission is
not purely an template-mismatch artifact. \Ha+\Nii\ emission was
detected also by \citet{Mac96}. 
Nuclear \Oiii\ emission was found also by \citet{Cap99}.
Patchy absorption features are visible on the north-western side of
the centre and $\sim$8\arcsec\ to the north-east.  Upon closer
inspection, the \HST\ image shows the presence of a very small ring of
dust ($r\!\!\sim\!\!0\farcs25$) surrounding the nucleus, consistent
with the analysis of \citet{Car97}.

\item[{\bf NGC 4564:}]
No significant emission is detected in this galaxy.

\item[{\bf NGC 4570:}] This object displays several scattered sources
of emission. The observed emission is characterised by narrow lines,
which appear to trace gas clouds coherently rotating in the same sense
of the stars.
We found no traces of dust in the \HST\ image, where the nuclear disk
studied by \citet{vdB98} is clearly visible.

\item[{\bf NGC 4621:}] Emission in this galaxy is found the centre and
in a number of isolated sources. Also in this case the scattered
sources broadly follow the main stellar stream. As in NGC~4552
template mismatch affects the central \Sg\ measurements, but in this
case the gas velocities do not show significant rotation and instead
appear to be offset from the central stellar velocities, which are
close to zero. The presence of nuclear emission should therefore be
confirmed. No \Ha+\Nii\ emission was detected by \citet{Gou94}

\item[{\bf NGC 4660:}] This galaxy shows clearly detected \Oiii\
emission, possibly originating from PNe, in a few spectra, in
particular from the southern part of the galaxy. A nuclear disk is
visible in the \HST\ image, as recognised also by \citet{Lau05}

\item[{\bf NGC 5198:}] 
The ionised-gas distribution in this object is characterised by a
central component and a filamentary structure extending from it to the
north. Near the centre the gas rotates rapidly and almost
perpendicularly with respect to the stars in the kinematically
decoupled core (Paper~III). 
No coherent pattern for the gas motions in the filaments is
recognised, except that the gas moves relatively slowly.
No evidence for dust is found, consistent with \citet{Tra01}.

\item[{\bf NGC 5308:}] No emission is detected in this galaxy, which
is almost certainly seen edge-on given the extreme inclination of its
nuclear stellar disk \citep{Kra04b}

\item[{\bf NGC 5813:}] The ionised-gas distribution in this object is
confined within a lane running along the galaxy minor axis. The
observed gas kinematics is rather complex outside the central
$\sim8$\arcsec, but in the central regions the gas show coherent
rotation in approximately the same sense of the stars.
The present \Vg-map is consistent with the measurements shown in
Paper~II, once the effect of spatial binning is accounted for.
Dust features take the form of filaments running towards the centre
from the south and the east. Absorption is also strong on the
north-western side of the nucleus. The emission-line fluxes have a
similar distribution near the centre. In particular the flux
distribution is lopsided along the major-axis direction.
Noteworthy are also the strong central \Sg\ peak and the rather
uniform distribution of intermediate values for the \Oiii/\Hb\ ratio.

\item[{\bf NGC 5831:}] This galaxy shows emission only close to the
centre, making it hard to comment on the gas kinematics. No trace of
dust absorption features is found.

\item[{\bf NGC 5838:}] Another S0 galaxy with a very regular dust
disk. Considering the small scale of the disk and the effects of
atmospheric blurring, the observed gas distribution and kinematics is
remarkably similar to that of NGC~4459 and NGC~4526.

\item[{\bf NGC 5845:}] Only weak \Oiii\ emission is detected towards
the centre of this galaxy. The \HST\ image shows a nuclear stellar
disk with an inner dust ring, as noticed by \citet{Tra01}.

\item[{\bf NGC 5846:}] Ionised-gas in this galaxy is found within
$\sim$10\arcsec and along a filament running to the north-east.
Although characterised by large fluctuations in some places, the \Vg\
field clearly indicates a nearly cylindrical rotation for the gas, in
the opposite sense of the stars. Remarkably, the zero-velocity curve
runs across the north-eastern filament, suggesting this is not a
separate component from the gas closer to the centre.
The \HST\ image shows dust along an almost parabolic lane swinging
around the nucleus. Upon closer inspection, the emission-line fluxes
follow very closely the dust distribution, in particular from \Hb.
Relatively intermediate \Oiii/\Hb\ values are observed across most of
the field.

\item[{\bf NGC 5982:}] 
As in NGC~5198, the gas in this bright elliptical is found in a
rapidly rotating central component and in a filament extending from
it, in this case to the south. In NGC~5982, however, the motions along
the filament are remarkably coherent. The gas velocities seem to
suggest that the gas, initially approaching the observer from the
southern end of the filament, progressively decelerates and plunges
back toward the centre following a curved trajectory.
The gas in the centre rotates in the same sense as the stars in the
well-known kinematically decoupled core \citep{Wag88}.
Absorption features are neither found in the V-band unsharp-masked
image nor in the V-H colour image of \citet{Qui00}.

\item[{\bf NGC 7332:}] 
\sauron\ maps for the ionised-gas distribution and kinematics for this
SB0 galaxy were already presented by \citet{Fal04}.
Here we note how, like in NGC~2549 and NGC~1023, the gas emission
remains relatively strong in the outer parts of the galaxy, where the
gas distribution is misaligned with respect to the galaxy major axis
and the gas kinematics is decoupled from that observed in the inner
regions. We also note the large variations in the \Oiii/\Hb\ values,
and the presence of dust in the \HST-WFPC image.

\item[{\bf NGC 7457:}] Another galaxy with strong and misaligned
emission in its outer parts. Near the centre the gas rotates in the
opposite sense to the gas in the outer parts and also against the
motion of the bulk of the stars in this object.
No dust is found in the \HST\ unsharp-masked image.

\end{description}

\section{Maps for independently derived \Oiii\ and \Hb\ kinematics}
\label{app:o3vhb}

In this appendix we show for 10 galaxies maps for independently
derived velocity and intrinsic velocity dispersion of the \Oiii\ and
\Hb\ lines.
These objects are the only ones in our sample for which, over most of
the regions where emission is observed, the measured \Hb\ kinematics do
not suffer from the template-mismatch effects described in
\S\ref{subsec:GasPaper_method}.
This was ensured by comparing the \Hb\ velocity maps to the stellar
velocity maps and by further noticing that spurious \Hb\ emission
lines tend to be considerably broader than the \Oiii\ lines.

Figures~\ref{fig:appendix5a}-\ref{fig:appendix5c} show the
independently derived \Oiii\ and \Hb\ velocity and intrinsic velocity
dispersion of these 10 galaxies.
Since among them, 3 are giant ellipticals and 4 show regular and
concentric dust lanes, Figure~\ref{fig:appendix5a} shows first the
maps for the remaining 3 galaxies, while figures~\ref{fig:appendix5b}
and \ref{fig:appendix5c} shows maps for these two groups.
In addition, Figure~\ref{fig:appendix5a} shows again the case of
NGC~3489, to highlight the main characteristics of an unreliable \Hb\
kinematics.

\placefigAppfoura
\placefigAppfourb
\placefigAppfourc

\label{lastpage}


\begin{thebibliography}{}
%
\bibitem[Afanasiev \& Sil'chenko(2002)]{Afa02} Afanasiev, 
V.~L., \& Sil'chenko, O.~K.\ 2002, \aj, 124, 706

\bibitem[Bacon et al.(2001)]{Bac01} Bacon R.\ et al., 2001, \mnras,
326, 23 (Paper~I)

\bibitem[Bertola et al.(1992)]{Ber92} Bertola, F., Buson, L.~M., \&
Zeilinger, W.~W.\ 1992, \apjl, 401, L79

\bibitem[Bertola et al.(1995)]{Ber95} Bertola, F., Cinzano, P.,
Corsini, E.~M., Rix, H., \& Zeilinger, W.~W.\ 1995, \apjl, 448, L13

\bibitem[Bertola \& Corsini(1999)]{Ber99} Bertola, F., \& Corsini,
E.~M.\ 1999, IAU Symp.~186: Galaxy Interactions at Low and High
Redshift, 186, 149

\bibitem[Binette et al.(1994)]{Bin94} Binette, L., Magris, C.~G.,
Stasinska, G., \& Bruzual, A.~G.\ 1994, \aap, 292, 13

\bibitem[Binney(2005)]{Bin05} Binney, J.\ 2005, \mnras, 863 
 
\bibitem[B{\" o}ker et al.(1999)]{Boe99} B{\" o}ker, T., et 
al.\ 1999, \apjs, 124, 95 

\bibitem[Burbidge \& Burbidge(1959)]{Bur59} Burbidge E.~M.~\&
Burbidge G.~R., 1959, \apj, 130, 20

\bibitem[Buson et al.(1993)]{Bus93} Buson, L.~M., et al.\ 
1993, \aap, 280, 409 

\bibitem[Cappellari et al.(1999)]{Cap99} Cappellari, M., 
Renzini, A., Greggio, L., di Serego Alighieri, S., Buson, L.~M.,
Burstein, D., \& Bertola, F.\ 1999, \apj, 519, 117

\bibitem[Cappellari \& Copin(2003)]{Cap03} Cappellari M., Copin Y.,
2003, \mnras, 342, 345

\bibitem[Cappellari \& Emsellem(2004)]{Cap04} Cappellari M., Emsellem
E., 2004, PASP, 116, 138

\bibitem[Carollo et al.(1997)]{Car97} Carollo, C.~M., Franx, 
M., Illingworth, G.~D., \& Forbes, D.~A.\ 1997, \apj, 481, 710

\bibitem[Caon et al.(2000)]{Cao00} Caon, N., Macchetto, D., 
\& Pastoriza, M.\ 2000, \apjs, 127, 39 

\bibitem[Ciardullo et al.(1989)]{Cia89} Ciardullo, R., 
Jacoby, G.~H., \& Ford, H.~C.\ 1989, \apj, 344, 715

\bibitem[Ciardullo et al.(1993)]{Cia93} Ciardullo, R., 
Jacoby, G.~H., \& Dejonghe, H.~B.\ 1993, \apj, 414, 454 

\bibitem[Cinzano \& van der Marel(1994)]{Cin94} Cinzano, P., 
\& van der Marel, R.~P.\ 1994, \mnras, 270, 325 

\bibitem[Debattista et al.(2002)]{Deb02} Debattista, V.~P., 
Corsini, E.~M., \& Aguerri, J.~A.~L.\ 2002, \mnras, 332, 65 

\bibitem[de Jong et al.(1990)]{dJo90} de Jong, T., Norgaard-Nielsen,
H.~U., Jorgensen, H.~E., \& Hansen, L.\ 1990, \aap, 232, 317

\bibitem[de Robertis \& Osterbrock(1986)]{dRo86} de Robertis, 
M.~M., \& Osterbrock, D.~E.\ 1986, \apj, 301, 727 

\bibitem[de Vaucouleurs et al.(1991)]{RC3} de Vaucouleurs G., de
Vaucouleurs A., Corwin H.~G., Buta R.~J., Paturel G., Fouque P., 1991,
Volume 1-3, XII, Springer-Verlag Berlin Heidelberg New York,

\bibitem[de Zeeuw et al.(2002)]{dZe02} de Zeeuw P.~T., et al., 2002,
\mnras, 329, 513 (Paper~II)

\bibitem[di Serego Alighieri et al.(1990)]{dSe90} di Serego 
Alighieri, S., Trinchieri, G., \& Brocato, E.\ 1990, ASSL Vol.~160:
Windows on Galaxies, 301

\bibitem[Dopita \& Sutherland(1995)]{Dop95} Dopita, M.~A., \& 
Sutherland, R.~S.\ 1995, \apj, 455, 468 

\bibitem[Dopita \& Sutherland(1996)]{Dop96} Dopita, M.~A., \& 
Sutherland, R.~S.\ 1996, \apjs, 102, 161 

\bibitem[Emsellem et al.(2003)]{Ems03} Emsellem, E., Goudfrooij, P.,
\& Ferruit, P.\ 2003, \mnras, 345, 1297

\bibitem[Emsellem et al.(2004)]{Ems04} Emsellem, E., et al.\ 
2004, \mnras, 352, 721 (Paper~III)

\bibitem[Eskridge et al.(1995a)]{Esk95a} Eskridge, P.~B., 
Fabbiano, G., \& Kim, D.\ 1995a, \apj, 442, 523 
 
\bibitem[Eskridge et al.(1995b)]{Esk95b} Eskridge, P.~B., 
Fabbiano, G., \& Kim, D.\ 1995b, \apjs, 97, 141 

\bibitem[Fabbiano(2003)]{Fabb03} Fabbiano, G.\ 2003, Advances 
in Space Research, 32, 2013 

\bibitem[Faber \& Gallagher(1976)]{Fab76} Faber, S.~M., \& 
Gallagher, J.~S.\ 1976, \apj, 204, 365 

\bibitem[Falc{\' o}n-Barroso et al.(2004)]{Fal04} Falc{\' 
o}n-Barroso, J., et al.\ 2004, \mnras, 350, 35

\bibitem[Ferrari et al.(1999)]{Fer99} Ferrari, F., Pastoriza, 
M.~G., Macchetto, F., \& Caon, N.\ 1999, \aaps, 136, 269

\bibitem[Finoguenov \& Jones(2001)]{Fin01} Finoguenov, A., \& 
Jones, C.\ 2001, \apjl, 547, L107

\bibitem[Filippenko \& Halpern(1984)]{Fil84} Filippenko, 
A.~V., \& Halpern, J.~P.\ 1984, \apj, 285, 458 

\bibitem[Franx et al.(1991)]{Fra91} Franx, M., Illingworth, G., \& de
Zeeuw, P.~T.\ 1991, \apj, 383, 112

\bibitem[Fried \& Illingworth(1994)]{Fri94} Fried, J.~W., \&
Illingworth, G.~D.\ 1994, \aj, 107, 992

\bibitem[Galletta(1987)]{Gal87} Galletta, G.\ 1987, \apj, 
318, 531

\bibitem[Goudfrooij et al.(1994)]{Gou94} Goudfrooij, P., Hansen, L.,
Jorgensen, H.~E., \& Norgaard-Nielsen, H.~U.\ 1994, \aaps, 105, 341

\bibitem[Goudfrooij(1999)]{Gou99} Goudfrooij, P.\ 1999, ASP 
Conf.~Ser.~163: Star Formation in Early Type Galaxies, 163, 55 

\bibitem[Greggio(1997)]{Gre97} Greggio, L.\ 1997, \mnras, 285, 151

\bibitem[Ho et al.(1995)]{Ho95} Ho, L.~C., Filippenko, 
A.~V., \& Sargent, W.~L.\ 1995, \apjs, 98, 477 

\bibitem[Ho et al.(1997a)]{Ho97a} Ho, L.~C., Filippenko, 
A.~V., \& Sargent, W.~L.~W.\ 1997a, \apjs, 112, 315 

\bibitem[Ho et al.(1997b)]{Ho97b} Ho, L.~C., Filippenko, 
A.~V., Sargent, W.~L.~W., \& Peng, C.~Y.\ 1997b, \apjs, 112, 391 

\bibitem[Ho et al.(1997c)]{Ho97c} Ho, L.~C., Filippenko, 
A.~V., \& Sargent, W.~L.~W.\ 1997c, \apj, 487, 568 

\bibitem[Ho et al.(2002)]{Ho02} Ho, L.~C., Sarzi, M., Rix, H.,
Shields, J.~C., Rudnick, G., Filippenko, A.~V., \& Barth, A.~J.\ 2002,
\pasp, 114, 137

\bibitem[Hoaglin et al.(1983)]{Hoa83} Hoaglin, D. C., Mosteller,
F. \& Tukey, J. W. 1983, Understanding Robust and Exploratory Data
Analysis (New York: Wiley)

\bibitem[Jones(1997)]{Jon97} Jones L. A., 1997, Ph.D. thesis, 
Univ. North Carolina, Chapel Hill

\bibitem[Kannappan \& Fabricant(2001)]{Kan01} Kannappan, S.~J., \&
Fabricant, D.~G.\ 2001, \aj, 121, 140

\bibitem[Kim(1989)]{Kim89} Kim, D.\ 1989, \apj, 346, 653

\bibitem[Krajnovi\'c(2004)]{Kra04a} Krajnovi\'c, D. 2004, Ph.D.~Thesis,
University of Leiden

\bibitem[Krajnovi{\' c} \& Jaffe(2004)]{Kra04b} Krajnovi{\' 
c}, D., \& Jaffe, W.\ 2004, \aap, 428, 877

\bibitem[Krajnovi{\' c} et al.(2005)]{Kra05} Krajnovi{\' c}, D.,
Cappellari, M., Emsellem, E., McDermid, R.~M., \& de Zeeuw, P.~T.\
2005, \mnras, 357, 1113

\bibitem[Laing \& Bridle(1987)]{Lai87} Laing, R.~A., \& 
Bridle, A.~H.\ 1987, \mnras, 228, 557 

\bibitem[Lauer et al.(2005)]{Lau05} Lauer, T.~R., et al.\ 
2005, \aj, 129, 2138

\bibitem[Macchetto et al.(1996)]{Mac96} Macchetto, F., Pastoriza, M.,
Caon, N., Sparks, W.~B., Giavalisco, M., Bender, R., \& Capaccioli,
M.\ 1996, A\&AS, 120, 463

\bibitem[Maraston et al.(2003)]{Mar03} Maraston, C., Greggio, 
L., Renzini, A., Ortolani, S., Saglia, R.~P., Puzia, T.~H., \&
Kissler-Patig, M.\ 2003, \aap, 400, 823

\bibitem[Martel et al.(2004)]{Mar04} Martel, A.~R., et al.\ 2004, \aj,
128, 2758

\bibitem[McDermid et al.(2005)]{McD05} McDermid, R.~M., et al.\ 2005, New Astr. Rev., 
in press, astro-ph/0508631

\bibitem[Morelli et al.(2004)]{Mor04} Morelli, L., et al.\ 2004, \mnras, 
354, 753

\bibitem[Osterbrock(1989)]{Ost89} Osterbrock, D.~E. 1989, 
Astrophysics of Gaseous Nebulae and Active Galactic
Nuclei. Univ. Sc. Books, USA

\bibitem[Phillips et al.(1986)]{Phi86} Phillips, M.~M., 
Jenkins, C.~R., Dopita, M.~A., Sadler, E.~M., \& Binette, L.\ 1986,
\aj, 91, 1062

\bibitem[Quillen et al.(2000)]{Qui00} Quillen, A.~C., Bower, G.~A., 
Stritzinger, M.\ 2000, \apjs, 85, 128

\bibitem[Raimond et al.(1981)]{Rai81} Raimond, E., Faber, 
S.~M., Gallagher, J.~S., \& Knapp, G.~R.\ 1981, \apj, 246, 708

\bibitem[Ravindranath et al.(2001)]{Rav01} Ravindranath, S., Ho,
L.~C., Peng, C.~Y., Filippenko, A.~V., \& Sargent, W.~L.~W.\ 2001,
\aj, 122, 653

\bibitem[Rix et al.(1992)]{Rix92} Rix, H., Franx, M., Fisher, 
D., \& Illingworth, G.\ 1992, \apjl, 400, L5

\bibitem[Roberts et al.(1979)]{Rob79} Roberts, W.~W., 
Huntley, J.~M., \& van Albada, G.~D.\ 1979, \apj, 233, 67

\bibitem[Rubin et al.(1992)]{Rub92} Rubin, V.~C., Graham, 
J.~A., \& Kenney, J.~D.~P.\ 1992, \apjl, 394, L9

\bibitem[Sadler \& Gerhard(1985)]{Sad85} Sadler, E.~M., \& 
Gerhard, O.~E.\ 1985, \mnras, 214, 177 

\bibitem[Sancisi et al.(1984)]{San84} Sancisi, R., van 
Woerden, H., Davies, R.~D., \& Hart, L.\ 1984, \mnras, 210, 497 

\bibitem[Sage \& Wrobel(1989)]{Sag89} Sage, L.~J., \& Wrobel, 
J.~M.\ 1989, \apj, 344, 204

\bibitem[Shields(1992)]{Shi92} Shields, J.~C.\ 1992, \apjl, 399, L27

\bibitem[Sil'chenko(2000)]{Sil00} Sil'chenko, O.~K.\ 2000, \aj, 120,
741

\bibitem[Sil'chenko \& Afanasiev(2004)]{Sil04} Sil'chenko, O.~K., 
\& Afanasiev, V. L.\ 2004, \aj, 127, 2641

\bibitem[Sparks et al.(1989)]{Spa89} Sparks, W.~B., 
Macchetto, F., \& Golombek, D.\ 1989, \apj, 345, 153

\bibitem[Sparks et al.(1993)]{Spa93} Sparks, W.~B., Ford, 
H.~C., \& Kinney, A.~L.\ 1993, \apj, 413, 531 

\bibitem[Sparks et al.(2004)]{Spa04} Sparks, W.~B., Donahue, 
M., Jord{\' a}n, A., Ferrarese, L., \& C{\^ o}t{\' e}, P.\ 2004, \apj, 607, 
294 

\bibitem[Steiman-Cameron \& Durisen(1982)]{Ste82} 
Steiman-Cameron, T.~Y., \& Durisen, R.~H.\ 1982, \apjl, 263, L51 

\bibitem[Storey \& Zeippen(2000)]{Sto00} Storey, P.~J., \& 
Zeippen, C.~J.\ 2000, \mnras, 312, 813 

\bibitem[Thomas et al.(2005)]{Tho05} Thomas, D., Maraston, 
C., Bender, R., \& de Oliveira, C.~M.\ 2005, \apj, 621, 673
 
\bibitem[Tomita et al.(2000)]{Tom00} Tomita, A., Aoki, K., 
Watanabe, M., Takata, T., \& Ichikawa, S.\ 2000, \aj, 120, 123 

\bibitem[Trager et al.(2000)]{Tra00} Trager, S.~C., Faber, 
S.~M., Worthey, G., \& Gonz{\'a}lez, J.~J.\ 2000, \aj, 120, 165

\bibitem[Tran et al.(2001)]{Tra01} Tran, H.~D., Tsvetanov, Z., Ford,
H.~C., Davies, J., Jaffe, W., van den Bosch, F.~C., \& Rest, A.\ 2001,
\aj, 121, 2928

\bibitem[Trinchieri \& Goudfrooij(2002)]{Tri02} Trinchieri, G., \& 
Goudfrooij, P.\ 2002, \aap, 386, 472

\bibitem[Turnbull et al.(1999)]{Tur99} Turnbull, A.~J., Bridges,
T.~J., \& Carter, D.\ 1999, \mnras, 307, 967

\bibitem[van den Bosch et al.(1994)]{vdB94} van den Bosch, 
F.~C., Ferrarese, L., Jaffe, W., Ford, H.~C., \& O'Connell, R.~W.\ 1994, 
\aj, 108, 1579 

\bibitem[van den Bosch et al.(1998)]{vdB98} van den Bosch, 
F.~C., Jaffe, W., \& van der Marel, R.~P.\ 1998, \mnras, 293, 343 

\bibitem[van Dokkum \& Franx(1995)]{vDo95} van Dokkum, P.~G., \&
Franx, M.\ 1995, \aj, 110, 2027

\bibitem[Vazdekis(1999)]{Vaz99} Vazdekis, A.\ 1999, \apj, 513, 224

\bibitem[Veilleux \& Osterbrock(1987)]{Vei87} Veilleux, S., 
\& Osterbrock, D.~E.\ 1987, \apjs, 63, 295 

\bibitem[Wagner, Bender \& Moellenhoff(1988)]{Wag88} Wagner
S.~J., Bender R., Moellenhoff C., 1988, \aap, 195, L5

\bibitem[Wiklind \& Henkel(2001)]{Wik01} Wiklind, T., \& 
Henkel, C.\ 2001, \aap, 375, 797

\bibitem[Worthey et al.(1992)]{Wor92} Worthey, G., Faber, 
S.~M., \& Gonzalez, J.~J.\ 1992, \apj, 398, 69

\bibitem[Yi et al.(2005)]{Yi05} Yi, S.~K., et al.\ 2005, \apjl, 619, L111

\bibitem[Young(2002)]{You02} Young, L.~M.\ 2002, \aj, 124, 788

\bibitem[Zeilinger et al.(1996)]{Zei96} Zeilinger, W.~W., et 
al.\ 1996, \aaps, 120, 257


\end{thebibliography}
\end{document}